\documentclass[twocolumn]{aastex631}
\usepackage{graphicx} % Required for inserting images
\usepackage[T1]{fontenc}
\usepackage{textcomp}
\usepackage{lmodern}
\usepackage{amsmath}
\usepackage{subfigure}
\usepackage{natbib}
\usepackage[]{hyperref}
\usepackage{gensymb}
\usepackage{booktabs}
\usepackage[version=4]{mhchem}

\received{27 September 2024}
\revised{23 January 2025}
\accepted{3 February 2025}

\begin{document}

%\title{Constraining the rotational temperature of methanol in the HD 100546 disk: multiple desorption mechanisms at play}

\title{ALMA reveals thermal and non-thermal desorption of methanol ice in the HD~100546 protoplanetary disk}

\author[0009-0006-1929-3896]{Lucy Evans}
\affiliation{School of Physics and Astronomy, University of Leeds, LS2 9JT, United Kingdom}
\email{l.e.evans@leeds.ac.uk}

\author[0000-0003-2014-2121]{Alice S. Booth}
\affiliation{Center for Astrophysics $\vert$ Harvard $\&$ Smithsonian, 60 Garden St., Cambridge, MA 02138, USA}
%\date{2024}

\author[0000-0001-6078-786X]{Catherine Walsh}
\affiliation{School of Physics and Astronomy, University of Leeds, LS2 9JT, United Kingdom}
\email{c.walsh1@leeds.ac.uk}

\author[0000-0003-1008-1142]{John D. Ilee}
\affiliation{School of Physics and Astronomy, University of Leeds, LS2 9JT, United Kingdom}

%\author{Mikhel Kama}
%\affiliation{Department of Physics and Astronomy, University College London, Gower Street, London, WC1E 6BT, UK}
%\affiliation{Tartu Observatory, University of Tartu, Observatooriumi 1, 61602 T$\tilde{o}$ravere, Tartumaa, Estonia}

\author[0000-0001-5849-577X]{Luke Keyte}
\affiliation{Astronomy Unit, School of Physics and Astronomy, Queen Mary University of London, London E1 4NS, UK}

\author[0000-0003-1413-1776]{Charles J. Law}
\altaffiliation{NASA Hubble Fellowship Program Sagan Fellow}
\affiliation{Department of Astronomy, University of Virginia, Charlottesville, VA 22904, USA}

\author[0000-0003-3674-7512]{Margot Leemker}
\affiliation{Dipartimento di Fisica, Universit\`{a} degli Studi di Milano, Via Celoria 16, 20133 Milano, Italy}

%\author{Hideko Nomura}
%\affiliation{Division of Science, National Astronomical Observatory of Japan, 2-21-1 Osawa, Mitaka, Tokyo 181-8588, Japan}

\author[0000-0003-2493-912X]{Shota Notsu}
\affiliation{Department of Earth and Planetary Science, Graduate School of Science, The University of Tokyo, 7-3-1 Hongo, Bunkyo-ku, Tokyo 113-0033, Japan}
\affiliation{Star and Planet Formation Laboratory, RIKEN Cluster for Pioneering Research, 2-1 Hirosawa, Wako, Saitama 351-0198, Japan}

\author[0000-0001-8798-1347]{Karin \"{O}berg}
\affiliation{Center for Astrophysics $\vert$ Harvard $\&$ Smithsonian, 60 Garden St., Cambridge, MA 02138, USA}

\author[0000-0002-7935-7445]{Milou Temmink}
\affiliation{Leiden Observatory, Leiden University, PO Box 9513, 2300 RA Leiden, the Netherlands}

\author[0000-0003-2458-9756]{Nienke van der Marel}
\affiliation{Leiden Observatory, Leiden University, 2300 RA Leiden, the Netherlands}

%\author[0000-0001-7591-1907]{Ewine F. van Dishoeck}
%\affiliation{Leiden Observatory, Leiden University, 2300 RA Leiden, the Netherlands}
%\affiliation{Max-Planck-Institut f\"{u}r Extraterrestrische Physik, Giessenbachstrasse 1, 85748 Garching, Germany}

%\maketitle

\begin{abstract}
Methanol (\ce{CH3OH}) and formaldehyde (\ce{H2CO}) are chemically coupled organic molecules proposed to act as an intermediate step between simple molecules and more complex prebiotic compounds.
Their abundance distributions across disks regulate the prebiotic potential of material at different disk radii. We
present observations of multiple methanol and formaldehyde transitions toward the Herbig~Ae disk HD~100546 obtained with ALMA, building upon the previous serendipitous detection of methanol in this source.
We find that methanol has a higher rotational temperature ($T_\mathrm{rot}$) than formaldehyde towards both the centrally concentrated emission component in the inner disk ($0-110$~au) and a radially separate dust ring farther out in the disk ($180-260$~au).
$T_\mathrm{rot}$ decreases for methanol and formaldehyde from the inner ($152^{+35}_{-27}$~K and $76^{+9}_{-8}$~K) to the outer disk ($52^{+8}_{-6}$~K and $31^{+2}_{-2}$~K), suggesting that we are tracing two different chemical environments.
$T_\mathrm{rot}$ for both species in the inner disk is consistent with thermal desorption as the origin, while the outer disk reservoir is driven by non-thermal desorption.
The \ce{CH3OH}/\ce{H2CO} column density ratio decreases from $14.6^{+5.2}_{-4.6}$ in the inner disk to $1.3^{+0.3}_{-0.2}$ in the outer disk, consistent with modelling predictions.
%Hence, in the two chemically and thermally distinct reservoirs of organic molecules that exist in this disk, one is driven by thermal desorption in the inner disk, the other by non-thermal desorption in the outer disk.
% and that is also consistent with the location of an outer dust ring
The \ce{CH3OH}/\ce{H2CO} column density ratio for the inner disk is consistent with the median value in the range of column density ratios compiled from Solar System comets which would have formed at a similar distance. This supports the notion that interstellar ice is inherited and preserved by protoplanetary disks around solar-mass and intermediate-mass stars as we are seeing 'fresh' ice sublimation, as well as providing more evidence for the presence of prebiotic precursor molecules in planet-forming regions.
\end{abstract}
\section{Introduction}
\label{introduction}

Complex organic molecules (COMs) are defined as carbon-containing molecules with at least six atoms \citep{Herbst2009}.
As such, these molecules are seen as a bridge between simple molecules and more complex, prebiotic molecules that are vital for the development of life as we know it \citep[e.g.,][]{Meinert2016}.
This is particularly relevant for the astrochemical study of star formation and many COMs have been detected towards objects at various stages of this process \citep[e.g.,][]{Herbst2009,Jorgensen20,Scibelli2020,Ceccarelli23}.
In addition to these observations, laboratory experiments have revealed the icy origin of many COM species.
The simplest COM, methanol (\ce{CH3OH}), is known to form in star-forming molecular clouds when dust grains are coated with ice \citep[e.g.,][]{Hiraoka94}.
When CO is frozen out onto the dust grains (typically at temperatures $\lesssim 20$~K), it undergoes hydrogenation (shown in Eqs.~\ref{eq.formaldehydeform} and \ref{eq.methanolform}), forming formaldehyde (\ce{H2CO}) in an intermediate reaction before continuing to form \ce{CH3OH} \citep{Hiraoka94,Watanabe2002,Fuchs2009}.
\begin{equation}
\ce{CO ->[H] HCO ->[H] H2CO}
\label{eq.formaldehydeform}
\end{equation}
\begin{equation}
\ce{H2CO ->[H] CH3O/CH2OH ->[H] CH3OH}
\label{eq.methanolform}
\end{equation}

Hydrogen abstraction reactions are also important for controlling the rate of \ce{H2CO} and \ce{CH3OH} ice formation \citep[e.g.,][]{Chuang16}.
A combination of hydrogenation and abstraction reactions can also lead to increased chemical complexity in the ice with only CO and H as the primary ingredients \citep[e.g.,][]{Fedoseev17} and without the need for energetic processing \citep[e.g.,][]{%Bennett07,
Oberg09,Chen13}. For example, methyl formate is a more complex COM that can form via hydrogenation \citep[]{Chuang17,Simons2020}.
It is now known that chemical complexity can already build in interstellar ices in advance of the formation of stars and their surrounding planetary systems (see \citet{McClure23}, \citet{Rocha2024} and \citet{Chen2024} for recent JWST results), although it remains an open question how much of this complexity is inherited and retained in the protoplanetary disk.

Evidence that planet formation is already likely underway at the protoplanetary disk (Class II) stage \citep[e.g.,][]{ALMA2015} has provided even stronger motivation for the detection of COMs in these objects \citep[see also][]{Ligterink2024}. 
Despite clear progress in detecting and characterising the complex inventory towards less evolved objects, it remains challenging to do so for planet-forming disks around young stars because their small angular size (typically less than a few arcseconds for the nearest star-forming regions) means that the ice sublimation zone is only a few percent of an arcsecond and thus it is difficult to resolve.
However, with the recent advent of facilities with sufficiently high spatial and spectral resolution, most notably the Atacama Large Millimeter/submillimeter Array (ALMA), the characterisation of the COM inventory in protoplanetary disks is now possible.
The simplest COM, \ce{CH3OH}, was first detected in the gas phase with ALMA towards the disk surrounding the nearby T Tauri star, TW Hya \citep{Walsh2016}.
Despite the confirmation of the presence of \ce{CH3OH} in a protoplanetary disk, subsequent observations failed to detect \ce{CH3OH} in targeted studies of the disks surrounding MWC~480 and LkCa15 \citep{Loomis2020,Yamato2024}, as well as HD~163296 \citep{Carney2019}, although dimethyl ether (\ce{CH3OCH3}) has been detected in MWC~480 \citep{Yamato2024}.
Gas-phase \ce{H2CO}, on the other hand, is (almost) ubiquitously observed towards both T Tauri and Herbig disks \citep[e.g.][]{Dutrey97,Thi04,Oberg11,Guilloteau13,Garufi21,Guzman21,Rampinelli2024}, with tentatively lower column densities measured towards the latter \citep{Pegues2020}.
This is due to \ce{H2CO} also having efficient gas-phase formation pathways, primarily the neutral-neutral reaction involving \ce{CH3} and O \citep[see Eq. \ref{eq.gasphaseh2co};][]{Fockenberg2002,Semenov2011,Loomis15}, which are not affected by dust temperature and thus proceed regardless of whether or not CO is frozen out.
\begin{equation}
\ce{CH_3 + O -> H_2CO + H}
    \label{eq.gasphaseh2co}
\end{equation}
With Herbig stars having typical luminosities up to four orders of magnitude higher than T Tauri stars \citep{Wichittanakom2020}, the temperature of the disk surrounding these stars was considered too warm for the in-situ formation of \ce{CH3OH} ice because of the inability of CO ice to freeze out in the warm ($\gtrsim 20$~K) midplanes of Herbig disks.

Given the proposed difficulty for in-situ \ce{CH3OH} ice formation in Herbig disks, the detection of gas-phase \ce{CH3OH} in the protoplanetary disk surrounding the Herbig Ae star, HD~100546, was unexpected and indicates inheritance of the \ce{CH3OH} ice reservoir from an earlier evolutionary stage \citep{Booth2021}.
Moreover, two distinct emission components were detected, one being centrally concentrated with the other coincident with the previously tentatively detected dust ring at approximately 220 au. This gives rise to the possibility that two distinct desorption mechanisms are at play in this disk, a consequence of the radial temperature gradient.

It is now known that disks around Herbig stars offer a unique insight into the composition of the ice reservoir.
Gas-phase \ce{CH3OH} has since been successfully detected towards HD~169142 and IRS~48.
Similar to that found for HD~100546, it is thought that \ce{CH3OH} (along with SO) is tracing oxygen-rich gas in these disks arising due to sublimation of the ice reservoir \citep{vanderMarel2021,Booth2021,Booth2023}. As well as HD~100546, the ices in the disks surrounding IRS~48 \citep[]{vanderMarel2021} and HD~169142 \citep[]{Booth2023} show evidence for inheritance, meaning that the ices formed prior to the current evolutionary stage. As a result of this earlier formation, the ices have undergone radial transport towards the dust trap location, where they are desorbed due to the warm nature of the dust traps in these particular disks. In other disks that have dust traps (regardless of whether they are T Tauri or Herbig) but where \ce{CH3OH} remains undetected, it is suggested that this is due to the dust trap being colder, therefore the desorption of \ce{CH3OH} and other organics does not take place \citep[see also][]{Temmink2023,vanderMarel2023}.

A more complete chemical inventory of both HD~100546 and IRS~48 has since been reported \citep{Booth2024HD,Booth2024IRS}.
In these works, 19 molecules have been detected towards HD~100546 and 16 towards IRS~48, including several higher-complexity COMs: \ce{CH3OCHO} towards both disks and \ce{CH3OCH3} and \ce{c-H2COCH2} towards IRS~48 \citep[see also][]{Brunken2022,Leemker2023}.
However, these works do not determine the rotational temperature of \ce{CH3OH} nor its predecessor, \ce{H2CO}, towards HD~100546.
The particular desorption mechanisms at play and their potential connection to an inherited origin of \ce{CH3OH} in this source, in which two separate emission components have been clearly identified, have not yet been investigated.
The rotational temperature of \ce{CH3OH} is important to constrain in these two emission components as it may reflect the particular desorption mechanism(s) at play across different disk radii.
%Therefore, to investigate this, multiple transitions of each molecule covering a sufficient range of excitation conditions need to be observed at a sufficient angular resolution so as to determine separate results for the separate emission components.

In this work, we present a detailed rotational diagram analysis of multiple molecular lines of \ce{H2CO} and \ce{CH3OH} detected towards the disk around HD~100546, observed during ALMA Cycles 7 and 8
% (for preliminary results from both of these datasets see \citealt{Booth2021} and \citealt{Booth2024HD}, respectively)
, with comparisons included between our results for HD~100546 and other objects across the various stages of the star formation process.
\ce{CH3OH} can only form efficiently on the grain surface and is subsequently desorbed into the gas phase; however, as previously mentioned, \ce{H2CO}, an intermediary of \ce{CH3OH} (refer to Eqs. \ref{eq.formaldehydeform} and \ref{eq.methanolform}), also has efficient gas-phase formation mechanisms \citep[][]{Fockenberg2002}. %in addition to the ice desorption pathway \citep[see e.g.,][]{Loomis15}.
\ce{H2CO} is therefore an important inclusion as complementary observations allow a more tight constraint on the formation and/or destruction mechanisms at play in this source.
%In total, we include ten unblended lines of \ce{CH3OH}, along with four lines of its precursor molecule, \ce{H2CO} (one of which is a combination of two blended lines). 
%We use a rotational diagram analysis to constrain the temperature of the two distinct \ce{CH3OH} and \ce{H2CO} emission components to discern the particular desorption mechanism(s) at play in setting the location and abundance of the observed gas-phase reservoir.
%We determine the rotational temperatures ($T_\mathrm{rot}$) and column densities ($N_\mathrm{T}$) for both species across the disk.
%We compare $T_\mathrm{rot}$ for methanol to that found for younger sources and towards the disk around TW~Hya, along with comparing the column density ratio for \ce{CH3OH}/\ce{H2CO} with that found towards younger objects and comets in the solar system.
Previous research has shown that lower \ce{CH3OH}/\ce{H2CO} ratios may be associated with chemical processing of accreting material as it is delivered to disks \citep{Podio2020}, can be a tracer of non-thermally desorbed material \citep{vantHoff2020} and also provide an indication of efficient gas-phase formation of \ce{H2CO} in the outer disk \citep[e.g.,][]{Loomis15,Carney17,TerwisschavanScheltinga2021,Temmink2023,HernandezVera2024}.

The paper is structured as follows.
We describe the source (HD~100546) and the Cycle 7 and 8 ALMA observations in Section~\ref{Observations} and the emission morphology of lines detected in these data in Section \ref{Emission_morphology}.
We explain the methodology used in the rotational diagram analysis in Section~\ref{Methodology} and present the results in Section \ref{Results}, which are subsequently discussed and compared with modelling predictions in Section \ref{Discussion}.
Finally, we state our conclusions in Section \ref{Conclusions}.

\section{Observations}
\label{Observations}

\subsection{The Disk Around HD~100546}
\label{source}

HD 100546 (R.A. (J2000)~$=$~11h33m25.3s, Decl. (J2000)~$=$~-70d11m41.2s) is a nearby ($110\pm1$ pc) Herbig Ae star of intermediate age ($4.8^{+1.8}_{-0.2}$~Myr), with a mass of $2.18^{+0.02}_{-0.17}$~M$_\odot$ and a stellar effective temperature of 10,000 K, that is surrounded by a warm, gas-rich disk \citep{Fairlamb2015,Wichittanakom2020}.
The position angle and inclination of the disk are 146$\degree$ and 44$\degree$, respectively \citep{Walsh2014}.
Previous continuum observations detected two dust rings centered at approximately $20-40$~au and $180-220$~au separated by a wide dust gap \citep{Pineda2019,Fedele2021}, supporting the evidence for two possible giant planets embedded in the disk at approximately 10 au and 60 au \citep{Quanz2013,Walsh2014,Brittain2014,Pinilla15}.
The warm, thermally-heated edge of the inner dust cavity is proposed to be the origin of compact SO, \ce{H2CO}, \ce{CH3OH} and \ce{CH3OCHO} emission observed with ALMA \citep{Booth2018,Booth2021,Booth2023a}.

The proximity of HD~100546 (see e.g., W\"{o}lfer et al. subm.; refer also to \citealt{Lindegren2016}) has facilitated many dedicated studies, leading to the detection of multiple simple species, including SO, O, OH, \ce{C+}, \ce{C}, \ce{H2O}, \ce{HCO+}, CS and multiple isotopologues of CO \citep{Fedele13,Wright2015,Kama16,Booth2018,Miley2019,Pirovano22,Keyte2023}.
This has allowed some constraints on the disk structure and global C/O ratio; \citet{Kama16} found this to be similar to solar, meaning that the disk gas is only moderately depleted in heavy elements such as C and O.
Recent thermochemical modelling supports evidence for radial C/O variations across the disk shaped by the sculpted dust rings.
A value for C/O of 0.8 inwards of the dust rings rising to 1.2 beyond the rings better reproduces the trends in the observations \citep[][]{Leemker24}.
A more recent chemical inventory reported by \citet{Booth2024HD} revealed that the disk has significant molecular substructure, with rings of HCN, CN, \ce{C2H}, CS, \ce{H2CS}, \ce{H2CO} and \ce{CH3OH} peaking in emission just beyond the location of the two dust rings, with oxygen-bearing species peaking both within the inner cavity and at or beyond the outer dust ring.

In summary, HD~100546 is a good target for this study as it is a nearby, well-studied disk with multiple simple organic and COM detections enabling the detailed excitation analysis of \ce{CH3OH} and \ce{H2CO}.
\subsection{ALMA Observations}
\label{almabos}

We use ALMA observations obtained during Cycle 8 targeting multiple lines of \ce{CH3OH} in Band 7 with baselines ranging from 15~m to 3000~m (2021.1.00738.S with A.~S.~Booth as PI).
The imaging of these data are already described in \citet{Booth2024HD} and we limit the description here to necessary details only.
All lines were imaged using the \texttt{tclean} task provided by the Common Astronomy Software Application (CASA)\footnote{\url{https://casa.nrao.edu/}; \citep{CASA2022}}, version 5.6.
We used Briggs weighting \citep{Briggs95} and a multiscale deconvolver with scale sizes of 0, 5, 10 and 20 and applied a uv taper of 0\farcs4 while cleaning down to a threshold of 3$\sigma$, where $\sigma$ was taken as the rms (root mean squared) measured using CASA in emission-free areas.
We applied a Keplerian mask using a Python script (first published in \citealt{Teague2020}) during imaging which had a radial extent of 260 au (the radial extent of the \ce{H2CO} emission) and a line width set to 200~m~s$^{-1}$ at 1\farcs0 varying radially with an exponent of $-$0.5.
We assumed a position angle of 146$\degree$, a disk inclination of 44$\degree$ \citep[]{Walsh2014} and distance of 110 pc \citep[]{Wichittanakom2020}.

The data were imaged using a Briggs robust parameter of 2.0 for the weaker lines to boost the signal-to-noise ratio (S/N) in the channel maps as measured using the rms value calculated from the emission-free channels of each spectral window (spw) as given by CASA.
The \ce{H2CO} line detected in the Cycle 8 data was strong enough to permit imaging with a robust parameter of 0.5 in order to improve spatial resolution. %The data were imaged using a range of robust parameters ranging from 0.5 to 2.0. 
%The stronger \ce{H2CO} line was well detected in the Cycle 8 images across this range; however, for the weaker \ce{H2CO} lines in the Cycle 7 data (see below), as well as all \ce{CH3OH} lines, we chose to use the images generated with a robust parameter of 2.0 to maximise the signal-to-noise ratio (S/N) in the channel maps. 
The resulting channel maps have a nominal beam size of 0\farcs4$\times$0\farcs3, equivalent to a physical scale of $33-44$~au, with a native spectral resolution of 0.9 km~s$^{-1}$.

We detect ten well-separated lines of \ce{CH3OH} at a S/N of at least three (according to peak intensity in the channel maps) in this dataset, as well as one transition of \ce{o-H2CO}. Table~\ref{table.trans} lists the %frequency (GHz), quantum numbers, upper and lower level energies ($E_\mathrm{u}$ and $E_\mathrm{l}$; K), the upper level degeneracy ($g_\mathrm{u}$) and Einstein A coefficient ($A_\mathrm{ul}$; s$^{-1}$) for each transition used in further analysis. 
%All 
detected lines used in this analysis along with their spectroscopic data as listed in %that provided in 
the Cologne Database for Molecular Spectroscopy (CDMS\footnote{\url{https://cdms.astro.uni-koeln.de/cdms/portal/}; \citet{Muller01}, \citet{Muller05} and \citet{Endres16}}), along with the imaging parameters for each spectral cube.
For \ce{CH3OH} the spectroscopic data in CDMS are from \citet{Xu2008}, whereas that for \ce{H2CO} are from \citet{Muller2017}.
Additional \ce{CH3OH} lines were detected in the dataset; however, many of these, particularly around 338~GHz, are affected by blending, either with other \ce{CH3OH} lines or with lines from other species; hence, we use only the lines with profiles that are clearly separated from other lines for our rotational diagram analysis. The upper energies ($E_\mathrm{u}$) of our selected lines of \ce{CH3OH} span from 16 to 260~K enabling the construction of a rotational diagram to determine the rotational temperature of gas-phase methanol.

We supplement our Cycle 8 data with ALMA Cycle 7 observations of \ce{o-H2CO} and \ce{p-H2CO} towards HD~100546 with baselines ranging from 15~m to 2517~m (program 2019.1.00193.S with A.~S.~Booth as PI).
These data are also described in \citet{Booth2021} and \citet{Booth2023a}.
The data were imaged in the same manner as described above and the resulting channel maps have a nominal beamsize of 0\farcs3$\times$0\farcs2, which is equivalent to a physical scale of $22-33$~au, with a spectral resolution of 0.25~km~s$^{-1}$.
To better match the image resolution of the Cycle 8 data, we convolve the Cycle 7 channel maps using \texttt{imsmooth} with a beam of 0\farcs4.
The JvM correction \citep{Jorsater95} was applied to all images produced from the Cycle 7 data; epsilon was calculated as 0.3. \citep[see][for further information]{Czekala2021}. No JvM correction was applied to the images produced from the Cycle 8 data because a single configuration only was used for these data.

Five \ce{H2CO} lines were detected in the Cycle 7 dataset at a S/N of at least three and the parameters for these are also listed in Table \ref{table.trans}.
Note that two of the \ce{H2CO} lines detected in the Cycle 7 observations at approximately 291.38 GHz are blended with each other.
We chose to keep these lines in our analysis to include a higher energy line ($E_\mathrm{u} = 141$~K) in the generated rotational diagram, which puts tighter constraints on our derived values for rotational temperature and column density.
%We explain how we deal with the line blending in Section~\ref{Methodology}.

\begin{table*}[hbt!]
    \caption{%Transition parameters for all 
    \ce{H2CO} and \ce{CH3OH} lines used in the rotational diagram analysis}% (see text for references).
    
    \begin{tabular}{c c c c c c c c c}
        \hline
         Molecule & Transition & Frequency & $E_\mathrm{u}$ & $g_\mathrm{u}$ & A$_\mathrm{ul}$ & Beam & Robust & Peak S/N$^{(a)}$ \\
         & & [GHz] & [K] & & [s$^{-1}$] & & & \\
         \hline
         \hline
         \multicolumn{9}{c}{2019.1.00193.S} \\
         \hline
         & & & & & & & & \\
         p-H$_2$CO & 4$_{2,3}$-3$_{2,2}$ & 291.238 & 82.1 & 9 & 5.21$\times$10$^{-4}$ & 0\farcs33$\times$0\farcs29 (28$\degree$) & 2.0 & 18 \\
         p-H$_2$CO & 4$_{2,2}$-3$_{2,1}$ & 291.948 & 82.1 & 9 & 5.25$\times$10$^{-4}$ & 0\farcs33$\times$0\farcs29 (28$\degree$) & 2.0 & 9 \\
         & & & & & & & & \\
         o-H$_2$CO$^{(b)}$ & 4$_{3,2}$-3$_{3,1}$ & 291.380 & 140.9 & 27 & 3.04$\times$10$^{-4}$ & 0\farcs33$\times$0\farcs29 (28$\degree$) & 2.0 & 4 \\ 
         o-H$_2$CO$^{(b)}$ & 4$_{3,1}$-3$_{3,0}$ & 291.384 & 140.9 & 27 & 3.04$\times$10$^{-4}$ & 0\farcs33$\times$0\farcs29 (27$\degree$) & 2.0 & 8 \\
         %CH$_3$OH-A & 6$_{2,4}$-5$_{2,3}$ & 290.264 & 86.5 & 9.50$\times$10$^{-5}$ & & \\
         %& & & & & & & \\
         %CH$_3$OH-E & 6$_{2,4}$-5$_{2,3}$ & 290.307 & 69.8 & 1.06$\times$10$^{-4}$ & & \\
         %CH$_3$OH-E & 6$_{1,5}$-5$_{1,4}$ & 290.249 & 71.0 & 38.8 & 52 & 9.35$\times$10$^{-5}$ & \\
         %& 6$_{2,5}$-5$_{2,4}$ & 290.308 & 74.7 & 9.46$\times$10$^{-5}$ & & \\
         & & & & & & & & \\
         \hline
         \multicolumn{9}{c}{2021.1.00738.S} \\
         \hline
         & & & & & & & & \\
         o-H$_2$CO & 5$_{1,5}$-4$_{1,4}$ & 351.768 & 62.5 & 33 & 1.20$\times$10$^{-3}$ & 0\farcs37$\times$0\farcs30 (27$\degree$) & 0.5 & 57 \\
         & & & & & & & & \\
         CH$_3$OH-A & 1$_{1,1}$-0$_{0,0}$ & 350.905 & 16.1 & 12 & 3.32$\times$10$^{-4}$ & 0\farcs47$\times$0\farcs37 (31$\degree$) & 2.0 & 7 \\
         & 7$_{0,7}$-6$_{0,6}$ & 338.408 & 65.0 & 60 & 1.70$\times$10$^{-4}$ & 0\farcs48$\times$0\farcs39 (32$\degree$) & 2.0 & 10 \\
         & 7$_{2,5}$-6$_{2,4}$ & 338.640 & 102.7 & 60 & 1.58$\times$10$^{-4}$ & 0\farcs48$\times$0\farcs39 (32$\degree$) & 2.0 & 8 \\
         %& 7$_{4,4}$-6$_{4,3}$ & 338.513 & 145.3 & 60.1 & 60 & 1.15$\times$10$^{-4}$ & $^{(b)}$ \\
         %& 7$_{4,3}$-6$_{4,2}$ & 338.513 & 145.3 & 60.1 & 60 & 1.15$\times$10$^{-4}$ & $^{(b)}$ \\
         %& 7$_{5,3}$-6$_{5,2}$ & 338.486 & 202.9 & 129.7 & 60 & 1.20$\times$10$^{-4}$ & $^{(b)}$ \\
         %& 7$_{5,2}$-6$_{5,1}$ & 338.486 & 202.9 & 129.7 & 60 & 1.20$\times$10$^{-4}$ & $^{(b)}$ \\
         & 13$_{0,13}$-12$_{1,12}$ & 355.603 & 211.0 & 108 & 1.27$\times$10$^{-4}$ & 0\farcs45$\times$0\farcs32 (58$\degree$) & 2.0 & 5 \\
         %& 13$_{1,12}$-12$_{0,13}$ & 342.730 & 227.5 & 146.7 & 108 & 4.23$\times$10$^{-4}$ & \\
         & 14$_{1,13}$-14$_{0,14}$ & 349.107 & 260.2 & 116 & 2.20$\times$10$^{-4}$ & 0\farcs47$\times$0\farcs37 (31$\degree$) & 2.0 & 11 \\
         %& 15$_{1,14}$-15$_{0,15}$ & 356.007 & 295.3 & 193.3 & 124 & 2.75$\times$10$^{-3}$ & \\
         & & & & & & & & \\
         %CH$_3$OH-E & 4$_{0,4}$-3$_{1,3}$ & 350.688 & 36.3 & 13.6 & 36 & 8.67$\times$10$^{-5}$ & $^{(b)}$ \\
         CH$_3$OH-E & 7$_{1,7}$-6$_{1,6}$ & 338.345 & 70.6 & 60 & 1.67$\times$10$^{-4}$ & 0\farcs48$\times$0\farcs39 (32$\degree$) & 2.0 & 8 \\
         & 7$_{0,7}$-6$_{0,6}$ & 338.124 & 78.1 & 60 & 1.70$\times$10$^{-4}$ & 0\farcs48$\times$0\farcs39 (32$\degree$) & 2.0 & 9 \\
         & 7$_{1,6}$-6$_{1,5}$ & 338.614 & 86.1 & 60 & 1.71$\times$10$^{-4}$ & 0\farcs48$\times$0\farcs39 (32$\degree$) & 2.0 & 9 \\
         & 7$_{3,5}$-6$_{3,4}$ & 338.583 & 112.7 & 60 & 2.69$\times$10$^{-4}$ & 0\farcs48$\times$0\farcs39 (32$\degree$) & 2.0 & 8 \\
         & 7$_{3,4}$-6$_{3,3}$ & 338.560 & 127.7 & 60 & 1.40$\times$10$^{-4}$ & 0\farcs48$\times$0\farcs39 (32$\degree$) & 2.0 & 6 \\
         %& 7$_{4,4}$-6$_{4,3}$ & 338.504 & 152.9 & 95.0 & 60 & 1.94$\times$10$^{-4}$ & $^{(b)}$ \\
         %& 7$_{4,3}$-6$_{4,2}$ & 338.530 & 161.0 & 100.6 & 60 & 1.90$\times$10$^{-4}$ & $^{(b)}$ \\
         %& 7$_{5,3}$-6$_{5,2}$ & 338.456 & 189.0 & 120.1 & 60 & 1.25$\times$10$^{-4}$ & $^{(b)}$ \\
         %& 7$_{5,2}$-6$_{5,1}$ & 338.475 & 201.1 & 128.5 & 60 & 1.20$\times$10$^{-4}$ & $^{(b)}$ \\
         & & & & & & & & \\
         \hline
         & & & & & & & & \\
    \end{tabular}
    \\$^{(a)}$As estimated from the channel maps.
    \\$^{(b)}$Sufficiently close to one another to %appear in the same spw and to 
    be blended in the inner disk.
    %\\$^{(b)}$Detected but not included in our analysis due to blending.
    \label{table.trans}
\end{table*}

\section{Emission morphology}
\label{Emission_morphology}

\begin{figure*}[hbt!]
    \centering
    \includegraphics[width=0.26\textwidth]{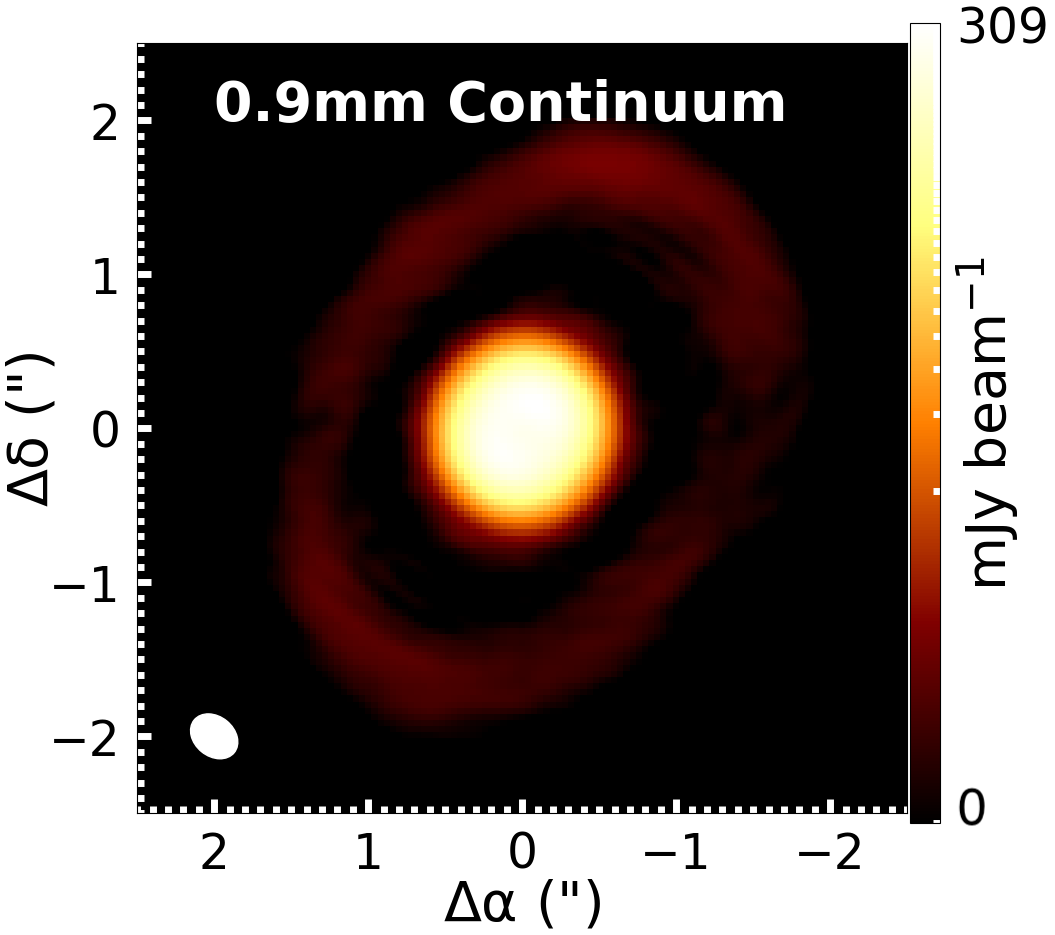}
    \includegraphics[width=0.26\textwidth]{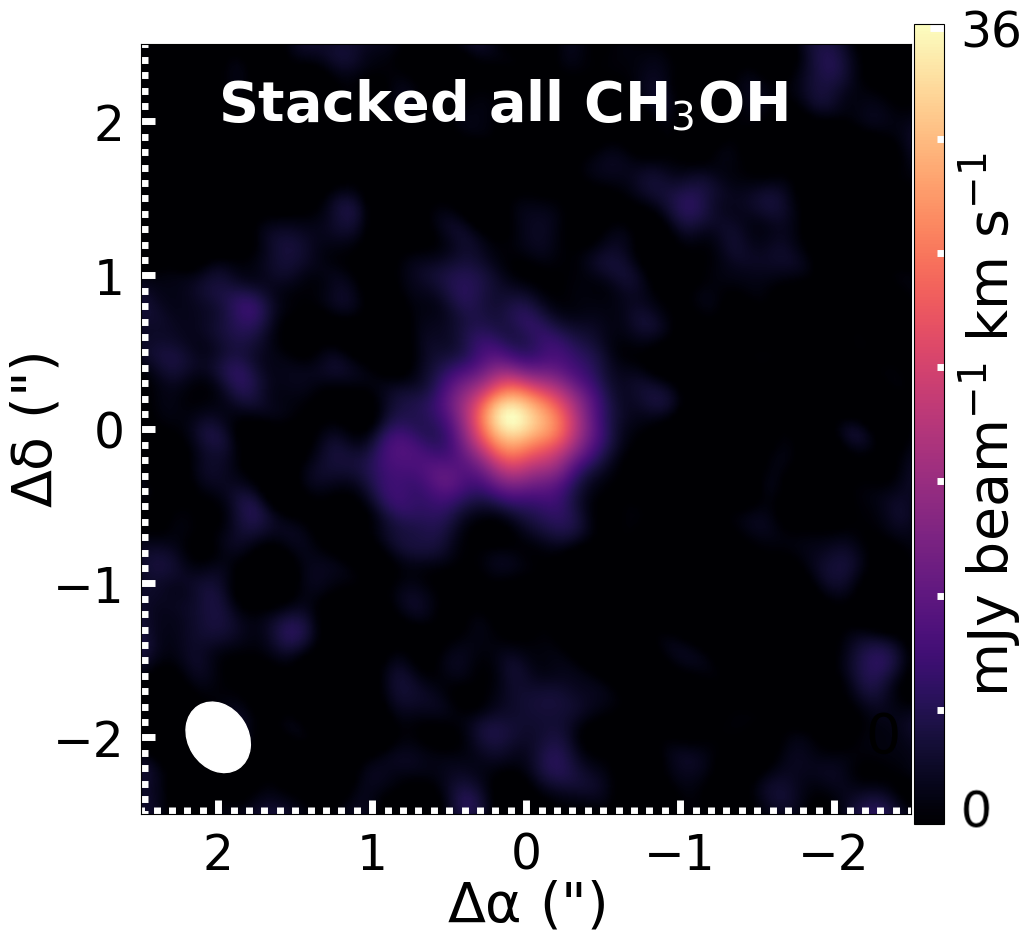}
    \includegraphics[width=0.26\textwidth]{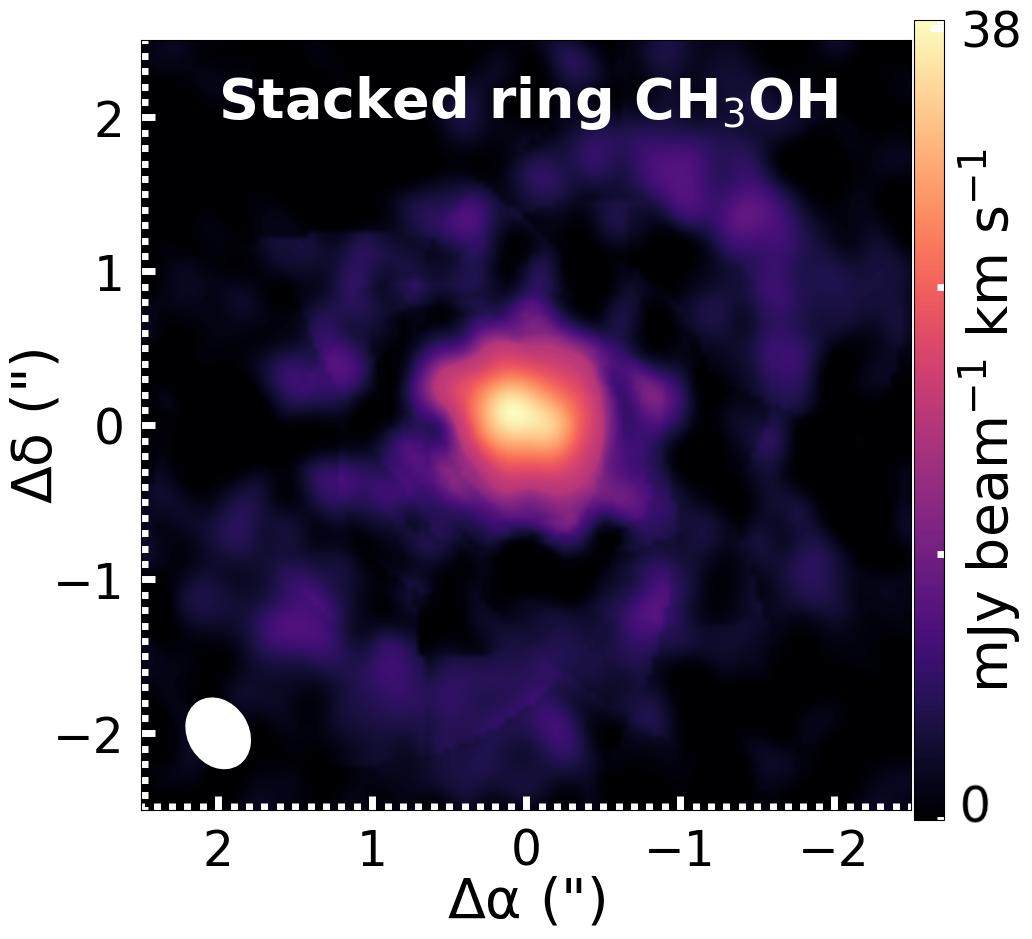}
    \includegraphics[width=0.24\textwidth]{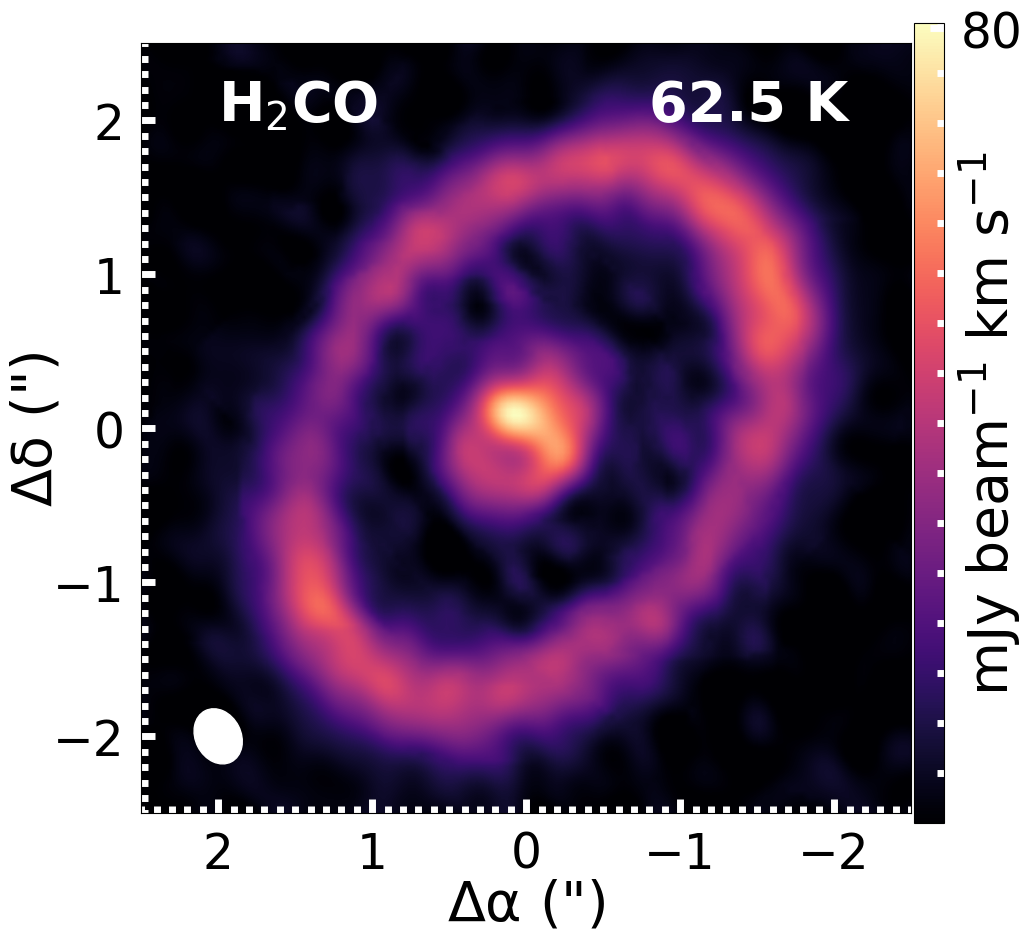}
    \includegraphics[width=0.24\textwidth]{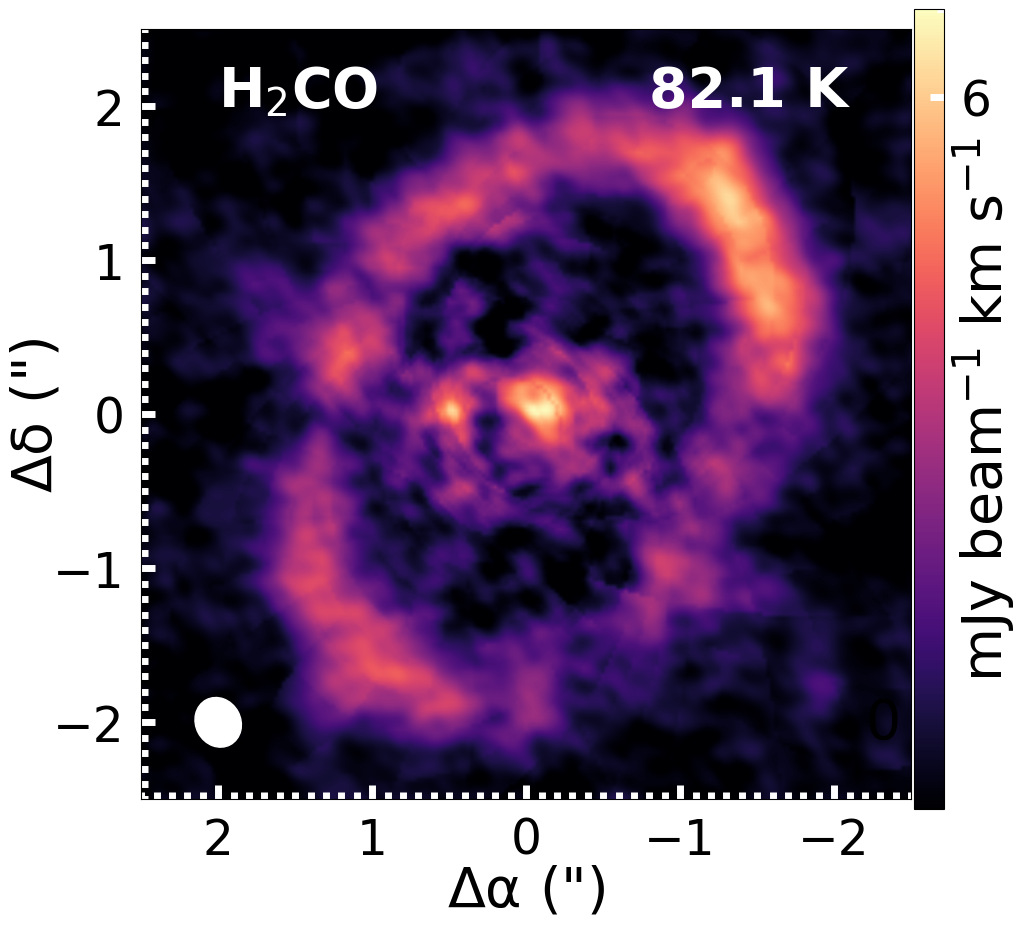}
    \includegraphics[width=0.24\textwidth]{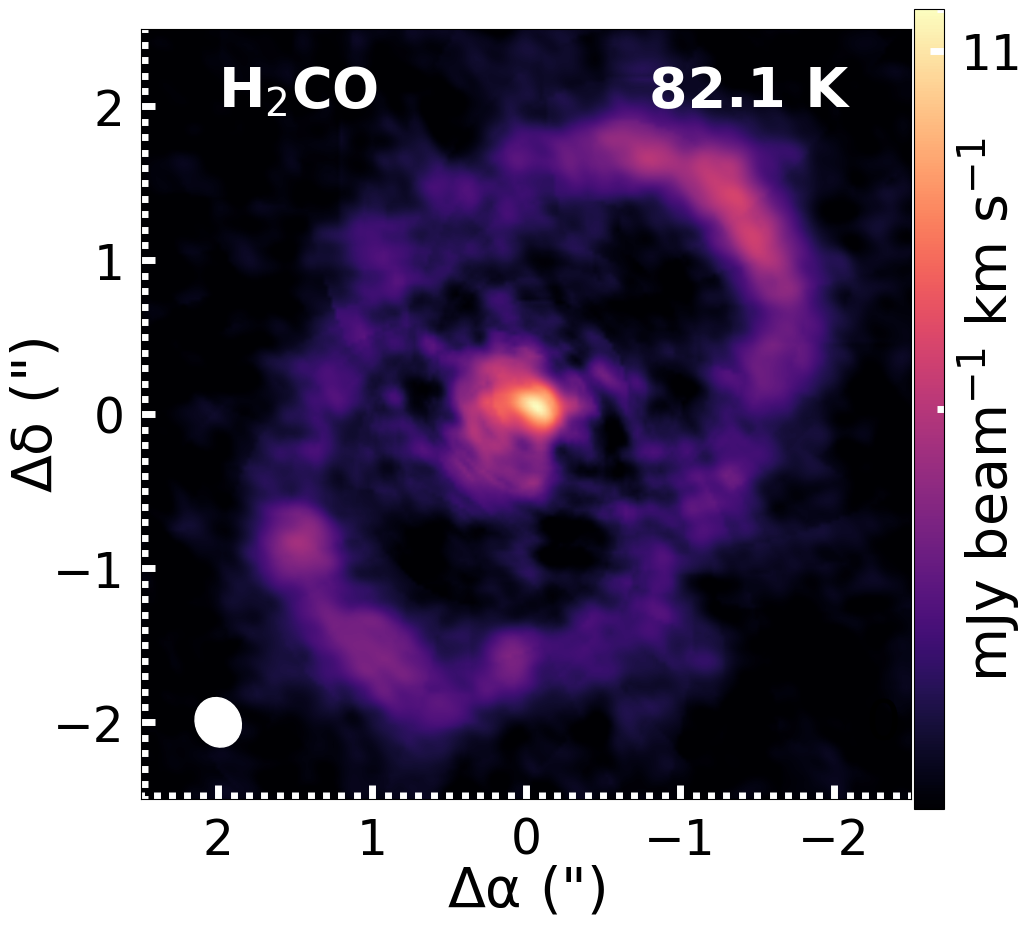}
    \includegraphics[width=0.24\textwidth]{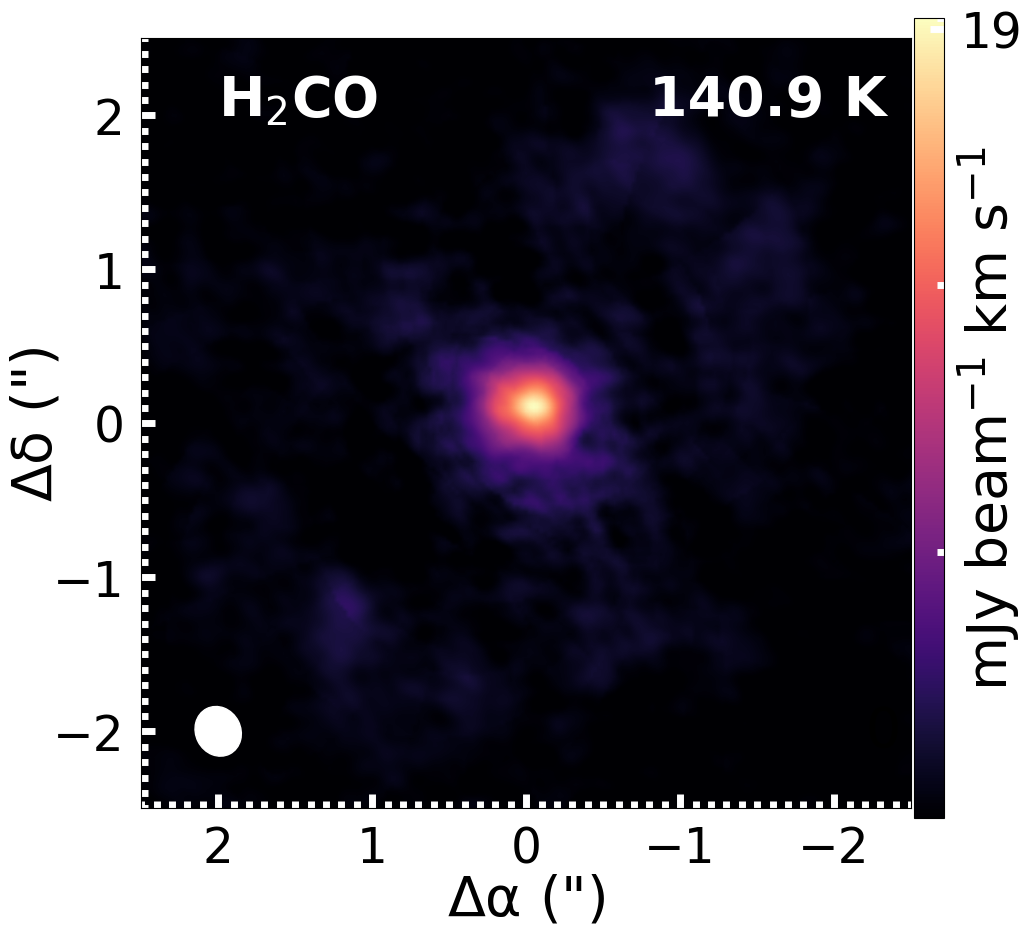}
\caption{Top-left panel: dust continuum at 0.9~mm from the HD~100546 protoplanetary disk first published in \citet{Booth2024HD} and plotted on a log scale to highlight the outer ring of continuum emission.
Top-middle panel: stacked integrated intensity map generated using all ten detected \ce{CH3OH} lines.
Top-right panel: stacked integrated intensity map combining only the four \ce{CH3OH} lines that are detected in the outer ring. These data are presented with a stretched colour scale to improve the visibility of the outer ring of emission.
Bottom row: integrated intensity maps of the four detected \ce{H2CO} lines for the coolest transition (left) to the hottest transition (right). Note that there are two \ce{H2CO} transitions at 82.1 K (4$_{2,3}$-3$_{2,2}$ and 4$_{2,2}$-3$_{2,1}$) and both are presented here.
The synthesised beam is represented by the white ellipse in the bottom left corner of each panel.}
    \label{fig.morph}
\end{figure*}

In Fig.~\ref{fig.morph} we present a selection of integrated intensity or moment 0 maps (generated using the aforementioned Keplerian mask and the \texttt{immoments} task in CASA with no noise clip) to show the emission morphology of the \ce{CH3OH} (top row) and \ce{H2CO} (bottom row) lines detected towards the HD~100546 disk in comparison with the dust continuum at 0.9~mm (top-left panel).
There are two components of molecular emission which follow the ringed dust substructures: one component which is compact and lying within the inner $\sim 100$ au, the other coincident with the previously observed outer dust ring centered at $\sim 220$~au.
Emission from the highest energy \ce{H2CO} line ($E_\mathrm{u} = 140.9$~K; bottom-right panel) is bright only in the central compact component, whereas emission from the lowest energy \ce{H2CO} line ($E_\mathrm{u} = 62.5$~K; bottom-left panel) also reveals emission from the outer ringed component.
The line emission morphologies demonstrate the expected decreasing temperature gradient from the inner to the outer disk.
%Some lines, notably \ce{H2CO} at 82.1 K, show azimuthally asymmetric emission in these maps which is generally coincident with the brightness asymmetry previously observed in the dust ring \citep[]{Fedele2021}; this low temperature emission is further evidence for non-thermal desorption from the outer icy grains.

\begin{figure*}[hbt!]
    \centering
    \includegraphics[width=0.82\textwidth]{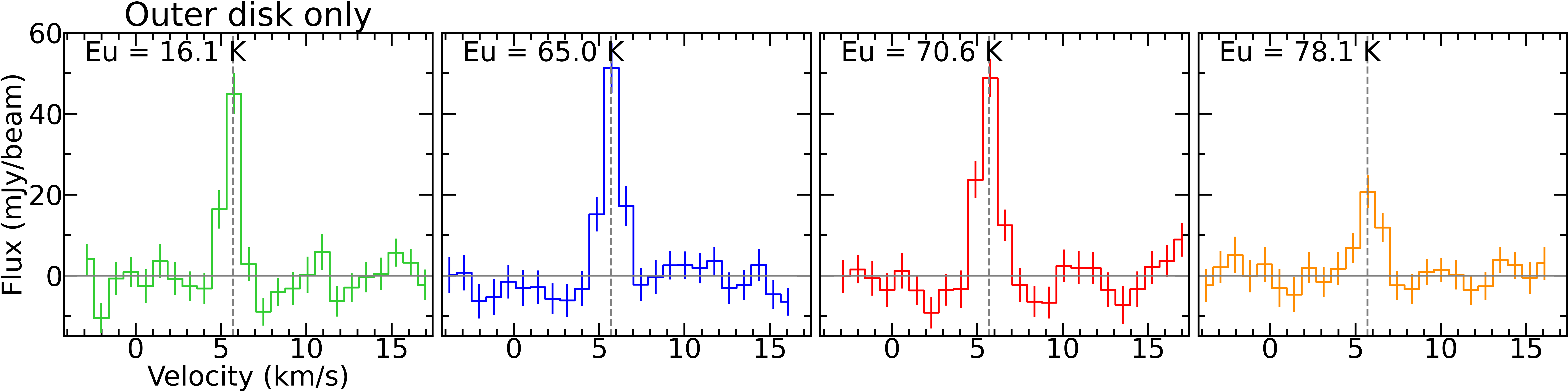}
    \includegraphics[width=\textwidth]{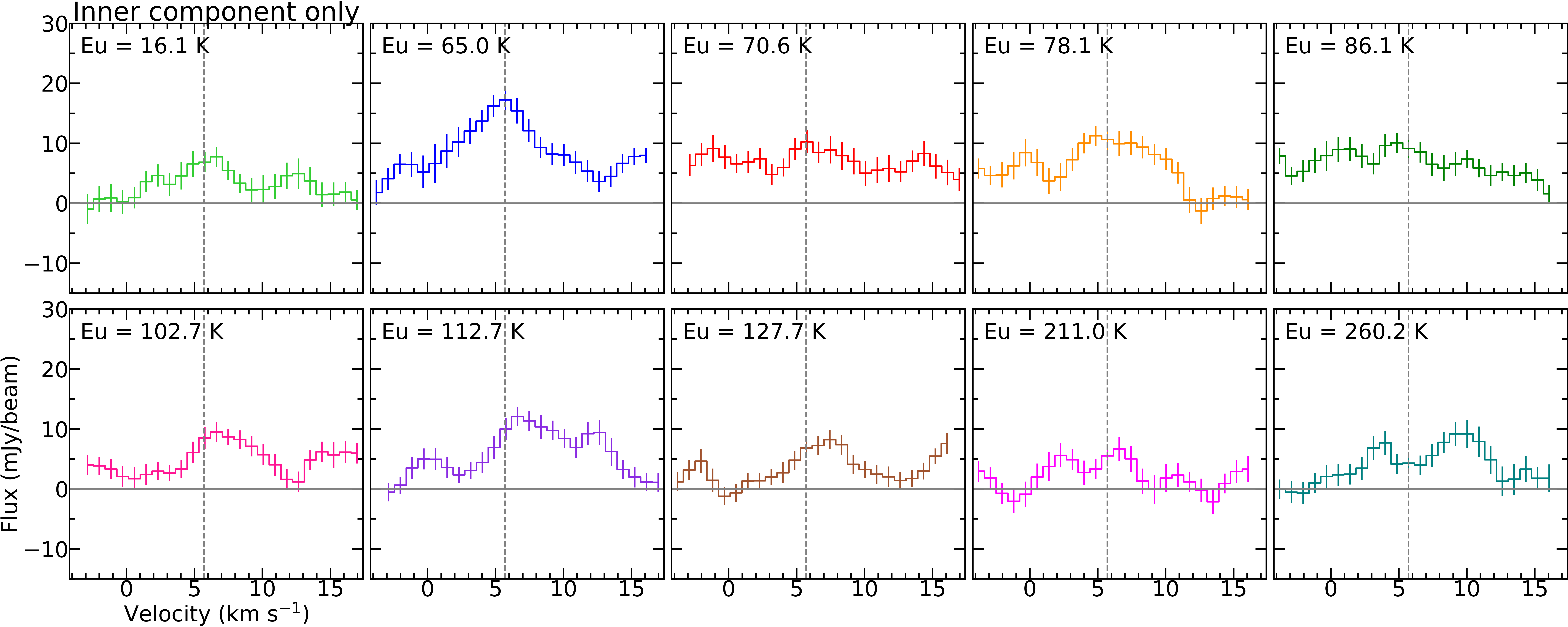}
    \caption{Top row: Spectral line profiles for the four \ce{CH3OH} lines that show %evidence of
    emission from the outer ringed component according to the radial profiles in Fig.~\ref{fig.azavouter}. These spectral lines have been generated by integrating the emission between $180-260$~au. 
    Bottom two rows: Spectral lines extracted for all ten detected \ce{CH3OH} lines generated from integrating the emission within 110~au. 
    The dashed line represents the $\mathrm{v}_\mathrm{LSR}$ of this source (5.7 km~s$^{-1}$).}
    \label{fig.ch3oh_outer_spec}
\end{figure*}

In Fig.~\ref{fig.morph} we also show two stacked \ce{CH3OH} maps, generated using the \texttt{immath} task in CASA, to boost the S/N in the images of the weaker \ce{CH3OH} lines.
The first stacked image (top-middle panel) was created using all ten detected lines (listed in Table~\ref{table.trans}) and better highlights the inner compact component. However, the outer component is faint in this image as not all transitions are excited in the outer ring. 
We checked for extended emission using the spectral stacking technique implemented by the \texttt{gofish} package (\citealt{Teague2019}). This method of azimuthally averaging the data increases the effective S/N. From this analysis, we find that four lines have significant emission peaking in a ring at $\sim$~220 au (see Fig.~\ref{fig.ch3oh_outer_spec}). These four lines also have the lowest upper energy values ($<$80~K) of the ten detected lines. To show the morphology of this ringed emission more clearly, we stack these four images only in the second stacked image (top-right panel). x%was created using the four lines with an upper energy level below approximately 80 K, as only these are well detected in the outer dust ring (as identified in the spectral line profiles shown in Fig.~\ref{fig.ch3oh_outer_spec}).
This map reveals that the gas-phase methanol also traces both the compact inner emission component as well as the one coinciding with the outer mm dust ring.
Separate moment 0 maps for all ten \ce{CH3OH} transitions can be found in Appendix A (Fig. \ref{fig.mom0_cycle8}).

\begin{figure*}[hbt!]
    \includegraphics[width=\textwidth]{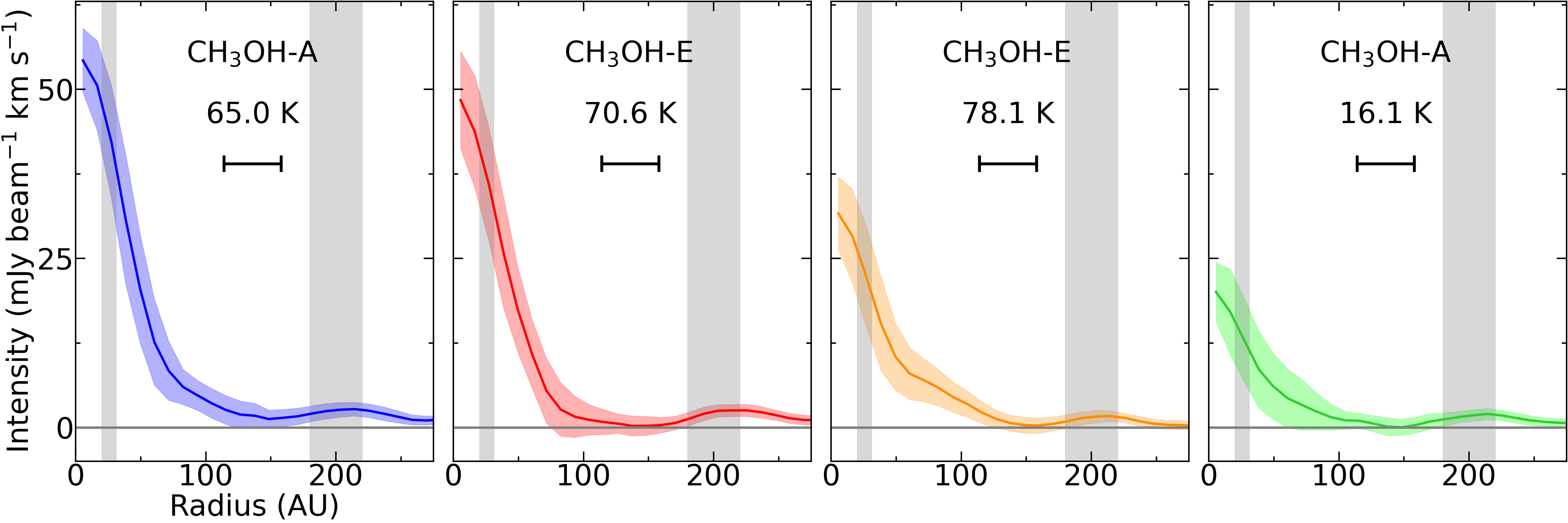}
    \includegraphics[width=\textwidth]{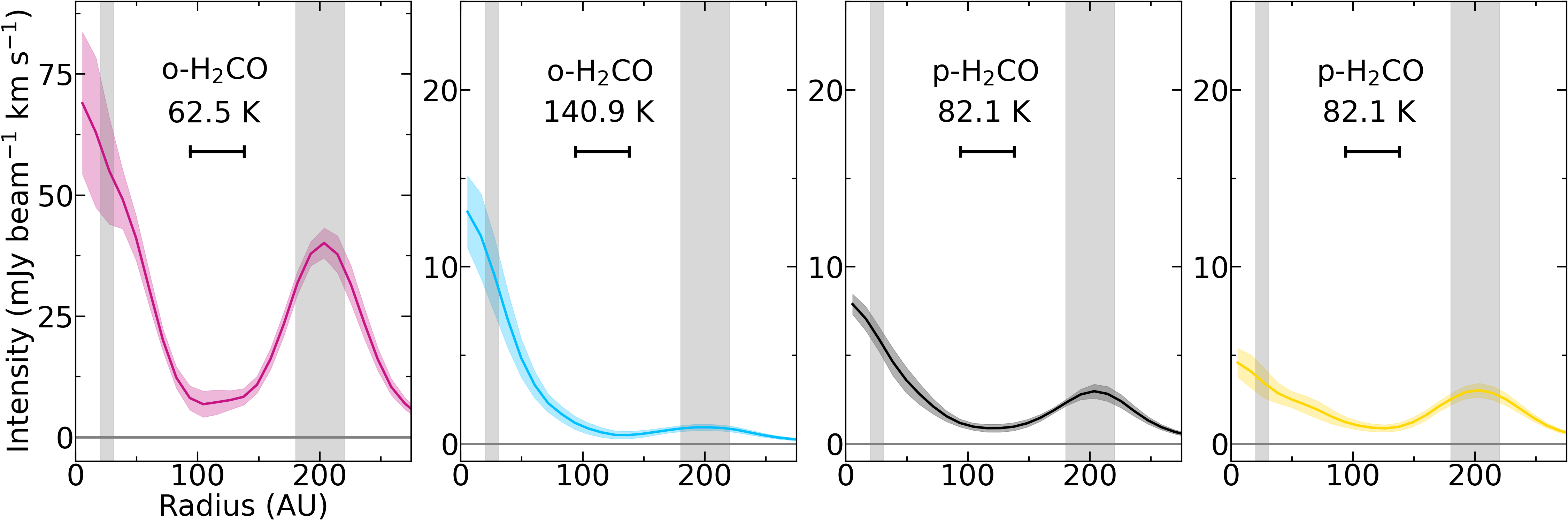}
    \caption{Azimuthally-averaged radial profiles for the four unblended \ce{CH3OH} lines (top row) and all \ce{H2CO} lines (bottom row) in order of decreasing peak intensity. 
    These four \ce{CH3OH} lines are those which show both an inner compact and outer ringed component of emission. 
    The latter coincides with the position of the dust ring, which peaks between approximately $180-220$ au; the radial ranges of the peaks of the two dust rings seen in the 0.9 mm continuum emission between $\sim 20-31$~au and $\sim 180-220$~au are denoted by the grey shaded regions. Note that the o-\ce{H2CO} profile at 231.38 GHz contains contributions from two lines with equal transition properties: one at 291.380 GHz and one at 291.384 GHz. 
    The horizontal bar shows the Full-Width at Half Maximum (FWHM) of the synthesised beam.}
    \label{fig.azavouter}
\end{figure*}

The top panels of Fig.~\ref{fig.azavouter} show the azimuthally-averaged radial profiles, generated by radially binning the emission (with bin size approximately equal to half of the beamsize) and then averaging (see \citealt[]{Booth2024HD}), for the four \ce{CH3OH} lines with detections of both emission components.
These transitions are those which have been stacked in the top-right panel in Fig.~\ref{fig.morph}.
The bottom panels of Fig.~\ref{fig.azavouter} show azimuthally-averaged radial profiles for all detected \ce{H2CO} lines.
The radial profiles for both molecules show the presence of two components of emission: a compact emission component which drops to a small or negligible value beyond approximately 110~au, as well as a ring peaking at around 200~au which is within the range of radii at which the emission peaks depending on semi-major axis projections (see \citealt{Fedele2021}). Appendix B (Fig. \ref{fig.radprof8}) shows the azimuthally averaged radial profiles for all \ce{CH3OH} lines in our sample.

\section{Methodology}
\label{Methodology}

\subsection{Flux Extraction and Error Estimation}

To investigate the two emission components seen in the integrated intensity maps shown in Sect.~\ref{Emission_morphology}, we perform the described rotational diagram analysis for two radial ranges.
For the inner compact emission component we adopt $r$ over the range $0-110$~au, while for the outer ringed emission component we adopt $r$ over the range $180-260$~au; according to Fig.~\ref{fig.azavouter}, in between these radial ranges, the flux from both \ce{H2CO} and \ce{CH3OH} drops to negligible values.
We calculate values of $7\times 10^{-11}$ sr and $2 \times 10^{-10}$ sr, for the solid angles of the inner and outer components, respectively, accounting for the fact that the outer component is an annulus.

Several of the spws of our Cycle 8 data have evidence of line emission from other species (with Keplerian motion evident in the channel maps), or possible multiple velocity components of \ce{CH3OH} due to e.g., a disk wind or outflow.
However, the weak S/N of these lines prevents us from readily distinguishing between these hypotheses and we leave this to future work.
To ensure flux extraction only from the line of interest, centered on the source velocity, we applied a velocity clip when extracting the line fluxes: $-$0.6 km~s$^{-1}$ to 12.0 km~s$^{-1}$ for the inner region and 4.5 to 6.9 km~s$^{-1}$ for the outer region.
For reference, the v$_{LSR}$ of this source is 5.7 km~s$^{-1}$ \citep[]{Walsh2017}. 
We adopt the same velocity ranges for the \ce{H2CO} lines, however, recall that there are two \ce{o-H2CO} lines at approximately 291.38 GHz that are separated only by approximately 5~km~s$^{-1}$.
This separation is sufficient for the lines to be blended in the spectrum extracted from the inner region (see Fig.~\ref{fig.h2co_inner_spec} in Appendix~C), while being well separated in the spectrum extracted from the outer region (see Fig.~\ref{fig.h2co_outer_spec} in Appendix C).
These two \ce{o-H2CO} transitions have identical $A_\mathrm{ul}$, $g_\mathrm{u}$ and $E_\mathrm{u}$ values (see Table~\ref{table.trans}) and so we assume that they contribute equally to the total flux extracted from the inner component using the same velocity clip as for the \ce{CH3OH} emission extracted from the inner region.
In the outer region, on the other hand, %the lines are well separated and of narrower linewidth, as shown in Appendix C (Fig.~\ref{fig.h2co_outer_spec}). 
we adopt velocity clips from 4.4 to 6.7 km~s$^{-1}$ and from 8.3 to 10.0 km~s$^{-1}$, to extract flux from the lines at 291.384 GHz and 291.380 GHz, respectively.

We use \texttt{gofish} to extract the line spectra for all \ce{H2CO} and \ce{CH3OH} lines for both the inner and outer components of emission from the disk.
To calculate the total line flux, we then integrate over the line spectra extracted using \texttt{gofish} using numerical integration.
Due to Keplerian line broadening and beam smearing in the inner disk, we cannot exploit the shift and stack method to extract the spectra here.
To test the accuracy of \texttt{gofish} for spectral (and flux) extraction for the inner unresolved emission, we performed parallel spectral extraction using the \texttt{specflux} tool within CASA, adopting an aperture of 1\farcs0~in radius to cover the radial extent (110~au) of the inner component of emission. Both methods return consistent results (within the uncertainties) thus we proceeded with \texttt{gofish} for all lines in order to use a consistent method.
The bottom panels of Fig.~\ref{fig.ch3oh_outer_spec} show the extracted spectra using \texttt{gofish} for all \ce{CH3OH} transitions observed in the inner disk.

We opt to use conservative estimates for the uncertainties in the line fluxes ($S_\nu\Delta \mathrm{v}$) because the statistical uncertainties provided by \texttt{gofish} (around $2-3$~mJy~km~s$^{-1}$) are likely to be underestimated. To do this we extract the rms noise, $\sigma$, from a line free channel of a non-contaminated spw (centred at 342.730 GHz) using \texttt{specflux} with an elliptical aperture of 4\farcs0~(covering the full disk) and calculate the uncertainty as $\sigma\sqrt{n_\mathrm{chan}}\Delta\mathrm{v}$
where $n_\mathrm{chan}$ is the number of channels integrated over for each spectral line and $\Delta \mathrm{v}$ is the velocity resolution (0.9 km~s$^{-1}$).
Our \ce{H2CO} lines come from two different ALMA datasets and so we also include a flux uncertainty of 10\% to account for any differences in the flux calibration between ALMA datasets.

The fluxes integrated over each emitting area ($S_\nu \Delta \mathrm{v}$) and their estimated uncertainties, along with the column densities in the upper energy level ($N_\mathrm{u}$) as calculated using Eq.~\ref{eq.nu} are listed in Table~\ref{table.rotdiagtrans} for all lines.

\subsection{Rotational Diagram Analysis}\label{Rotdiag_analysis}
As discussed in the previous section, we have detected ten unblended transitions of \ce{CH3OH} spanning $E_\mathrm{u}$ values from 16 to 260~K and four transitions of \ce{H2CO} with $E_\mathrm{u}$ values ranging between 63 and 141~K.
Hence, we can construct a rotational diagram and estimate the rotational temperature and column density for the two components of emission (inner and outer) for both species under the assumption of local thermal equilibrium (LTE) and optically thin emission \citep{Goldsmith99}.
Following the approach of \citet{Loomis2018} and \citet{Ilee2021}, we use Eq.~\ref{eq.nu} to calculate the column density in the upper energy level, $N_\mathrm{u}^\mathrm{thin}$ (cm$^{-2}$),
\begin{equation}
    N_\mathrm{u}^\mathrm{thin}=\frac{4\pi S_\nu\Delta \mathrm{v} }{A_\mathrm{ul}\Omega hc}.
    \label{eq.nu}
\end{equation}
In Eq.~\ref{eq.nu}, $A_\mathrm{ul}$ is the Einstein A coefficient for spontaneous emission (s$^{-1}$), while $h$ and $c$ represent Planck's constant and the speed of light, respectively, both in cgs units. $S_\nu \Delta \mathrm{v}$ represents the integrated flux density (measured in Jy km~s$^{-1}$ and converted to cgs units) and $\Omega$ is the solid angle (steradians) subtended by the emission, where the surface brightness $I_\nu=S_\nu/\Omega$.
The solid angle is calculated using $\Omega=(\pi r^2)/d^2$, where $r$ is the radial extent of the emission and $d$ is the distance to the source, taken as 110 pc \citep{Wichittanakom2020}, both converted into centimetres.
The column density in the upper energy level, $N_\mathrm{u}^\mathrm{thin}$, can then be related to the total column density, $N_\mathrm{T}$, assuming a Boltzmann distribution,
\begin{equation}
    \frac{N_\mathrm{u}^\mathrm{thin}}{g_\mathrm{u}} = \frac{N_\mathrm{T}}{Q(T_\mathrm{rot})} e^{-\frac{E_\mathrm{u}}{k_\mathrm{B}{T_\mathrm{rot}}}},
    \label{eq.boltz}
\end{equation}
where $Q(T_\mathrm{rot})$ is the partition function at the rotational temperature and $k_\mathrm{B}$ is Boltzmann's constant% and all other parameters have previously been defined
.
We use precomputed partition functions as a function of temperature interpolated from CDMS \citep{Muller01,Muller05,Endres16} for both species. Note that this database assumes an ortho-to-para ratio of 3 for \ce{H2CO}.
%For temperatures above those for which partition function values are available on CDMS (500~K and 1000~K for \ce{H2CO} and \ce{CH3OH}, respectively), we use the high temperature approximation, 
%\begin{equation}
%    Q(T_\mathrm{rot})=\frac{\sqrt{\pi}}{\sigma}\sqrt{\frac{(k_\mathrm{B}T_\mathrm{rot})^3}{ABCh^3}},
%    \label{eq.partitionfunction}
%\end{equation}
%where $\sigma$ represents the rotational symmetry number and $A$, $B$ and $C$ represent rotational constants (MHz) of the relevant molecule, which are also obtained from CDMS.
Taking the logarithm of Eq.~\ref{eq.boltz} converts this expression into a linear form,
\begin{equation}
    \ln \frac{N_\mathrm{u}^\mathrm{thin}}{g_\mathrm{u}} = \ln N_\mathrm{T} - \ln Q(T_\mathrm{rot}) - \frac{E_\mathrm{u}}{k_\mathrm{B}T_\mathrm{rot}}
    \label{eq.rotdiag}
\end{equation}
%and a fit to a semi-log plot of the data ($ \ln (N_\mathrm{u}/g_\mathrm{u})$ plotted against $E_\mathrm{u}$) 
allowing extraction of the rotational temperature ($T_\mathrm{rot}$) %(related to the gradient)
and the total column density ($N_\mathrm{T}$)%(related to the intercept)
.
To do this, we use an MCMC fitting procedure using the \texttt{emcee} Python module \citep[]{Foreman2013} with our measured S$_\nu\Delta$v values as input values.
We also test our assumption of optically thin emission by applying an optical depth correction factor,
\begin{equation}
   C_\tau = \frac{\tau}{1 - e^{-\tau}}, 
\end{equation}
where $\tau$ is the line optical depth %(not known \emph{a priori}) 
such that the true column density in the upper energy level is $N_\mathrm{u} = N_\mathrm{u}^\mathrm{thin} C_\tau$.
This modifies the left-hand side of Eq.~\ref{eq.rotdiag} to $\ln N_\mathrm{u}/g_\mathrm{u} + \ln C_\tau$ which is also fit in the MCMC routine. For this routine, our N$_T$ priors are set to values between 10$^6$~cm$^{-2}$ and 10$^{14}$~cm$^{-2}$ for \ce{H2CO} in the outer region, between 10$^6$~cm$^{-2}$ and 10$^{15}$~cm$^{-2}$ for \ce{H2CO} in the inner region and between 10$^6$~cm$^{-2}$ and 10$^{18}$~cm$^{-2}$ for both components of \ce{CH3OH}.
For the inner region, the $T_\mathrm{rot}$ priors are set to values between 15 and 300 K for both molecules, while for the outer region, they are set to values between 5 and 100 K.
We run the MCMC for 1500 steps total, with 1000 removed as burn-in. We run this with 300 walkers in all cases, with the likelihood function built using Eq.~\ref{eq.rotdiag}, which generates posterior probability distributions; the best fit value is chosen as the median of this distribution, while the 16$^\mathrm{th}-$84$^\mathrm{th}$ percentile forms the uncertainty range.

\section{Results}
\label{Results}

\begin{table*}[hbt!]
    \caption{Integrated line fluxes ($S_\nu \Delta \mathrm{v}$), %, extracted using \texttt{gofish}, 
    column density in the upper energy level ($N_\mathrm{u}$) and line opacity ($\tau$) for the inner (0$-$110 au) and outer (180$-$260 au) components of emission.}
    \resizebox{\textwidth}{!}{\begin{tabular}{c c c c c c c c c c}
        & & & & & & & & & \\
        \hline
        Molecule & Transition & Frequency [GHz] & $n_\mathrm{CR}$ [cm$^{-3}$] & \multicolumn{2}{c}{$S_\nu\Delta \mathrm{v}$ [mJy km~s$^{-1}$]} & \multicolumn{2}{c}{$N_\mathrm{u}$ [cm$^{-2}$]} & \multicolumn{2}{c}{$\tau$} \\
         \cmidrule(lr){5-6}\cmidrule(lr){7-8} \cmidrule(lr){9-10}
        & & & & Inner & Outer & Inner & Outer & Inner & Outer \\
        \hline
        & & & & & & & & & \\
        \ce{p-H2CO}$^{(a)}$ & 4$_{2,3}$-3$_{2,2}$ & 291.238 & 6.2$\times$10$^6$ & 59 $\pm$ 6 & 94 $\pm$ 9 & (9.8 $\pm$ 1.0)$\times$10$^{10}$ & (5.3 $\pm$ 0.5)$\times$10$^{10}$ & 0.008 & 0.03 \\
        & 4$_{2,2}$-3$_{2,1}$ & 291.948 & 6.3$\times$10$^6$ & 46 $\pm$ 5 & 121 $\pm$ 12 & (7.7 $\pm$ 0.8)$\times$10$^{10}$ & (6.9 $\pm$ 0.7)$\times$10$^{10}$ & 0.008 & 0.03 \\
        & & & & & & & & & \\
        \ce{o-H2CO}$^{(a)}$ & 4$_{3,2}$-3$_{3,0}$ & 291.380 & 7.4$\times$10$^6$ & 38 $\pm$ 4$^{(b)}$ & 26 $\pm$ 3$^{(c)}$ & (1.1 $\pm$ 0.1)$\times$10$^{11}$ & (2.5 $\pm$ 0.3)$\times$10$^{10}$ & 0.006 & 0.007 \\
        & 4$_{3,1}$-3$_{3,0}$ & 291.384 & 7.4$\times$10$^6$ & 38 $\pm$ 4$^{(b)}$ & 31 $\pm$ 3 & (1.1 $\pm$ 0.1)$\times$10$^{11}$ & (3.0 $\pm$ 0.3)$\times$10$^{10}$ & 0.006 & 0.007 \\
        & 5$_{1,5}$-4$_{1,4}$ & 351.768 & 2.4$\times$10$^6$ & 498 $\pm$ 50 & 1991 $\pm$ 199 & (3.6 $\pm$ 0.4)$\times$10$^{11}$ & (4.9 $\pm$ 0.5)$\times$10$^{11}$ & 0.06 & 0.3 \\
        & & & & & & & & & \\
        \hline
        & & & & & & & & & \\
        \ce{CH3OH-A} & 1$_{1,1}$-0$_{0,0}$ & 350.905 & --- & 45 $\pm$ 19 & 73 $\pm$ 8 & (1.2 $\pm$ 0.5)$\times$10$^{11}$ & (6.5 $\pm$ 0.7)$\times$10$^{10}$ & 0.003 & 0.007 \\
        & 7$_{0,7}$-6$_{0,6}$ & 338.408 & 4.0$\times$10$^6$ & 137 $\pm$ 19 & 67 $\pm$ 8 & (6.9 $\pm$ 1.0)$\times$10$^{11}$ & (1.2 $\pm$ 0.1)$\times$10$^{11}$ & 0.006 & 0.008 \\
        & 7$_{2,5}$-6$_{2,4}$ & 338.640 & 3.6$\times$10$^6$ & 69 $\pm$ 19 & --- & (3.8 $\pm$ 1.1)$\times$10$^{11}$ & --- & 0.004 & --- \\
        %& 7$_{4,4}$-6$_{4,3}$ & 338.513$^{(e)}$ & 37 $\pm$ 22 & --- & (2.8 $\pm$ 1.2)$\times$10$^{11}$ & --- & --- & --- \\
        %& 7$_{4,3}$-6$_{4,2}$ & 338.513$^{(e)}$ & 37 $\pm$ 22 & --- & (2.8 $\pm$ 1.2)$\times$10$^{11}$ & --- & --- & --- \\
        & 13$_{0,13}$-12$_{1,12}$ & 355.603 & 3.2$\times$10$^7$ & 51 $\pm$ 19 & --- & (3.5 $\pm$ 1.3)$\times$10$^{11}$ & --- & 0.003 & --- \\
        %& 13$_{1,12}$-12$_{0,13}$ & 342.730 & 32 $\pm$ 22 & --- & (6.5 $\pm$ 3.9)$\times$10$^{10}$ & --- & --- & --- \\
        & 14$_{1,13}$-14$_{0,14}$ & 349.107 & --- & 75 $\pm$ 19 & --- & (2.9 $\pm$ 0.8)$\times$10$^{11}$ & --- & 0.004 & --- \\
        %& 15$_{1,14}$-15$_{0,15}$ & 356.007 & 69 $\pm$ 22 & --- & (2.2 $\pm$ 0.6)$\times$10$^{10}$ & --- & --- & --- \\
        %CH$_3$OH-E & 4$_{0,4}$-3$_{1,3}$ & 350.688 & 127 $\pm$ 22 & 112 $\pm$ 19 & (1.3 $\pm$ 0.5)$\times$10$^{12}$ & (3.8 $\pm$ 1.6)$\times$10$^{11}$ & 0.004 & 0.009 \\
        & & & & & & & & & \\
        \ce{CH3OH}-E & 7$_{1,7}$-6$_{1,6}$ & 338.345 & 1.3$\times$10$^8$ & 96 $\pm$ 19 & 84 $\pm$ 8 & (4.9 $\pm$ 1.0)$\times$10$^{11}$ & (1.5 $\pm$ 0.1)$\times$10$^{11}$ & 0.006 & 0.007 \\
        & 7$_{0,7}$-6$_{0,6}$ & 338.124 & 1.3$\times$10$^8$ & 105 $\pm$ 19 & 38 $\pm$ 8 & (5.3 $\pm$ 1.0)$\times$10$^{11}$ & (6.8 $\pm$ 1.4)$\times$10$^{10}$ & 0.006 & 0.006 \\
        & 7$_{1,6}$-6$_{1,5}$ & 338.614 & 1.3$\times$10$^8$ & 110 $\pm$ 19 & --- & (5.5 $\pm$ 1.0)$\times$10$^{11}$ & --- & 0.005 & --- \\
        & 7$_{3,5}$-6$_{3,4}$ & 338.583 & 1.1$\times$10$^8$ & 103 $\pm$ 19 & --- & (3.3 $\pm$ 0.6)$\times$10$^{11}$ & --- & 0.007 & --- \\
        & $7_{3,4}$-$6_{3,3}$ & 338.560 & 1.1$\times$10$^8$ & 51 $\pm$ 19 & --- & $(3.1 \pm 1.2) \times 10 ^{11}$ & --- & 0.003 & --- \\
        & & & & & & & & & \\
        \hline
        \end{tabular}}
\label{table.rotdiagtrans}
\footnotesize{$^{(a)}$The uncertainty values for the \ce{H2CO} lines have been calculated assuming an uncertainty of 10\%\\ 
$^{(b)}$These two Cycle 7 \ce{o-H2CO} lines are blended in the inner region but not the outer region; hence, the inner region fluxes for these two lines were assumed to have equal contributions to the total measured flux.\\ 
$^{(c)}$The blended \ce{o-H2CO} lines are well separated in the outer region but appear in the same spw, therefore a separate velocity clip between 8.3 and 10.0 km~s$^{-1}$ is used to measure the flux of this second line.}
\end{table*}

We use the MCMC code and methodology described in Sect.~\ref{Methodology} for both species and for both the inner and outer components of emission, which results in four best-fit rotational diagrams constraining the total column density ($N_\mathrm{T}$), rotational temperature ($T_\mathrm{rot}$) and line opacity for each case.
We find that all of our detected lines are optically thin (see Table~\ref{table.rotdiagtrans}).
%However, the \ce{H2CO} transition at 351.768 GHz, which has an optical depth more than one order of magnitude higher (0.3 vs $\lesssim 0.03$) than that for the other \ce{H2CO} lines, which may have affected the rotational diagrams (see Table \ref{table.rotdiagtrans}). Given that this is the transition with the lowest E$_\mathrm{u}$, this is expected to be the case, however this optical depth is still low enough that our aforementioned correction factor is valid.
The rotational diagrams and fitted parameters are shown in Fig.~\ref{fig.tempconstrain}, while the corner plots generated from the MCMC fitting are shown in Appendix D (Fig.~\ref{fig.cornerplots}).
Note that the fluxes for the \ce{o-H2CO} lines at approximately 291.38 GHz are both included in the rotational diagrams, although they appear as a single point due to their identical $E_\mathrm{u}$ values (141~K).
\begin{table}
    \centering
    \caption{$T_\mathrm{rot}$ values as obtained using the rotational diagram analysis compared to gas temperatures measured in the disk midplane and atmosphere by \citet{Kama16}}.
    \begin{tabular}{c c c}
        \hline
         Species & Inner $T_\mathrm{rot}$ [K] & Outer $T_\mathrm{rot}$ [K] \\
        \hline
        & & \\
         \ce{CH3OH} & $152^{+35}_{-27}$ & $52^{+8}_{-6}$ \\
         & & \\
         \ce{H2CO} & $76^{+9}_{-8}$ & $31^{+2}_{-2}$ \\
         & & \\
         \hline
         Region & Inner $T_\mathrm{rot}$ [K] & Outer $T_\mathrm{rot}$ [K] \\
         \hline
         & & \\
         Midplane$^{(a)}$ & 50$-$600 & 20$-$30 \\
         & & \\
         Atmosphere$^{(a)}$ & 100$-$600 & $\sim$100 \\
         & & \\
         \hline
    \end{tabular}
    \\$^{(a)}$\citet{Kama16}
    \label{tab.trot}
\end{table}
The MCMC results reveal a strong gradient in the temperature from the inner to the outer disk and show that the \ce{CH3OH} emission is hotter than the relative \ce{H2CO} emission in both components.
In addition, we note the lack of scatter in both \ce{CH3OH} rotational diagrams, suggesting that this emission is in LTE up to the highest critical density transitions; from this it would be possible to directly estimate a lower limit for the gas density. We use Eq. \ref{eq.ncrit}, with Einstein coefficients ($A_\mathrm{ul}$) from the LAMDA molecular database 
\citep[]{Schoier2005} and collisional coefficients ($C_\mathrm{ui}$) from \citet{Rabli2010} for \ce{H2CO} and \citet{Wiesenfeld2013} for \ce{CH3OH}, to calculate the critical density ($n_\mathrm{CR}$) for each transition; we list these values in Table \ref{table.rotdiagtrans}.
\begin{equation}
n_{CR}=\frac{A_{ul}}{\Sigma_i C_{ui}}
    \label{eq.ncrit}
\end{equation}
Here, $A_\mathrm{ul}$ (s$^{-1}$) specifically refers to the Einstein coefficient between upper ($u$) and lower ($l$) levels, while $\Sigma_i C_\mathrm{ui}$ (cm$^3$~s$^{-1}$) refers to the sum of collisional coefficients from an upper level ($u$) to all other lower levels ($i$). From our calculations, we obtain a lower limit for the gas density in the midplane of HD~100546 of order 10$^6$~cm$^{-3}$; this is compatible with the models presented in \citet{Kama16} and \citet{Leemker24}, in which the midplane gas density varies between 10$^8-$10$^{10}$~cm$^{-3}$.

As shown in Table \ref{tab.trot}, T$_\mathrm{rot}$ for \ce{CH3OH} is $152^{+35}_{-27}$~K in the inner region and decreases to $52^{+8}_{-6}$~K in the outer region.
T$_\mathrm{rot}$ for \ce{H2CO} is $76^{+9}_{-8}$~K in the inner region and falls to $31^{+2}_{-2}$~K in the outer region.
We also compare these values with gas temperatures in the midplane and atmosphere of the disk as obtained by \citet{Kama16}.
\ce{CH3OH} has stronger binding to a multilayer surface than \ce{H2CO} with measured binding energies of 3820~K and 3260~K \citep{Penteado2017}, respectively.
Hence, we expect that \ce{CH3OH} will desorb at a higher temperature than \ce{H2CO} and that it should have a higher rotational temperature if the bulk of the emission is arising from near the snowlines (or snow surfaces) for both species.
In Appendix E (Fig. \ref{fig.desorption}), we calculate the expected desorption temperatures for \ce{CH3OH} and \ce{H2CO}, assuming a range of values for the gas density at the midplane of HD~100546 as modelled by \citet{Kama16}. We find that \ce{CH3OH} will desorb above a temperature of 109~K for a gas density of 10$^{12}$~cm$^{-3}$ \citep[]{Kama16} and \ce{H2CO} will desorb above a temperature of 93~K at the same density. Using a higher binding energy for \ce{CH3OH} (e.g. 4931~K, as measured by \citealt{Brown07}) leads to the desorption temperature of methanol changing to 140~K. We also find that, for the range of gas densities assumed, \ce{CH3OH} will desorb at a higher temperature than \ce{H2CO} by 14$-$16 K \citep[]{Collings04,Noble12,Penteado2017}. The fact that \ce{CH3OH} has a higher T$_\mathrm{rot}$ overall than \ce{H2CO} could be explained if \ce{CH3OH} is surviving in the inner disk at higher temperatures than \ce{H2CO}% or that the binding energy is higher than that used in our model
. Note that the computation of thermal desorption rates and desorption temperatures from binding energy values measured in the laboratory is more complex than is commonly assumed \citep[]{Minissale2022,Ligterink2023}.%% 73-82 K (UMIST)

The T$_\mathrm{rot}$ values in the inner component are explainable solely by thermal desorption, however, this is not the case for the lower $T_\mathrm{rot}$ values measured in the outer component.
Here, the origin of the \ce{CH3OH} emission is likely non-thermal desorption from the ice mantles on the dust grains, because gas-phase \ce{CH3OH} does not have efficient gas-phase formation routes at low temperatures \citep[][]{Garrod06,Geppert06,Walsh2014}.
Sources of non-thermal desorption in the outer regions of protoplanetary disks are cosmic-rays, X-rays, UV photons (from the star and surrounding interstellar medium) and excess energy in grain-surface reactions (so-called reactive or chemical desorption; \citealt{Walsh2014,Cuppen17}). It should be noted, however, that in the case of non-thermal desorption at low temperatures, it has been suggested by experimental studies that \ce{CH3OH} does not desorb intact which therefore inhibits the photodesorption yield of intact \ce{CH3OH} under these conditions \citep[]{Bertin2016,Cruz2016,Notsu2021}. We take this fragmentation of methanol ice upon photodesorption into account in our models \citep[see][]{Bertin2016}.
On the other hand, cold gas-phase \ce{H2CO} has two potential origins: non-thermal desorption and gas-phase formation \citep{Loomis15,Carney17,TerwisschavanScheltinga2021}.

The best-fit rotational diagrams show that the column density for \ce{CH3OH} is higher than that for \ce{H2CO} in the inner component.
The \ce{CH3OH} column density is also higher in the inner region than in the outer region ($1.4^{+0.4}_{-0.3}\times 10^{14}$ ~cm$^{-2}$ versus $1.2^{+0.1}_{-0.1}\times 10^{13}$ ~cm$^{-2}$); however, the column density for \ce{H2CO} is similar across the disk with values of $9.6^{+0.5}_{-0.5}\times10^{12}$ and $9.1^{+1.0}_{-0.9} \times 10^{12}$~cm$^{-2}$ in the inner and outer components respectively.

Using our fitted column densities, we can now constrain values for the \ce{CH3OH}/\ce{H2CO} ratio across the disk surrounding HD 100546.
For the inner region we obtain a value of $14.6^{+5.2}_{-4.6}$, while for the outer region we obtain a value of $1.3^{+0.3}_{-0.2}$.
This order of magnitude difference in ratio supports the presence of two COM reservoirs in the disk surrounding HD 100546 with different chemical origins. 
%We note here that 
Previously, \citet{Booth2024HD} estimated the column density 
ratio for \ce{CH3OH}/\ce{H2CO} to be $18\pm4$ in the inner disk and $1.1\pm0.6$ for the outer ringed component, which is %.Those results were obtained using the peak flux of only one transition and assuming that the emitting area equated the beamsize of their observations; nevertheless, these values are 
in good agreement with the results of the more quantitative analysis presented here.

\begin{figure*}[hbt!]
    \centering
    \includegraphics[width=0.34\textwidth]{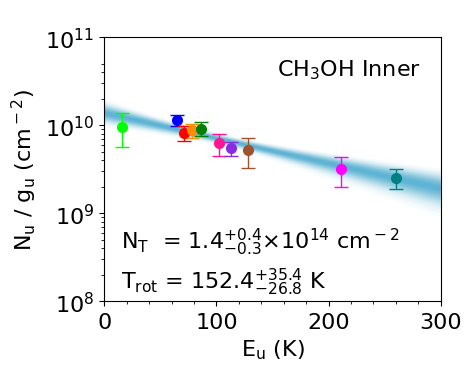}
    \includegraphics[width=0.34\textwidth]{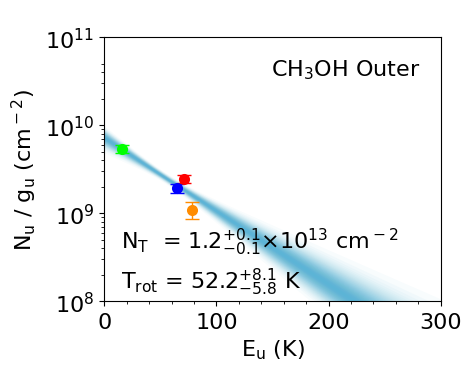}
    \includegraphics[width=0.34\textwidth]{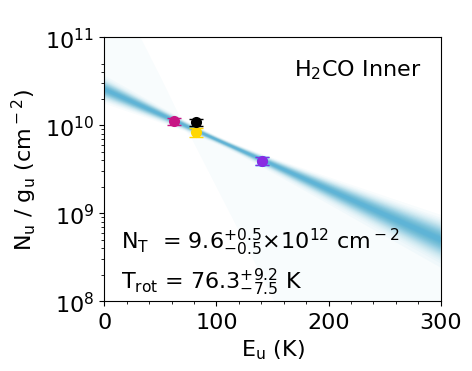}
    \includegraphics[width=0.34\textwidth]{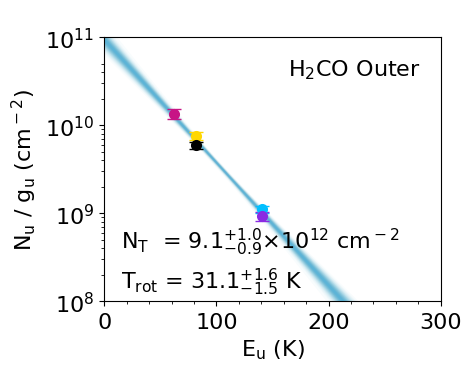}
    \caption{Best-fit rotational diagrams for \ce{CH3OH} (top row) and \ce{H2CO} (bottom row) for the inner (left panels) and outer (right panels) components of emission in the disk around HD 100546. The colour of each point corresponds to the colour of each transition as denoted in Figs.~\ref{fig.ch3oh_outer_spec} and \ref{fig.azavouter}.
    The two \ce{H2CO} transitions at 291.38 GHz are well separated in the outer region as shown in Appendix C (Fig.~\ref{fig.h2co_outer_spec}) and are thus denoted separately by blue and purple points in the bottom right panel; however, they are blended in the inner region and are thus represented by a single blue point in the bottom left panel.}
    \label{fig.tempconstrain}
\end{figure*}

\section{Discussion}
\label{Discussion}

\subsection{Rotational Temperature of \ce{CH3OH}}
\label{Rotdiag}

The analysis presented in Sect.~\ref{Results} shows that $T_\mathrm{rot}$ for \ce{CH3OH} ranges from $152^{+35}_{-27}$~K in the inner disk to $52^{+8}_{-6}$~K in the outer ring.
The rotational temperatures measured for the inner component for both species are consistent with what would be expected if the origin of the emission here were due to the thermal desorption from the ice phase to the gas phase, while the lower temperatures in the outer component are likely indicative of non-thermal desorption. 
Recently Ilee et al.~(in prep.) conducted a multi-line analysis in the disk around TW Hya and measured a $T_\mathrm{rot}$ for \ce{CH3OH} of $36^{+25}_{-10}$~K.  
This is also in line with a source of \ce{CH3OH} gas arising from non-thermal desorption from icy dust grains, as suggested by astrochemical models.
The lower rotational temperature for \ce{CH3OH} in TW Hya when compared to HD~100546 might reflect its generally cooler disk temperature \citep[see, e.g.,][]{Kama16}.
The only other disk in which $T_\mathrm{rot}$ has been empirically measured for \ce{CH3OH} is the disk around IRS~48.
\citet{vanderMarel2021} report a $T_\mathrm{rot}$ for \ce{CH3OH} of $103^{+6}_{-5}$~K for IRS~48 compared with $173^{+11}_{-9}$~K for \ce{H2CO}.  
This is the opposite trend to that seen in HD~100546 and is a trend which persists in a follow-up analysis published in \citet{Temmink2024}, although it was noted in that work that there is large scatter in the $T_\mathrm{rot}$ for \ce{CH3OH} which is likely due to sub-thermal excitation.
In the IRS~48 disk, the \ce{CH3OH} and \ce{H2CO} emission are both strongly associated with a highly asymmetric dust trap thought to be caused by a vortex. The complex dynamics in this system may influence the thermal and chemical structure of the disk and give rise to the relative differences in $T_\mathrm{rot}$ for these two molecules when compared to HD~100546.
Indeed, the $T_\mathrm{rot}$ value reported for \ce{H2CO} in IRS~48 is very high when compared to literature; our value of $31^{+2}_{-2}$~K measured towards the outer region of HD~100546 is much more in line with values reported towards the outer regions of other disks.
For example, \citet{HernandezVera2024} measured a \ce{H2CO}  rotational temperature of approximately 20~K for the disk around HD~163296, while the \ce{H2CO} disk survey by \citet{Pegues2020} found values between 10-40~K in four sources.

%The situation in IRS~48 may differ from that in HD~100546 as the emission from both \ce{CH3OH} and \ce{H2CO} are strongly associated with a highly asymmetric dust trap thought to be caused by a vortex. 
%The complex dynamics in IRS~48 could be driving this difference in the trends in $T_\mathrm{rot}$ between IRS~48 and HD~100546, with the latter source appearing a more quiescent system.}  

In Appendix F, we compare and discuss our results for $T_\mathrm{rot}$ with those reported in the literature for objects at different evolutionary stages of star and planet formation.
%and  the main takeaway from this comparison. }

\subsection{\ce{CH3OH}/\ce{H2CO} Column Density Ratio}
\label{Ratios}

\begin{figure*}[hbt!]
    \centering
    \includegraphics[width=0.8\textwidth]{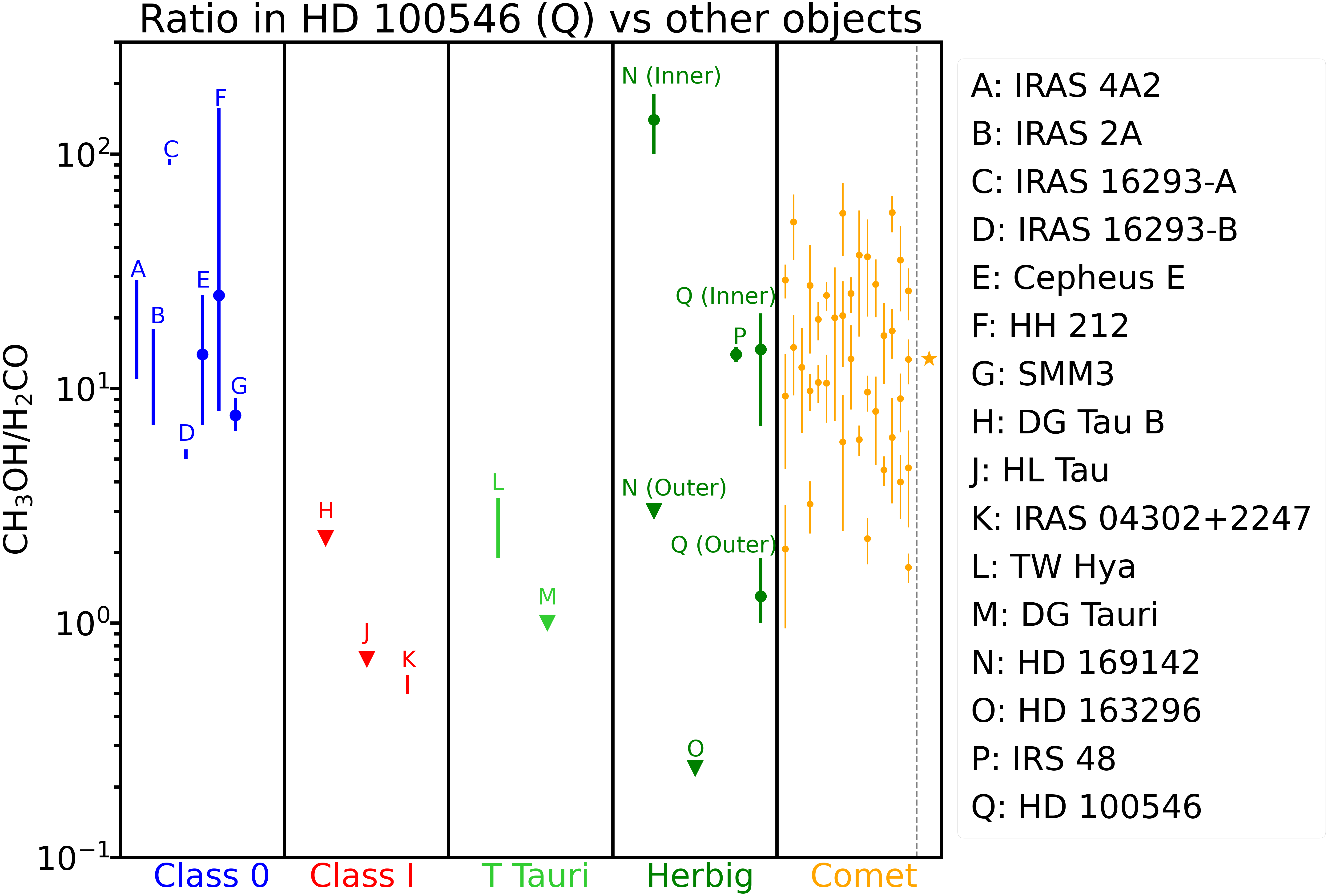}
    \caption{\ce{CH3OH}/\ce{H2CO} ratios measured towards the inner and outer regions of HD~100546 (Q) compared to values measured towards objects of varying evolutionary stages. Letters have been used to denote each object; refer to Appendix F (Table \ref{tab.refs}) for the reference(s) associated with each source.
    Note that circles represent values with error bars, bars represent a range, triangles represent upper limit values for the \ce{CH3OH}/\ce{H2CO} ratios for DG Tau B (H), HL Tau (J), HD~163296 (O) and the outer region of HD~169142 (N), while the star represents the median value of measurements across a sample of 38 Solar System comets \citep[]{Lippi2024}.}
    \label{fig.coldenscomp}
\end{figure*} 

%The ratio of \ce{CH3OH} to \ce{H2CO} has been suggested to trace the degree of chemical processing undergone by icy material as it is delivered to and processed within the protoplanetary disk \citep[see, e.g.,][]{Podio2020}. 
It is important to note that our observations are purely tracing the gas phase and so there is no guarantee that they are directly tracing the ice ratios. The inner region is thermally desorbed and so it is likely that the ice ratios are being traced here. However, in the outer region we have non-thermal desorption in the form of photodesorption, for which the efficiencies are different for each species \citep{Martin-Domenech2016,Bertin2016}. Furthermore, we have the added complication of gas-phase \ce{H2CO} formation occurring in the outer region which also reduces the likelihood that we can directly trace the ice ratios here.
From our results, the column density for \ce{CH3OH} decreases by approximately one order of magnitude going from the inner to outer disk of HD~100546, while \ce{H2CO} has a similar column density in the two emission components.
This is reflected in the \ce{CH3OH}/\ce{H2CO} column density ratio which ranges from $14.6^{+5.2}_{-4.6}$ in the inner disk to $1.3^{+0.3}_{-0.2}$ in the outer disk. 
We compare our measured ratios for both components of emission in the disk around HD~100546 with representative values from younger objects, T Tauri and other Herbig objects, along with comets in Fig.~\ref{fig.coldenscomp}.

We reiterate that the successful detection of gas-phase \ce{CH3OH} in the HD~100546 disk was first reported in \citet{Booth2021}.
The values that we derive here are consistent with the radial column density ratio in that work, which were calculated assuming the same rotational temperature for both \ce{H2CO} and \ce{CH3OH} (which in turn was derived from a rotational diagram analysis for \ce{H2CO}).
Our results for the rotational temperature for \ce{H2CO} are also consistent with this previous work.
The column densities and rotational temperatures derived in the present work have significantly smaller error bars due to the higher sensitivity and higher spatial resolution of these data, as well as the detection of multiple lines of both species spanning a wider upper energy level range.
Lower measured \ce{CH3OH}/\ce{H2CO} ratios have previously been attributed to more efficient gas-phase formation of \ce{H2CO}; this has been observed for the T Tauri disk surrounding TW Hya \citep{TerwisschavanScheltinga2021} and a low upper limit to the ratio has been constrained for HD~163296 \citep{Carney2019}.
Efficient gas-phase formation of \ce{H2CO} in the outer region of HD~100546 
is a possible explanation for the constant column density observed across the disk, while the aforementioned low photodesorption rates of \ce{CH3OH} at low temperatures \citep[]{Bertin2016,Cruz2016,Notsu2021} may provide an explanation for the lower column density measured in the outer disk.
Despite also being a Herbig Ae disk, HD~163296 is cold compared to HD~100546 and possesses a CO snowline at $\sim 60$~au (\citealt{Qi2011}; \citealt{Zhang19}\; \citeyear{Zhang2021}).
%There are also indications that HD~163296 is a flat %, shadowed
%disk \citep[e.g.,][]{Garufi14}, meaning that it captures less radiation from its host star than the flared disk around HD~100546 \citep[e.g.,][]{Garufi16}, which will weaken the ability of the former to non-thermally desorb ice in the outer regions of the disk. 
Modelling based on \textit{Herschel} observations of the inner part of HD~163296 reveal high abundances of gas phase \ce{H2O}, in contrast to HD~100546, where higher energy transitions of \ce{H2O} remain undetected \citep[see][]{Pirovano22}, which is proposed to be a consequence of differing chemical histories of the two disks, including pebble drift in HD~163296.%Moreover, it has tentatively been shown by \citet{Pegues2020} that Herbig disks possess lower amounts of \ce{H2CO} than T Tauri disks, which is possibly due to less CO freeze-out. The same study also concluded that low \ce{H2CO} temperatures of roughly 20-50 K across the full range of disks in their sample (including Herbig Ae and T Tauri) suggest a warm molecular layer origin of \ce{H2CO} with a small contribution from the colder midplane.

Comparing our column density ratios to those measured towards other Herbig disks, the ratio towards IRS~48 has been measured as $14 \pm 1$, which is close to the value measured towards the inner region of HD 100546.
However, if the emission from \ce{H2CO} towards IRS 48 is optically thick, as is suggested by H$_2^{13}$CO observations \citep[]{Booth2024IRS}, as well as analysis of more recent \ce{H2CO} observations \citep[]{Temmink2024}, the ratio may be as low as 1, which is closer to our measured value for the outer region of HD 100546.
Similar to HD 100546, the \ce{CH3OH}/\ce{H2CO} ratio also decreases from the inner to outer disk of HD~169142 \citep{Booth2023}.
The ratio measured towards the inner region of HD~169142 is an order of magnitude higher than that for HD 100546; however, the outer region upper limit 
(due to the non-detection of \ce{CH3OH}) is consistent with our value for HD~100546. 
An upper limit ratio of $< 0.24$ was measured towards the disk surrounding HD~163296 \citep{Carney2019}, which is lower than that seen towards the outer region of HD 100546 in the present work.
Meanwhile, an outwards decreasing trend is also seen in the Horsehead PDR, which has been shown to exhibit similar desorption behaviour to protoplanetary disks, wherein the \ce{CH3OH}/\ce{H2CO} ratio decreases from $\sim 1$ in the thermally desorbed core to $\sim 0.5$ in the non-thermally desorbed PDR itself \citep{Guzman2013}.

The differences in the ratios between different Herbig disks may be related to their structure as discussed above for HD~163296.
IRS~48, HD~100546 and HD~169142 all possess significantly-sized inner cavities that allows direct exposure of the ice reservoir in the midplane to the stellar radiation driving thermal sublimation of the ices, thus revealing the presence of \ce{CH3OH}. However, \ce{CH3OH} was not detected in HD~142527, despite the presence of a cavity; this is likely due to the fact that, unlike other transition disks around Herbig stars, the dust trap (and the ice) is located much farther from the host star (peaking at $\sim$175~au) and so the conditions are too cold $\sim$40~K) for thermal sublimation to take place \citep[]{Casassus2013,Temmink2023}.
Whether or not an outer cooler component of \ce{CH3OH} emission is present will depend on the efficiency of non-thermal desorption in the outer regions of the disk.
%CO snowline being located beyond the dust trap ruling out in-situ formation on the grain surface \citep[]{Temmink2023}. This result may reflect the aforementioned dichotomy between warm and cold dust traps impacting the presence of gas-phase \ce{CH3OH} in disks.} %, with disks that are flared (like HD~100546) able to capture more radiation in their outer regions than those than are %shadowed and 
%flat (like HD~163296).

Comparing the column density ratios for HD~100546 with values measured towards younger objects, as well as towards Solar System comets (see~Fig.~\ref{fig.coldenscomp}), the inner region of HD 100546 has a higher column density ratio than that found in the Class I objects IRAS 04302 ($0.5-0.6$; \citealt{Podio2020}), DG Tau B ($< 1.5-2.3$; \citealt{Garufi2020}) and HL Tau ($< 0.7$; \citealt{Garufi2022}): these ratios are more similar to that measured towards the outer region of HD 100546, pointing to a similar chemical origin (non-thermal desorption and gas-phase formation of \ce{H2CO}). %In particular, HL Tau has shown a much higher excitation temperature for \ce{H2O} than its desorption temperature \citep{Facchini2024}, which would mean that \ce{CH3OH} should thermally desorb in the inner regions of this disk.
On the other hand, abundance ratios measured in hot corinos surrounding Class 0 objects NGC 1333-IRAS 2A \citep{Taquet2015}, HH212 \citep{Lee2022}, Cep E-mm \citep{Ospina-Zamudio2018}, the protobinary IRAS 16293-2422 \citep{Persson2018,Manigand2020}, as well as the potential hot corino surrounding SMM3 (a protostellar core which is embedded in Orion B9; \citealt{Miettinen2016}), range from $1-2$ orders of magnitude higher than that measured towards HD~100546, while the shock-driven \ce{CH3OH} enhancement in the Class 0/I object BHB07-11, which does not host a hot corino, leads to a ratio that is $3-4$ orders of magnitude higher than HD~100546 \citep{Vastel2022,Evans2023}.
These differences could indicate substantial chemical modification for both \ce{CH3OH} and \ce{H2CO} and/or diversity in chemical conditions in hot corinos and shocked regions compared with the inner regions of HD~100546.
%Moreover, previous modelling results have shown that the presence of an embedded disk lowers the temperature of the emitting environment considerably \citep{Nazari2022}. It should be noted, however, that optical depth effects are much more present at the protostellar stage for the main isotopologue of \ce{CH3OH}, which can hide COM emission from detection; isotopologues such as $^{13}$CH$_3$OH and CH$_3^{18}$OH are thus required to accurately constrain methanol abundances in such objects, see discussion in \citet{vanGelder2022}.
The inner region of HD~100546 has a more similar ratio to the median value measured across a sample of 38 Solar System comets compiled by \citet{Lippi2024}.
That the ratio in the recently sublimated ice in the inner regions of HD~100546 is similar to that in cometary ices further supports the hypothesis that a significant fraction of interstellar ice has been inherited and has survived transport through the HD~100546 protoplanetary disk.

\subsection{Chemical Modelling}
\label{Modelling}

\citet{Booth2021} supplemented their observations with a chemical model of the disk to investigate the chemical origins of \ce{CH3OH}.
They focused on whether it is possible that the observed \ce{CH3OH} formed in situ or was inherited from a previous evolutionary stage of the object.
They assumed a 2D physical structure and number density of hydrogen nuclei (n$_H$), gas and dust temperature, UV radiation field and X–ray ionisation rate as in \citet{Kama16}, with a cosmic-ray ionisation rate of 5$\times$10$^{-17}$~s$^{-1}$. 
They found that it was not possible to produce \ce{CH3OH} in situ in this disk under their assumption of atomic initial conditions.
Moreover, the abundance of \ce{CH3OH} was found to decay in the disk over time because it cannot reform due to the warm temperature of the dust throughout the disk ($\gtrsim 20$~K) that inhibits freeze-out of CO \citep[]{Kama16}.
Based on these results, they concluded that the \ce{CH3OH} observed towards HD~100546 was inherited from an earlier evolutionary phase.
In addition, recent results from physicochemical modelling of the dust rings also agree that the disk surrounding HD~100546 is warm \citep[]{Leemker24}, however, the CO map from the same paper suggests that freeze-out can occur in the outer regions of the disk, so there is the possibility for methanol reformation on the dust grains at the location of the outer dust ring, which should be tested in future gas-grain models.

Our constrained rotational temperatures for \ce{CH3OH} and \ce{H2CO} show two components of emission: one inner component driven by thermal desorption of ices and one outer component driven by non-thermal desorption, as well as a gradient in the \ce{CH3OH}/\ce{H2CO} column density ratio across the disk.
Here, we revisit the models presented in \citet{Booth2021} % (see that work for full details)
in light of the results reported here.
Note that this physical model for HD~100546 comes from the thermochemical modelling presented in \citet{Kama16}, which assumes a dust-to-gas mass ratio of 100 and uses the chemical network described in \citet{Walsh15} and references therein. A more detailed description of the chemical model can also be found in Appendix G.

We show in Fig. \ref{fig.modellingratio} the resultant predictions from our models for the radial column densities (top row) of both \ce{CH3OH} (left) and \ce{H2CO} (right) along with the resultant \ce{CH3OH}/\ce{H2CO} column density ratio (bottom row) for the gas (left) and ice (right) phases. To show a fair comparison to our observational results, the modelled column densities are first interpolated over a denser linear grid with 1~au resolution in radius, then averaged over 15~au radial bins. This is necessary because the original grid was constructed in log space with higher resolution in the inner regions versus the outer regions. The results from the model at its original grid resolution can be found in Appendix G (Fig. \ref{fig.modellingratiofull}).
 
The column densities in the top row of Fig. \ref{fig.modellingratio} are extracted at a time of 1 (solid lines), 2 (dashed lines) and 5 (dotted lines) Myr for both the gas-phase (brown colours) and ice phase (blue colours).
The model predicts a decreasing outwards column density trend (although a slight increase is seen in the 5 Myr case within 25 au) towards a minimum ($< 10^{11}$~cm$^{-2}$) beyond $30 - 40$~au, before rising again to reach a maximum ($\sim 10^{12}$~cm$^{-2}$) in the outer disk co-located with the outer dust ring.
The \ce{CH3OH} ice retains a substantial column density of $\sim 10^{16} - 10^{18}$~cm$^{-2}$ between radii of $\sim 20 - 220$~au; however, it does decrease with time.
The primary source of gas-phase methanol in the inner disk is thermal desorption and the origin of the emission in the outer disk is photodesorption.
This is supported by the distribution of the number density of gas-phase and ice-phase \ce{CH3OH} shown in Appendix G (Fig.~\ref{fig.abundance_ch3oh}) as a function of radius, $r$ and height divided by the radius, $z/r$.
The gas-phase \ce{CH3OH} peaks in abundance just inside its snow surface at a temperature above 150~K, as shown in the top-right panel of Appendix G (Fig.~\ref{fig.abundance_ch3oh}).
This is in excellent agreement with the derived rotational temperature for \ce{CH3OH} in the inner disk.
On the other hand, the spatial extent of ice-phase \ce{CH3OH} is set by the strength of the UV radiation field in the disk, with \ce{CH3OH} ice surviving only where the field strength is $\lesssim 0.01$ times that of the interstellar radiation field.
Gas-phase methanol reaches a reasonable abundance only at temperatures $\lesssim 50$~K in the outer disk, also in excellent agreement with the rotational temperature derived from the observations.

The column density for \ce{H2CO} is predicted to have a much more shallow radial abundance gradient, reflecting its efficient formation in both gas and ice.
Similar to \ce{CH3OH}, its column density peaks just within the dust cavity reaching a maximum of $\sim 10^{16}$~cm$^{-2}$ and declines steadily with radius before falling to low values ($< 10^{11}$~cm$^{-2}$) beyond $\sim 500$~au.
We can also see that what we can detect in the gas phase is only a small fraction of the total \ce{H2CO} in the disk, with the rest locked up in the form of ice in the outer disk.
We also note that the physical disk model used here has not accounted for the gas and dust substructure that we now know is present in the disk around HD~100546.
Accounting for the presence of a gas cavity between $40 - 175$~au as done in the recent models by \citet{Leemker24} would likely shape the radial column density of \ce{H2CO} to produce a ring-like structure in the outer disk and we will test this in future models.

Appendix G (Fig.~\ref{fig.abundance_h2co}) shows the abundance of gas-phase (left) and ice-phase (right) \ce{H2CO} at times of 1 (top), 2 (middle) and 5 (bottom) Myr, as a function of disk radius, $r$ and height divided by the radius, $z/r$.
The distributions of gas-phase and ice-phase \ce{H2CO} are more extended than that for \ce{CH3OH} supporting the notion that multiple chemical origins are contributing to their distribution.
The abundance of gas-phase \ce{H2CO} peaks at temperatures $\gtrsim 70$~K in the inner disk, in excellent agreement with the derived rotational temperature in this work.
We also see good agreement between our results and the predicted snowline locations for both \ce{CH3OH} and \ce{H2CO} according to the model (150 and 70 K, respectively) in the inner component. However, while this prediction is also in good agreement for the outer component of \ce{CH3OH}, the obtained T$_\mathrm{rot}$ for the outer component of \ce{H2CO} suggests a deeper origin closer to the colder midplane, which is not in agreement with the modelling predictions presented in Appendix G (Fig. \ref{fig.abundance_h2co}). As well as this, gas-phase \ce{H2CO} in the outer disk is distributed over a greater range of temperatures than that for \ce{CH3OH}, mainly at temperatures $\gtrsim 40$~K, which is also somewhat in disagreement with the rotational temperature for \ce{H2CO} derived from the observations for the outer disk ($31^{+2}_{-2}$~K). This could be as a result of the gas-phase formation of \ce{H2CO} being more efficient deeper in the disk than the non-thermal photodesorption of \ce{H2CO} that has been formed via the grain surface, or the outer disk being cooler in the outer regions than suggested by the model from \citet{Kama16}; more tailored models are needed to investigate this further.
In addition, the lowest E$_\mathrm{u}$ \ce{H2CO} transition has an optical depth that is an order of magnitude higher than the other transitions in our sample (see Table \ref{tab.trot}), so this may result in a higher than expected value for $T_\mathrm{rot}$.
Indeed, more recent models of the HD~100546 disk by \citet{Leemker24} suggest that the temperature in the disk midplane at the location of the outer dust ring reaches temperatures as low as 20~K which is $10-15$ K colder than the models used here. This is indicative of potential CO freeze-out in this region.

In the bottom row of Fig.~\ref{fig.modellingratio} we show the column density ratio (\ce{CH3OH}/\ce{H2CO}) for the gas phase (left) and the ice phase (right) at timesteps of 1, 2 and 5 Myr as predicted by the chemical model.
The ratio for the gas in the inner disk within the cavity ranges from $\sim 10^{-4}$ (5 Myr) to 10 (1 Myr) and that for the outer disk and co-located with the dust ring is $\sim 0.1$.
The model therefore predicts a similar trend in \ce{CH3OH}/\ce{H2CO} with radius, albeit with a higher ratio in the inner disk and a lower ratio in the outer disk at 1~Myr.
However, given that this model has not been tailored nor adapted in any way, this is considered reasonable agreement with the observations, especially as the inner disk ratio is decreasing with time and the outer disk ratio is increasing with time.
At a time of 2~Myr, the peak ratio in the inner region is $\sim 0.7$ and the peak outer ratio is $\sim 0.6$. Compared to the observed values, which are shown as grey shaded regions in the bottom-left panel of Fig.~\ref{fig.modellingratio}, the resultant peak values in the inner region are of the same order of magnitude at 1~Myr but a factor of 20 underestimated at 2~Myr. Meanwhile, the peak ratio in the outer region is more consistent across the different times of extraction and within a factor of 2 of the observations.
The radial behaviour of the ice-phase column density ratio shows a similar shape and trend with an overall decrease throughout the inner region and a peak co-located with the outer dust ring.
%Interestingly, the ice-phase column density ratio between 1 and 2 Myr in the inner region is consistent at $\sim $ and from $0.2 - 0.4$ in the outer region. %values that are also close to the ratios derived from the observations.
That the ice- and gas-phase ratios are similar shows that we are likely primarily tracing the desorption of ices in our observations, as well as that the main desorption mechanisms are those suggested by the model: thermal desorption in the inner region and photodesorption in the outer region.
The agreement with the cometary ratio in the inner region (see Fig.~\ref{fig.coldenscomp}) strongly supports that we are seeing `fresh' ice sublimation here. It should be noted, however, that the column density and column density ratio plots in Appendix G (Fig. \ref{fig.modellingratiofull}), which are not radially binned, show peaks within the inner component which are averaged out in our observations and the radially binned model.

HD~100546 is also one of the few disks within which cold water was detected with \emph{Herschel} \citep{Pirovano22}.
The fractional abundance of cold water emission from the outer disk was estimated using a forward model to be $3\times 10^{-9}$ in the region where photodesorption is expected to be the chemical origin.
The peak fractional abundance of \ce{CH3OH} predicted by the model here in the same region and at a time of 1~Myr is a few times $10^{-11}$ (see~Fig.~\ref{fig.abundance_ch3oh}); however, this model does under-predict the column density in the outer ring by around a factor of 10.
This implies a peak fractional abundance of \ce{CH3OH} closer to a few times $10^{-10}$, leading to a \ce{CH3OH}/\ce{H2O} gas-phase abundance ratio of order 0.1 (10\%).
This estimation lies at the higher end of the ratio of \ce{CH3OH}/\ce{H2O} seen in comets \citep[typically of order 0.1\% - 10\%; see, e.g.,][]{Bockelee-Morvan17}.
That the gas-phase abundance is estimated to be close to the higher end of the observed ratio for cometary ice suggests that we could be directly probing the composition of the ice reservoir in the outer ring, despite the origin being photodesorption.
However, as previously mentioned, experiments have suggested that the efficiency of intact \ce{CH3OH} photodesorption is around 100 to 1000 lower than that of \ce{H2O} which would suggest a gas-phase \ce{CH3OH}/\ce{H2O} of no higher than $\sim 0.001$ (0.1\%) assuming an ice-phase ratio as high as 0.1 \citep[10\%;][]{Oberg09b,Bertin2016}.
Indeed the model results for gas-phase \ce{H2O} show a higher fractional abundance and greater radial and vertical extent than that for gas-phase \ce{CH3OH}, reflecting the different photodesorption efficiencies for both species (not shown here).

There are multiple possible explanations for this apparent discrepancy.
Either \ce{CH3OH} is able to be synthesised efficiently in the colder outer regions thereby increasing the \ce{CH3OH}/\ce{H2O} ice (or gas) ratio to values higher than the cometary value, or there is another non-thermal desorption mechanism dominating the gas-phase abundances which has similar efficiencies for both \ce{CH3OH} and \ce{H2O}, e.g., chemical or reactive desorption.
Future gas-grain astrochemical models tailored to the dust and gas structure of HD~100546 are needed to test these theories.

%\begin{figure}
 %   \centering
  %  \begin{subfigure}{}
        
   % \end{subfigure}
    %\begin{subfigure}{}
        
   % \end{subfigure}
    %\caption{}
    %\label{fig.ncol_model}
%\end{figure}

\begin{figure*}
    \centering
    \includegraphics[width=0.45\textwidth]{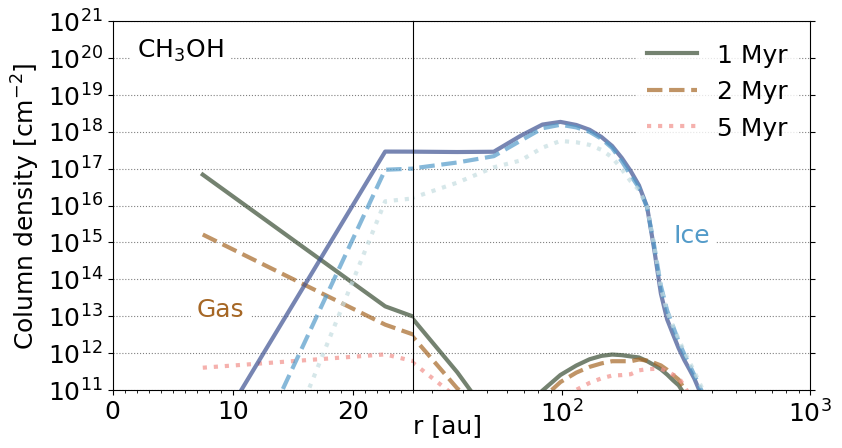}
    \includegraphics[width=0.45\textwidth]{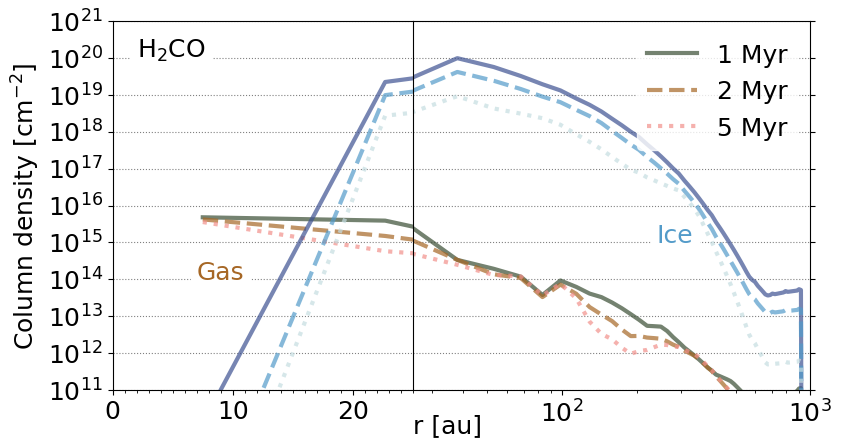}
    \includegraphics[width=0.45\textwidth]{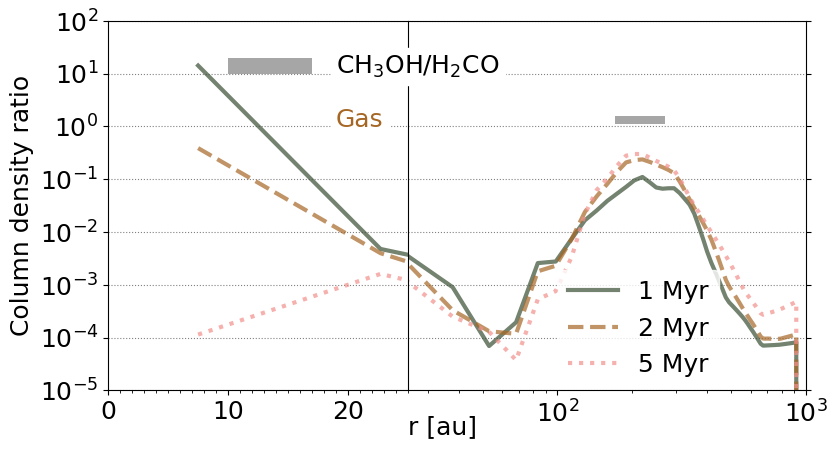}
    \includegraphics[width=0.45\textwidth]{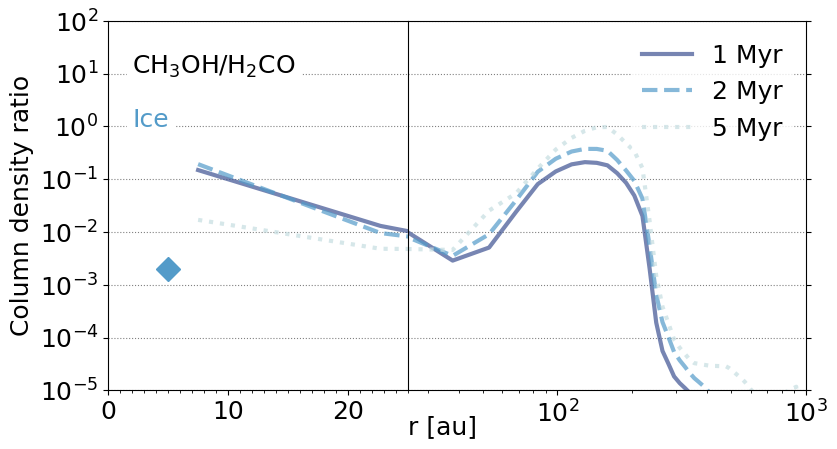}
    \caption{Top row: Column density calculated as a function of radius averaged over 15 au wide radial bins for the disk around HD~100546 for \ce{CH3OH} (left panel) and \ce{H2CO} (right panel) at 1 (solid), 2 (dashed) and 5 (dotted) Myr. Bottom row: \ce{CH3OH}/\ce{H2CO} column density ratio calculated for the gas (left panel) and ice (right panel) phase, at 1 (solid), 2 (dashed) and 5 (dotted) Myr. The brown colours denote gas phase molecules while the blue colours represent the ice phase. Note that the \ce{CH3OH}/\ce{H2CO} models plotted for the ice phase (bottom right panel) are restricted to radii $\ge 15$ au as the column density in the ice phase is negligible within this radius. The models within 25~au have been plotted on a linear scale to highlight the structure in the inner disk.
    The grey shaded regions on the bottom-left panel indicate the observed ratios derived in this work for the inner ($15^{+5}_{-5}$) and outer ($1.3^{+0.3}_{-0.2}$) components (note that the radial extent of the shaded regions is arbitrary and is set to guide the eye only). The diamond marker on the bottom-right panel indicates the initial abundance ratio for \ce{CH3OH}/\ce{H2CO} ice adopted in the chemical model ($2\times10^{-3}$).}
    \label{fig.modellingratio}
\end{figure*}

\section{Conclusions}
\label{Conclusions}

In this work, we analysed emission from multiple rotational transitions of \ce{CH3OH} and \ce{H2CO} arising from the structured disk around the Herbig Ae star HD~100546 as observed with ALMA.
Our aim was to measure the rotational temperature and column density of each molecule and ascertain its chemical origin. The main conclusions of this paper can be summarised as follows:
\begin{itemize}
    \item We detect the organic molecules \ce{H2CO} (three ortho and two para transitions) and \ce{CH3OH} (ten transitions).
    Both molecules have a centrally concentrated component within 110~au of the star and a ring of molecular emission centered at 220~au, which is associated with a faint ring of mm dust.
    \item Using a rotational diagram, we find that \ce{CH3OH} decreases in column density from the inner ($1.4^{+0.4}_{-0.3}\times10^{14}$ cm$^{-2}$) to the outer ($1.2^{+0.1}_{-0.1}\times10^{13}$ cm$^{-2}$) region, whereas the column density of \ce{H2CO} is constant across the two emission components at approximately 9$\times$10$^{12}$~cm$^{-2}$.
    \item The \ce{CH3OH/H2CO} ratio varies from $14.6^{+5.2}_{-4.6}$ in the inner disk to $1.3^{+0.3}_{-0.2}$ in the outer disk.
    This could be due to a contribution of gas-phase formation of \ce{H2CO} in the outer disk, or that there is a higher abundance of \ce{CH3OH} ice than \ce{H2CO} ice in the disk midplane which is traced in the thermally sublimated reservoir generating the inner component.
    \item %From the same rotational diagram, w
    We find a significant decrease in the rotational temperatures, $T_\mathrm{rot}$, across the disk for both species.
    $T_\mathrm{rot}$ for \ce{CH3OH} has values of $152^{+35}_{-27}$~K 
    and $52^{+8}_{-6}$~K in the inner and outer disk, respectively, whereas the equivalent values for \ce{H2CO} are $76^{+9}_{-8}$~K and $31^{+2}_{-2}$~K.
    The values for the inner disk are consistent with the inner component arising from thermal desorption of ices at each species' respective snowline or snow surface, whereas those for the outer disk are consistent with a non-thermal desorption origin (with gas-phase formation also possibly contributing to the \ce{H2CO} abundance here).
    The coincidence of the \ce{CH3OH} and \ce{H2CO} outer components of emission with the faint mm dust ring suggests that the dust grains sculpted into this ring are icy.
    \item The higher temperature seen in the inner region of HD 100546 is comparable to temperatures seen towards younger objects, suggesting a similar chemical origin (i.e., thermal desorption from the ice phase).
    However, \ce{CH3OH} in the outer region of HD~100546 has a more similar rotational temperature to that for the disk around TW~Hya where photodesorption is the most probable chemical origin of gas-phase \ce{CH3OH}.
    We also note that a similar trend in \ce{CH3OH}/\ce{H2CO} ratio across the HD~100546 disk has been seen for the disk around the Herbig Ae star, HD~169142.
    \item Comparing our results with predictions from a chemical model of the HD~100546 disk, we find that the model results are consistent with the observed radial column density trends in \ce{CH3OH} and \ce{H2CO}, in that both decrease going from the inner to outer regions.
    The model predicts a steeper decrease in the column density of \ce{CH3OH} compared with that for \ce{H2CO}, which is also consistent with the observations. 
    The peak column density ratio (\ce{CH3OH}/\ce{H2CO}) in the inner region predicted at 1~Myr is of the same order of magnitude as that seen in the observations, while the peak value predicted at 2~Myr is approximately $1-2$ orders of magnitude lower.
\end{itemize}

The original detection of gas-phase methanol in the disk around HD~100546 was serendipitous and led to the conclusion that a significant fraction of the ice reservoir is inherited from an earlier evolutionary phase \citep{Booth2021}. Our results support this notion, as they suggest that we are seeing `fresh' ice sublimation in this disk as shown by the similar \ce{CH3OH}/\ce{H2CO} ratio seen in HD~100546 to that seen in Solar System comets.
%This is because the temperature of this Herbig disk is deemed too high ($\gtrsim 20$~K) to allow the in-situ formation of methanol ice via hydrogenation of CO ice \citep{Kama16}. 
However, since the original detection of \ce{CH3OH} in HD 100546, it has also been detected in the Herbig disks around HD 169142 \citep{Booth2023} and IRS 48 \citep{vanderMarel2021}.
Given the lack of gas-phase \ce{CH3OH} detected towards the Herbig disks, MWC 480~and HD~163296 \citep{Loomis2018,Carney2019,Yamato2024}, there is an apparent chemical diversity among Herbig objects that is likely related to their structure: the disks where gas-phase \ce{CH3OH} is detected all possess inner cavities that allow direct irradiation of the icy midplane thus revealing the ice composition.
These class of disks are now providing valuable insight into the composition of the ice reservoir which still remains elusive to observe in colder disks, including in the disks around T~Tauri stars.
The results reported here for HD~100546 show that the building blocks of prebiotic molecules (e.g., \ce{CH3OH}) are present during the epoch of planet formation. 
However, further high sensitivity and potentially longer wavelength observations of COMs targeting multiple Herbig disks are required in order to determine if this conclusion is generally applicable.

\section*{Acknowledgements}
We thank Ewine van Dishoeck for useful feedback on a draft of this manuscript. L.~Evans acknowledges financial support from the Science and Technology Facilities Council (grant number ST/X001016/1).
A.~S.~Booth is supported by a Clay Postdoctoral Fellowship from the Smithsonian Astrophysical Observatory.
C.~Walsh acknowledges financial support from the Science and Technology Facilities Council and UK Research and Innovation (grant numbers ST/X001016/1 and MR/T040726/1). J.~D.~Ilee acknowledges support from an STFC Ernest Rutherford Fellowship (ST/W004119/1) and a University Academic Fellowship from the University of Leeds. L.~Keyte is funded by UKRI guaranteed funding for a Horizon Europe ERC consolidator grant (EP/Y024710/1) Support for C.~J.~Law was provided by NASA through the NASA Hubble Fellowship grant No. HST-HF2- 51535.001-A awarded by the Space Telescope Science Institute, which is operated by the Association of Universities for Research in Astronomy, Inc., for NASA, under contract NAS5-26555. S.~Notsu is grateful for support from Grants-in-Aid for JSPS (Japan Society for the Promotion of Science) Fellows Grant Number JP23KJ0329 and MEXT/JSPS Grants-in-Aid for Scientific Research (KAKENHI) Grant Numbers JP20H05845, JP20H05847, JP23K13155 and JP24K00674. M.~Temmink acknowledges support from the ERC grant 101019751 MOLDISK. M.~Leemker is funded by the European Union (ERC, UNVEIL, 101076613). Views and opinions expressed are however those of the author(s) only and do not necessarily reflect those of the European Union or the European Research Council. Neither the European Union nor the granting authority can be held responsible for them.
This paper makes use of the following ALMA data: 2019.1.00193.S and 2021.1.00738.S. ALMA is a partnership of ESO (representing its member states), NSF (USA) and NINS (Japan), together with NRC (Canada), NSC and ASIAA (Taiwan) and KASI (Republic of Korea), in cooperation with the Republic of Chile. The Joint ALMA Observatory is operated by ESO, AUI/NRAO and NAOJ. The data used are publicly available on the ALMA Archive (https://almascience.nrao.edu/aq/) with project codes 2019.1.00193.S and 2021.1.00738.S. The model used is available on request from C.~Walsh. The data products, images, spectra and modelling results are available on request from L.~Evans.

\bibliographystyle{aasjournal}
\bibliography{biblio}

\appendix

\section{Moment 0 Maps}
\begin{figure*}[hbt!]
    \centering
    \includegraphics[width=0.24\textwidth]{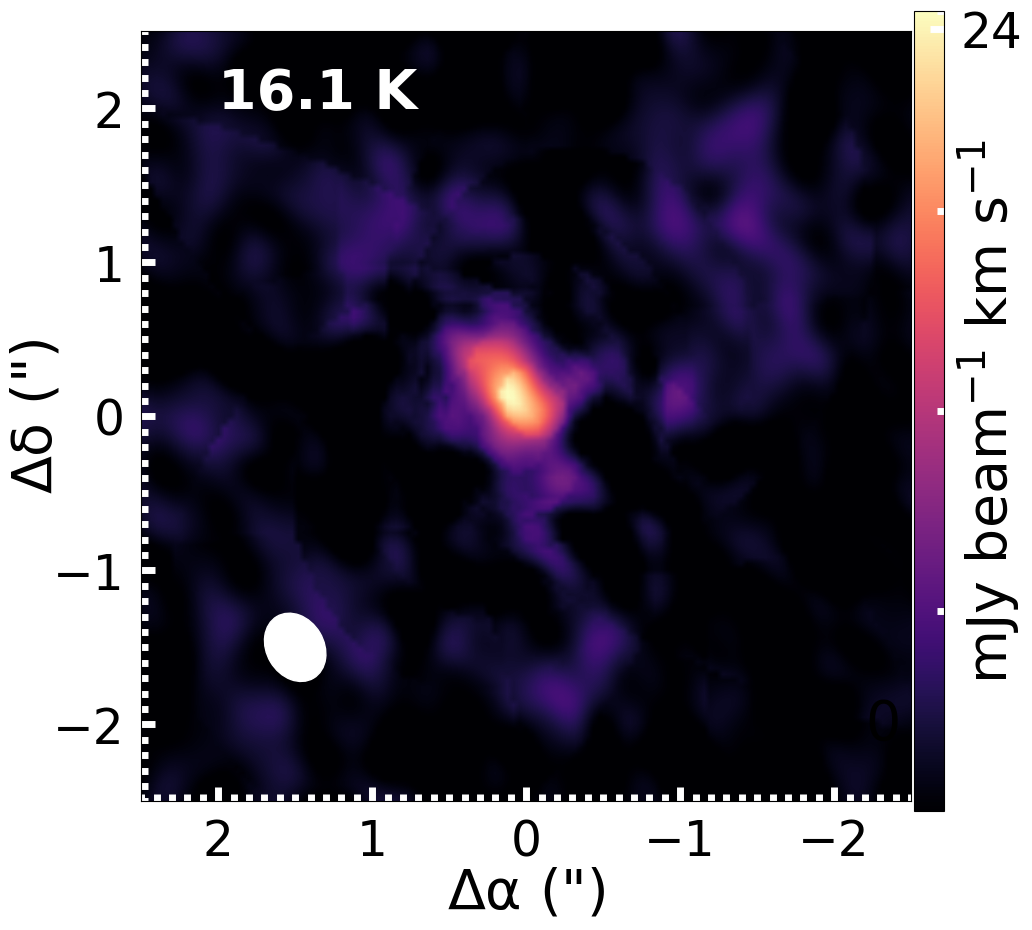}
    \includegraphics[width=0.24\textwidth]{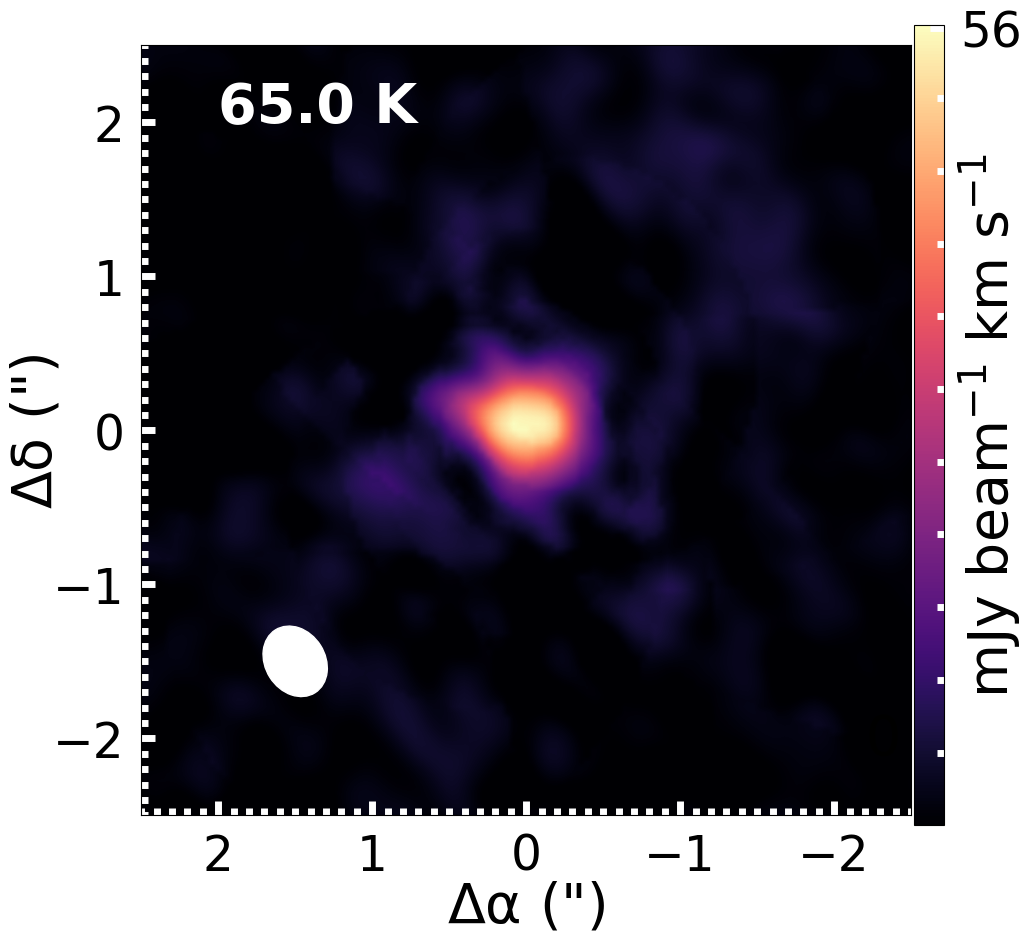}
    \includegraphics[width=0.24\textwidth]{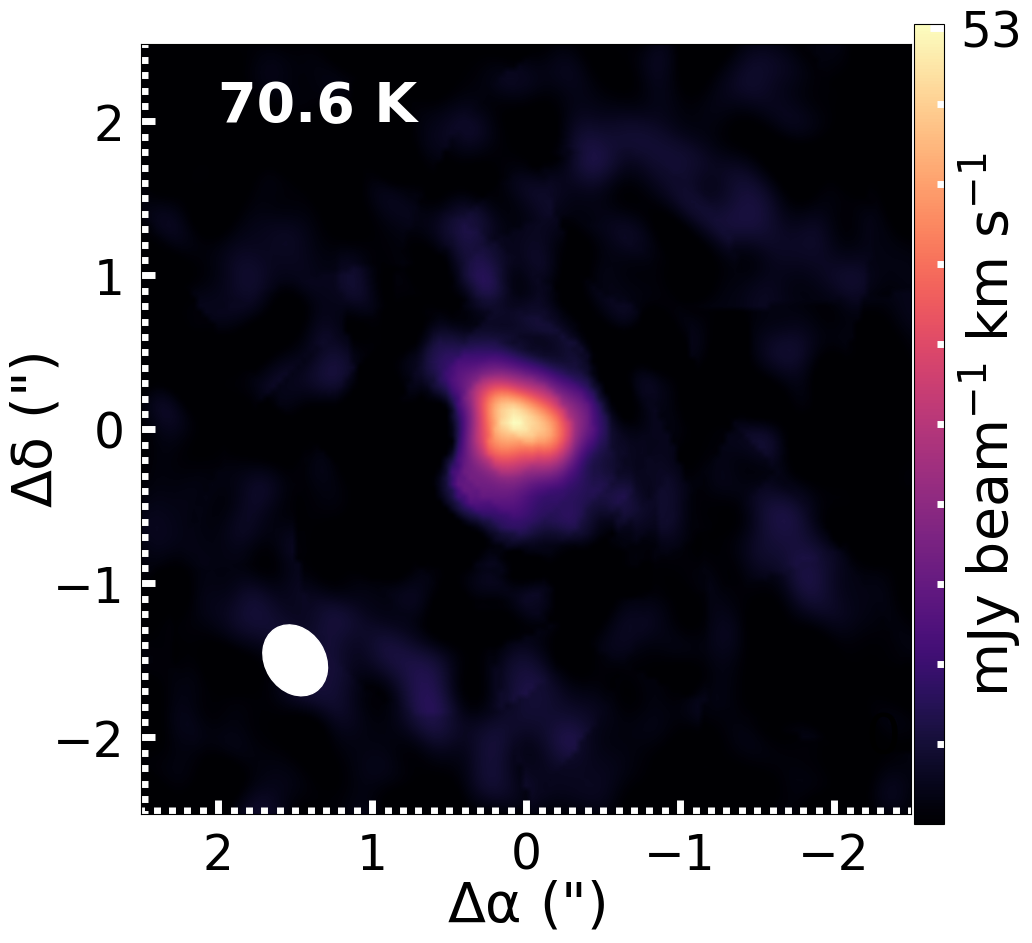}
    \includegraphics[width=0.24\textwidth]{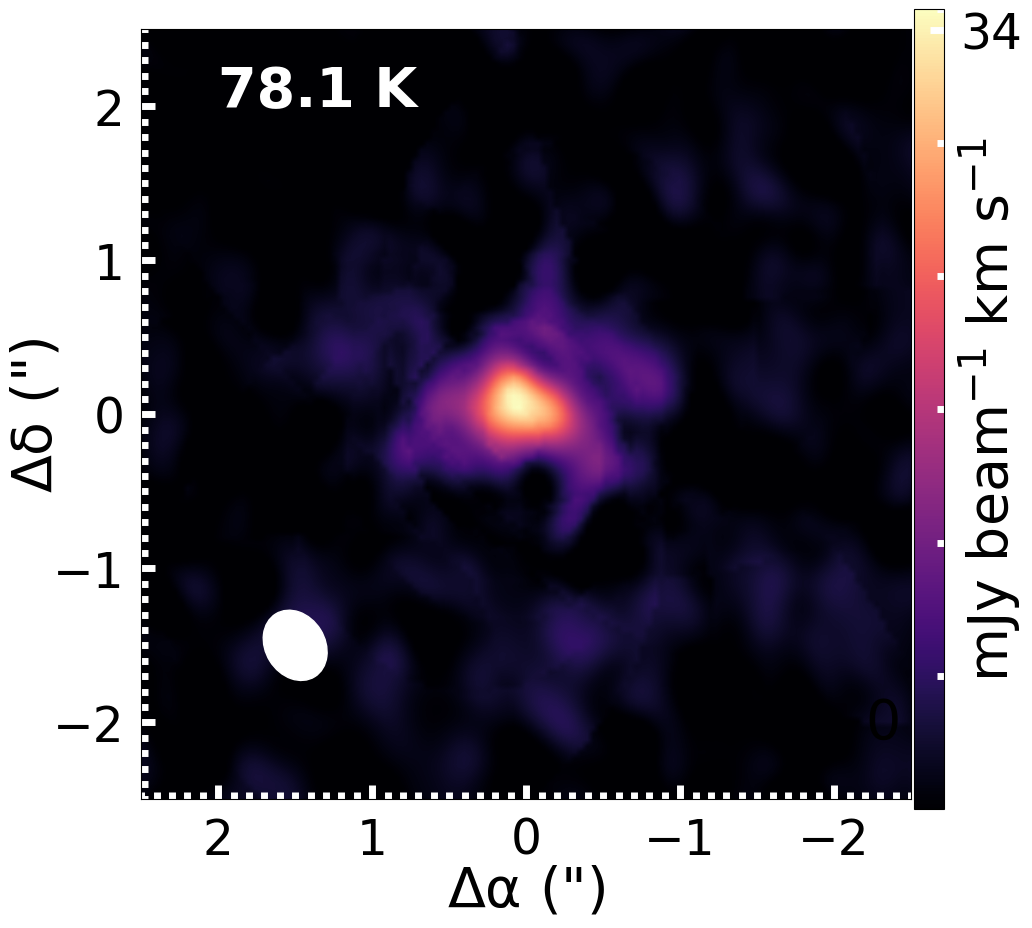}
    \includegraphics[width=0.24\textwidth]{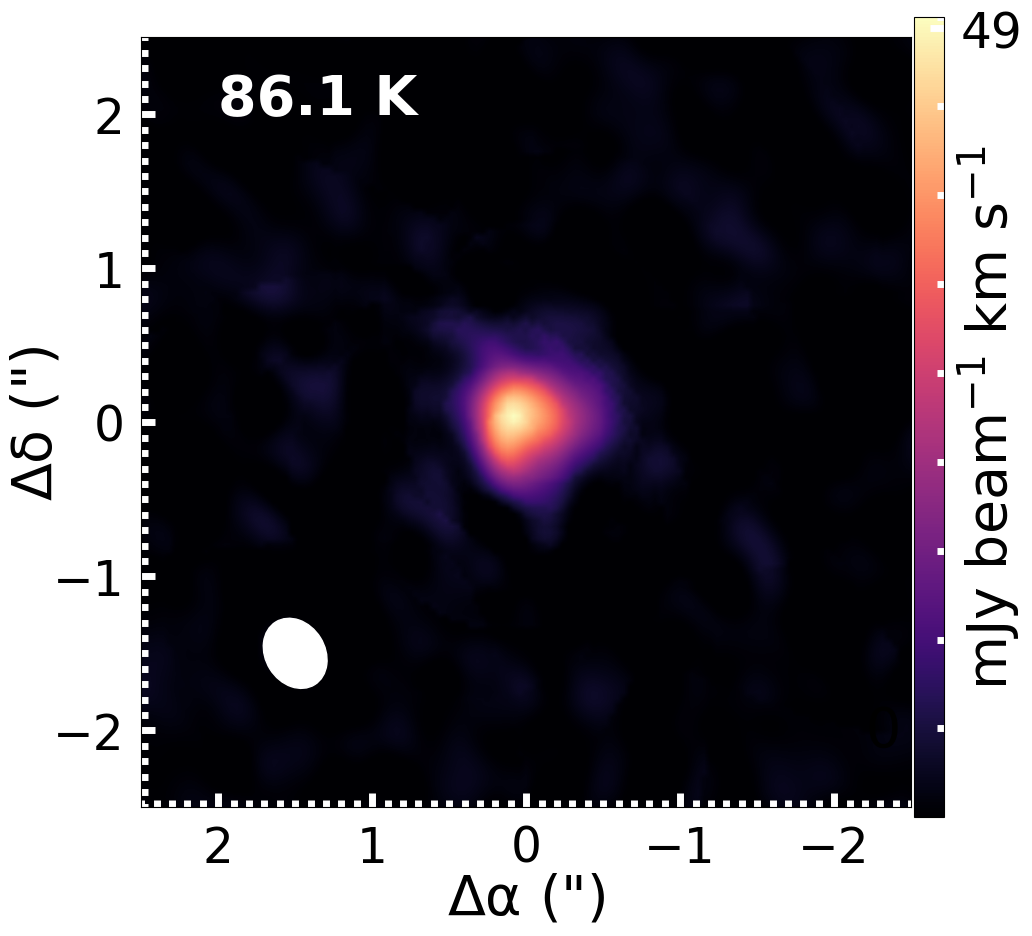}
    \includegraphics[width=0.24\textwidth]{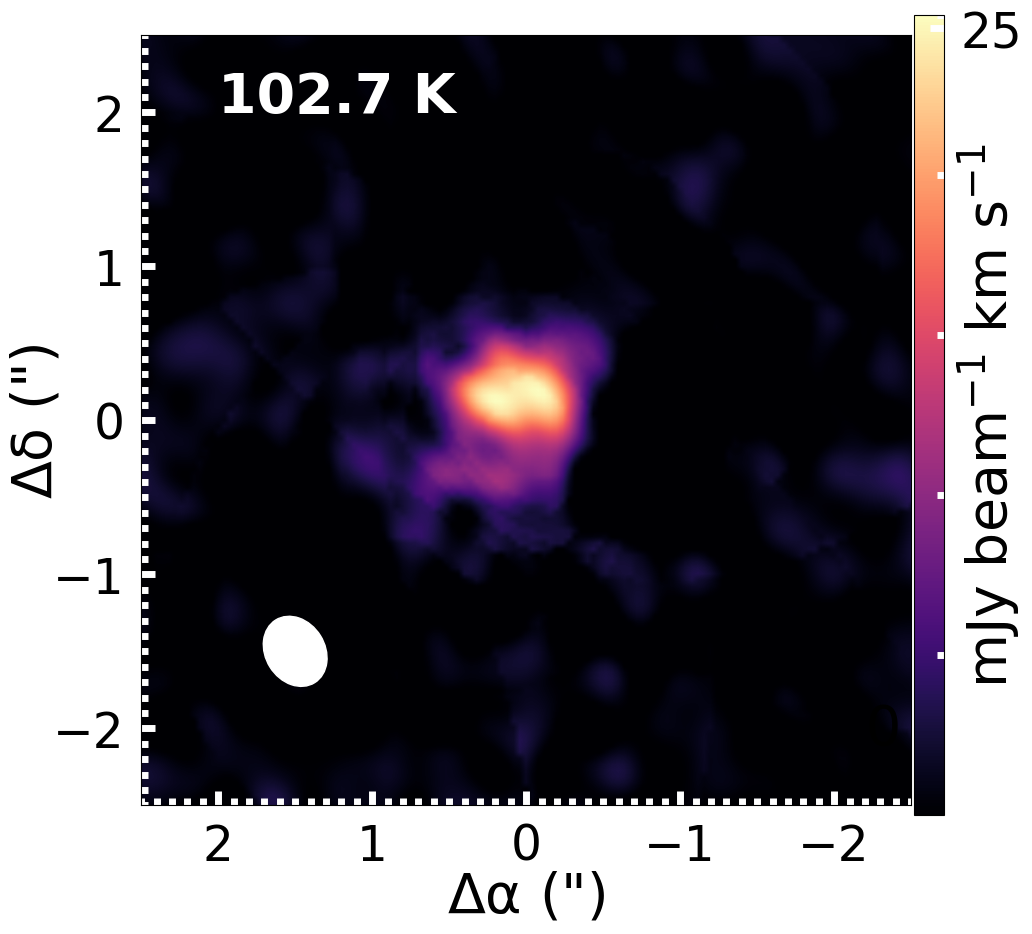}
    \includegraphics[width=0.24\textwidth]{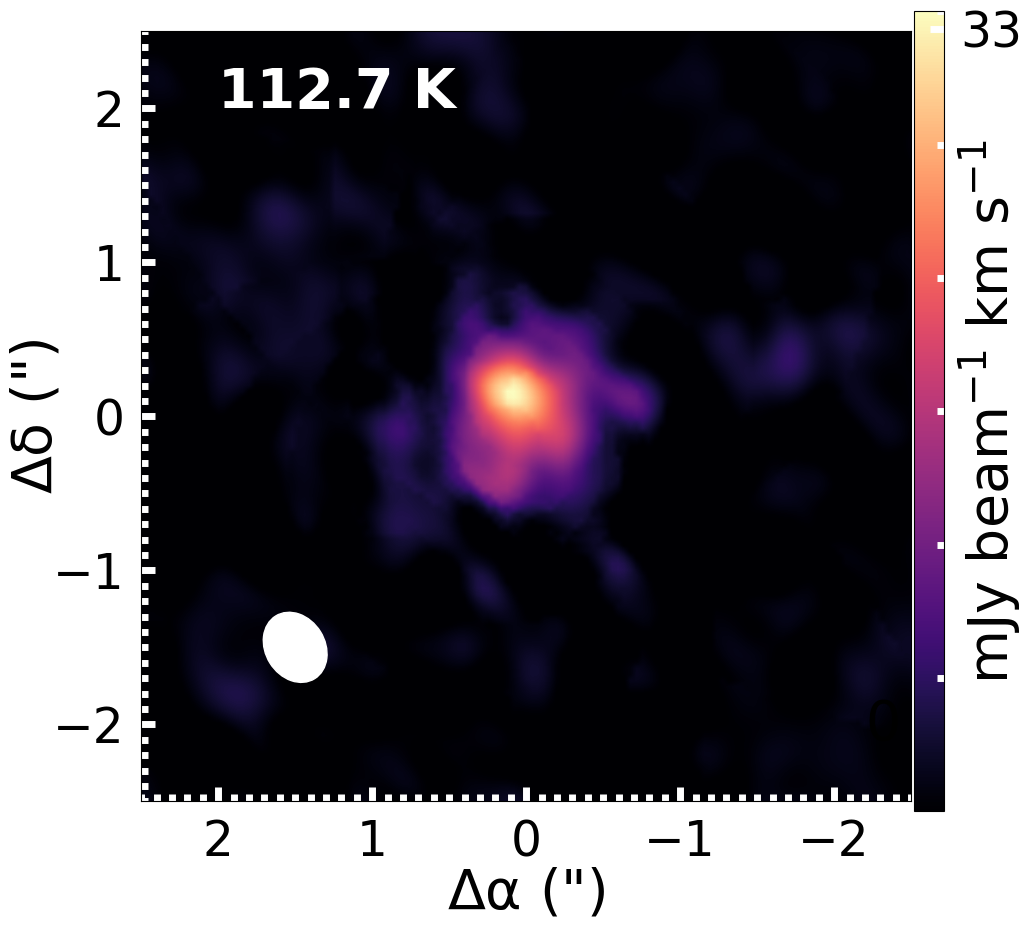}
    \includegraphics[width=0.24\textwidth]{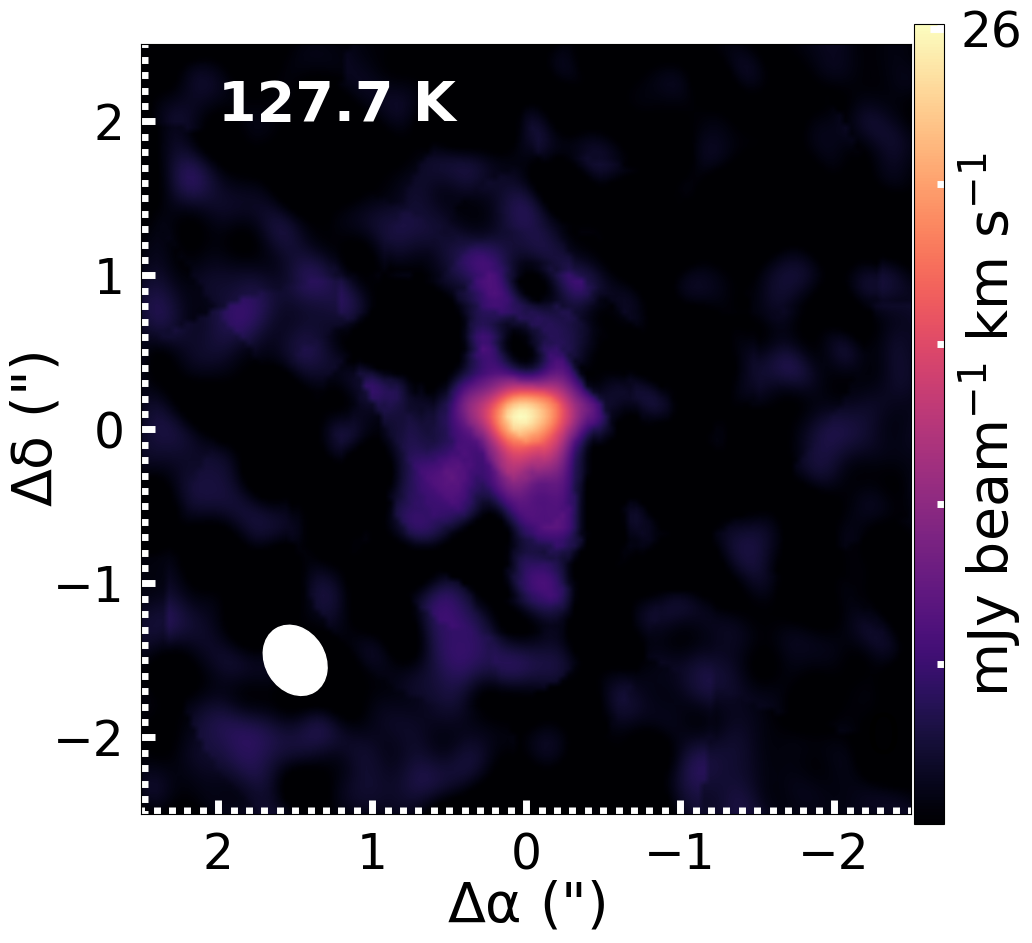}
    \includegraphics[width=0.24\textwidth]{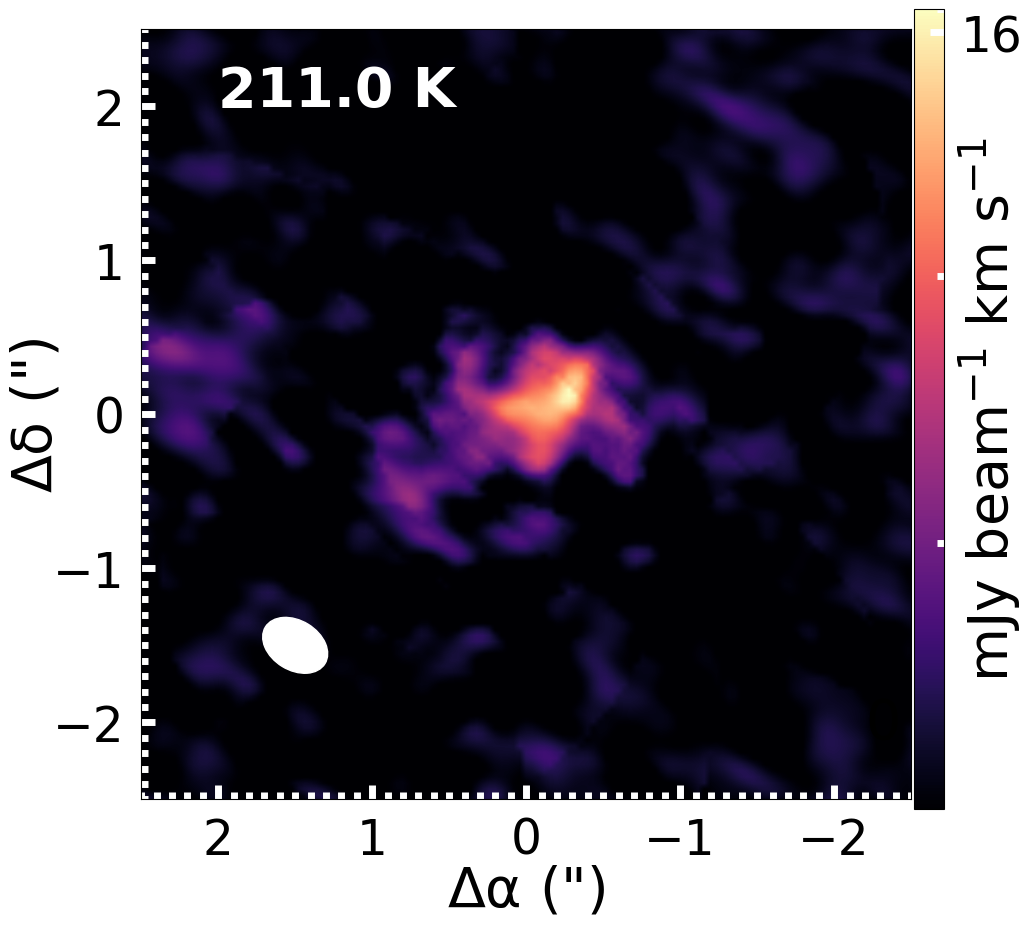}
    \includegraphics[width=0.24\textwidth]{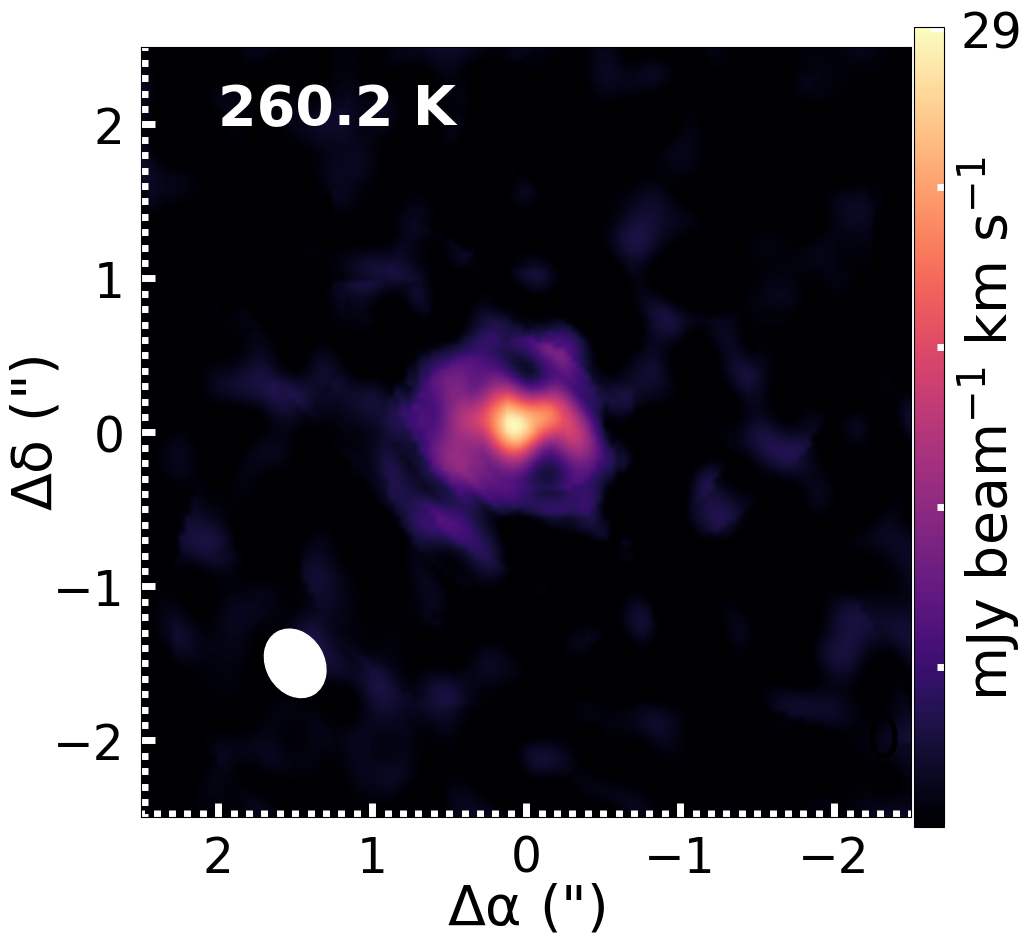}
    \caption{Moment 0 maps for all ten Cycle 8 \ce{CH3OH} transitions used in the flux extraction and rotational diagram analysis, in ascending order of $E_\mathrm{u}$. The synthesised beam is represented by the white ellipse in the bottom left corner.}
    \label{fig.mom0_cycle8}
\end{figure*}
\clearpage
\section{Radial Profiles}
\begin{figure*}[hbt!]
    \centering
    \includegraphics[width=\textwidth]{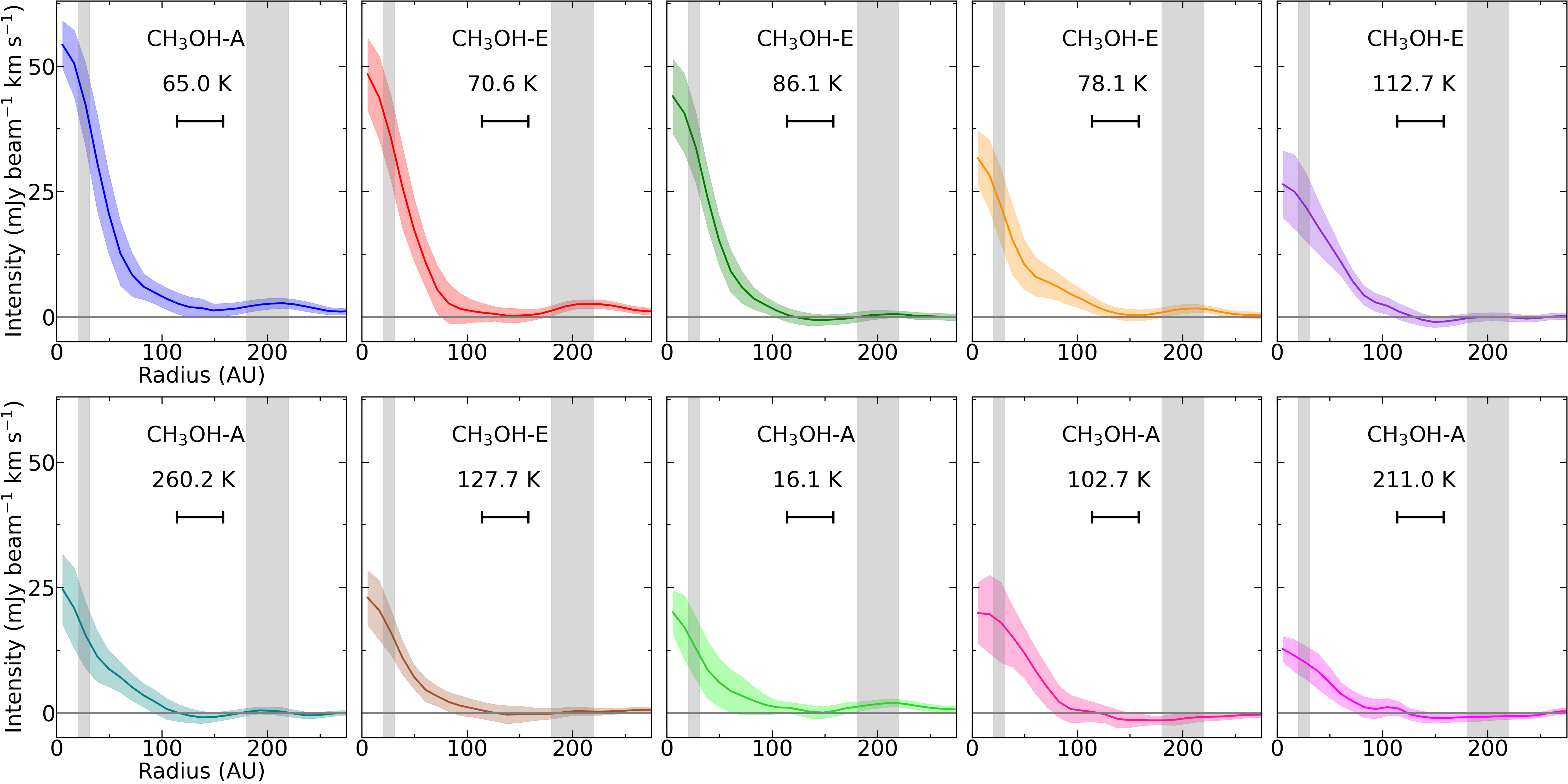}
    \caption{Azimuthally-averaged radial profiles for all of the Cycle 8 transitions of \ce{CH3OH} 
    along with associated error bars in order of decreasing peak intensity. The radial ranges of the peaks of the two dust rings seen in the 0.9 mm continuum emission between $\sim 20-31$~au and $\sim 180-220$~au are denoted by the grey shaded regions. The horizontal bars show the FWHM of the synthesised beam.}
    \label{fig.radprof8}
\end{figure*}
\clearpage
\section{Spectra}
\begin{figure*}[hbt!]
    \centering
    \includegraphics[width=0.8\textwidth,height=8cm]{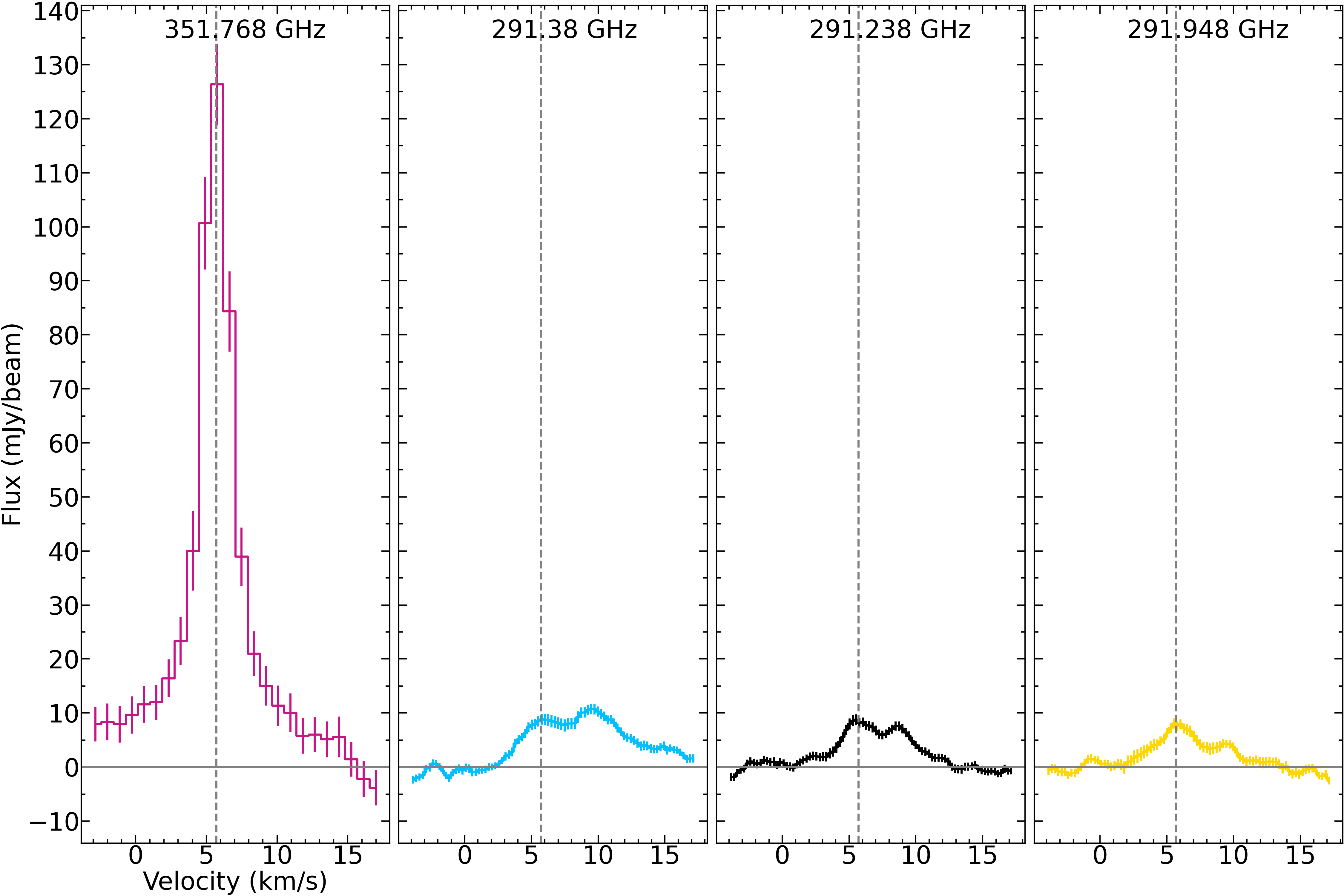}
    \caption{\ce{H2CO} spectral lines extracted using \texttt{gofish} witin 110 au for the full velocity range of the spw ($-$4.2 to 17.4 km~s$^{-1}$). The grey dashed line indicates the source velocity at 5.7 km~s$^{-1}$.}
    \label{fig.h2co_inner_spec}
\end{figure*}
\begin{figure*}[hbt!]
    \centering
    \includegraphics[width=0.8\textwidth,height=8cm]{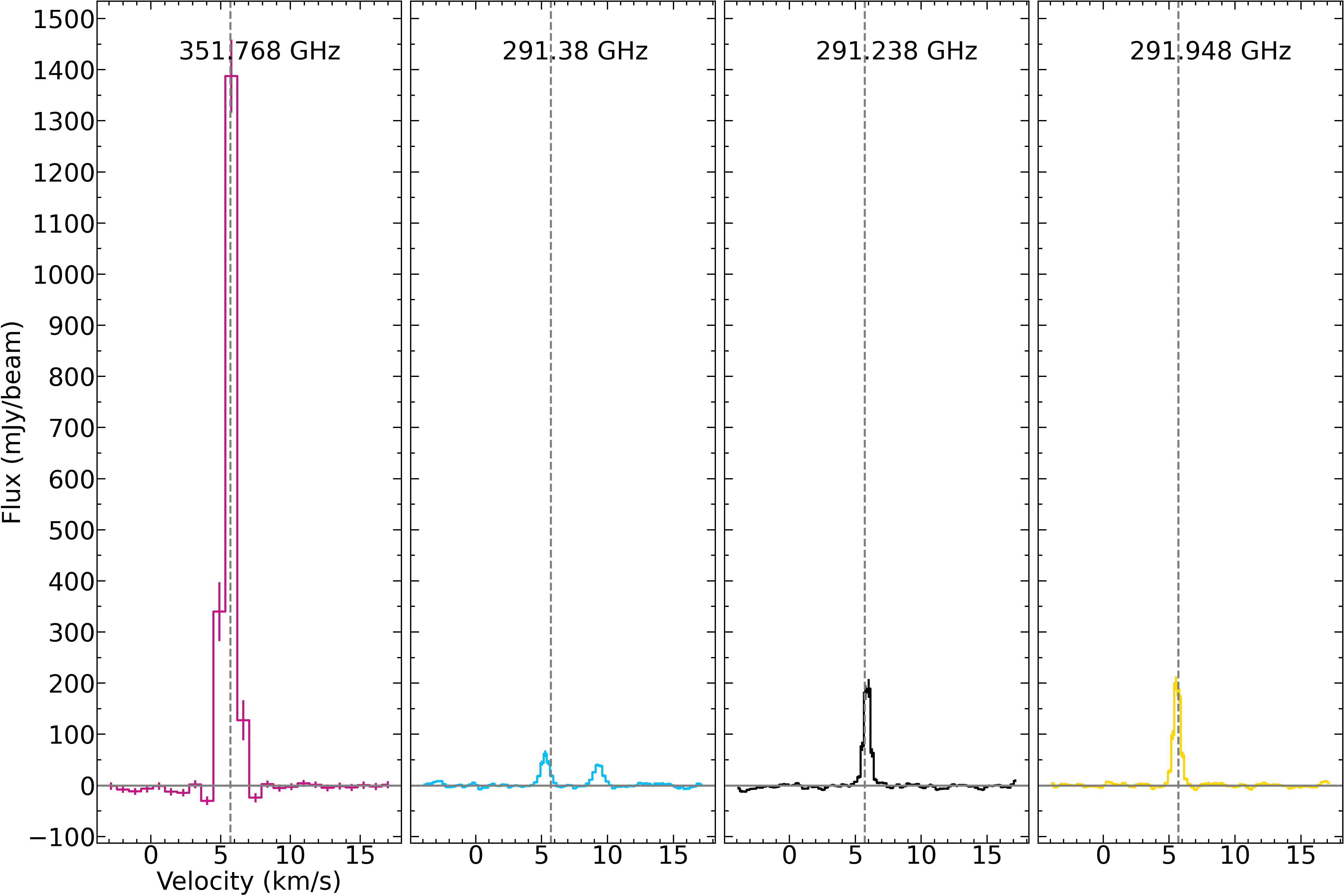}
    \caption{\ce{H2CO} spectral lines extracted using \texttt{gofish} between 180 and 260 au for the full velocity range of the spw ($-$4.2 to 17.4 km~s$^{-1}$). The grey dashed line indicates the source velocity at 5.7 km~s$^{-1}$.}
    \label{fig.h2co_outer_spec}
\end{figure*}
\clearpage
\section{Corner Plots}
\begin{figure*}[hbt!]
    \centering
    \includegraphics[width=0.45\textwidth]{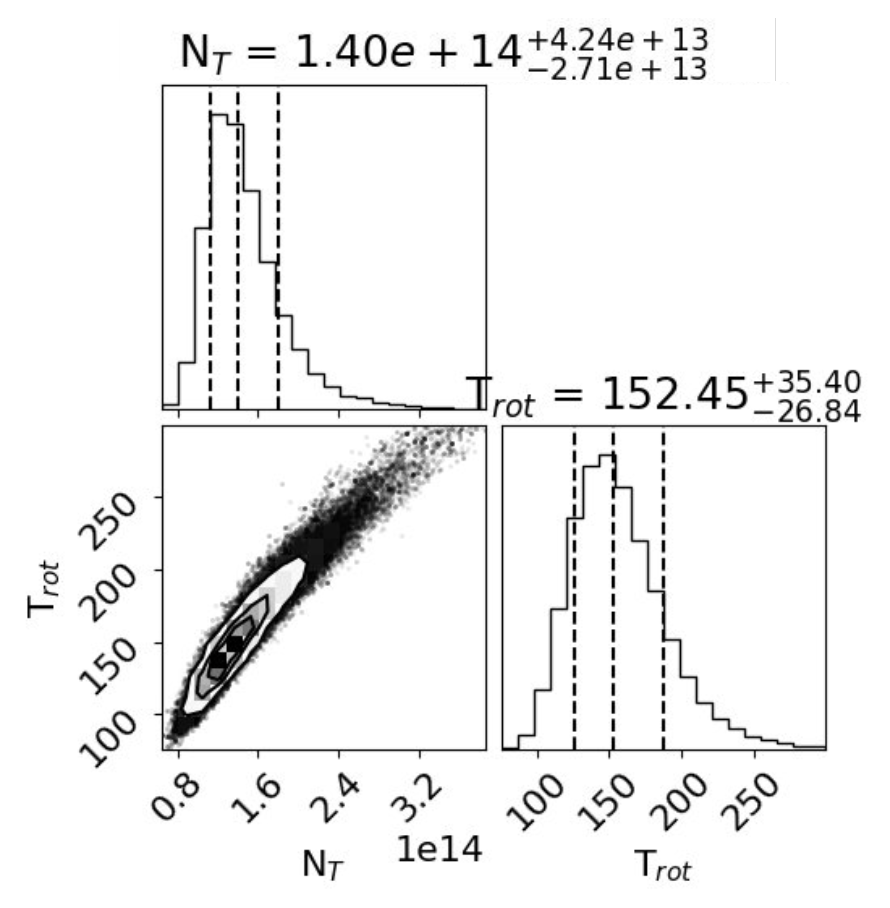}
    \includegraphics[width=0.45\textwidth]{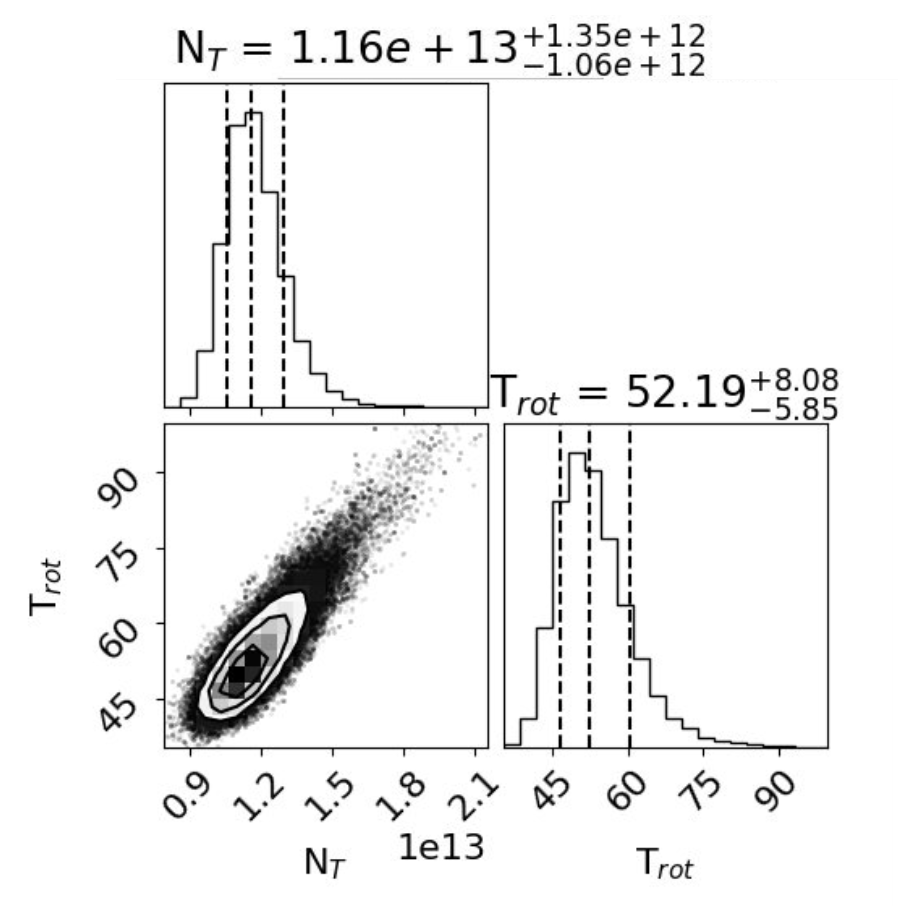}
    \includegraphics[width=0.45\textwidth]{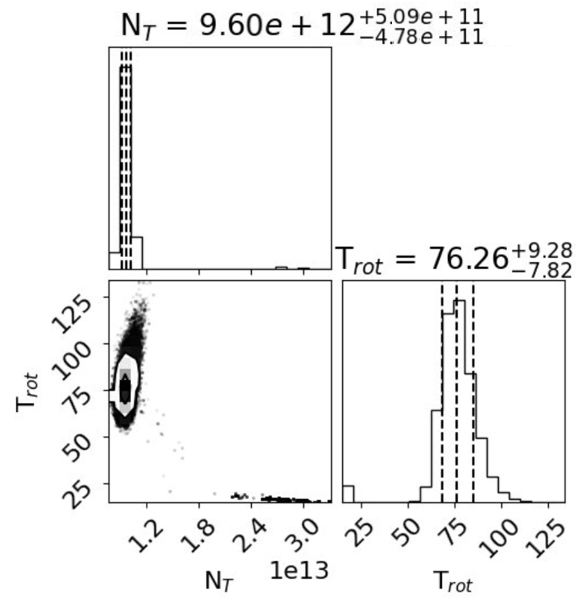}
    \includegraphics[width=0.45\textwidth]{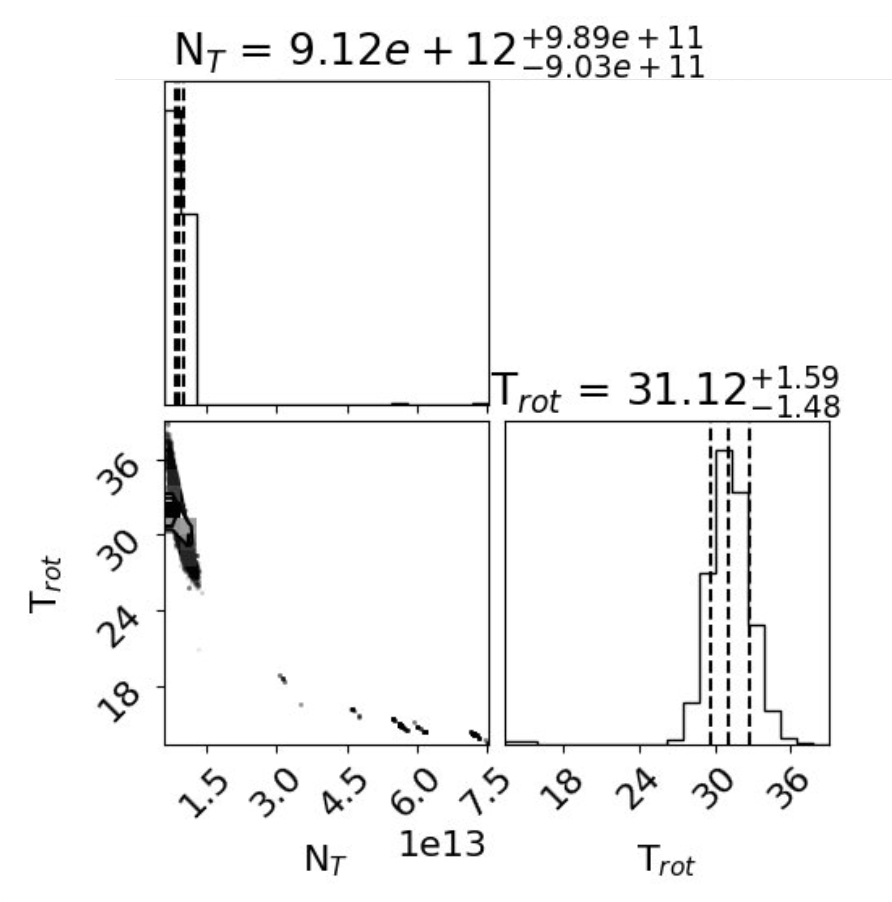}
    \caption{Corner plots generated from the MCMC best fits used to produce the rotational diagrams shown in Fig.~\ref{fig.tempconstrain}. Clockwise from top left: \ce{CH3OH} inner component, \ce{CH3OH} outer component, \ce{H2CO} outer component and \ce{H2CO} inner component.}
    \label{fig.cornerplots}
\end{figure*}
\clearpage
\section{Desorption Temperature Calculation}
Here we detail the methodology used to calculate the desorption temperatures of \ce{CH3OH} and \ce{H2CO} in the midplane of HD~100546. The rate of desorption ($k_\mathrm{d}$) per second from dust grains can be calculated using Eq. \ref{eq.desorption} \citep[]{Hasegawa1992},
\begin{equation}
\rm{k_d=\nu_0.exp\Big(-\frac{E_d}{T_d}\Big).}
\label{eq.desorption}
\end{equation}
In Eq. \ref{eq.desorption}, $E_\mathrm{d}$ represents the binding energy of the species and $T_\mathrm{d}$ represents the dust grain temperature, both in K, while $\nu_\mathrm{0}$, the characteristic vibrational frequency of the adsorbed species in its potential well, is given by
\begin{equation}
\rm{\nu_0=\sqrt{\frac{2n_sE_d}{\pi^2m}},}
    \label{eq.desparttwo}
\end{equation}
where $n_\mathrm{s}$ is the number density of surface sites on each dust grain (cm$^{-2}$), $E_\mathrm{d}$ is the binding energy (erg) and m is the mass of the species (g). We assume that the binding energies of \ce{CH3OH} and \ce{H2CO} are 3820~K and 3260~K, %4930 and 2050
respectively, as recommended in \citet{Penteado2017} using data from \citet{Collings04} and \citet{Noble12}. Meanwhile, the rate of freeze-out per second is given by
\begin{equation}
\rm{k_f=<v>\sigma_dn_dS,}
\label{eq.freezeout}
\end{equation}
where $\sigma_d$ is the cross sectional area (cm$^2$) of a dust grain with radius $a$ (hence $\sigma$ = $\pi a^2$), $n_\mathrm{d}$ is the number density of the dust grains (cm$^{-3}$), $S$ is the sticking coefficient (for this case it is safe to assume a value of $S=1$ for both species) and $<\mathrm{v}>$ is the thermal velocity (cm~s$^{-1}$) of the species of mass $m$ (g) and at gas temperature $T_\mathrm{g}$ (K) and is given by
\begin{equation}
\rm{<v>=\sqrt{\frac{8k_BT_g}{\pi m}},}
\label{eq.freezeparttwo}
\end{equation}
where $k_\mathrm{B}$ is the Boltzmann constant in cgs units. We assume a range of values for the density of the gas in the midplane, $n_\mathrm{gas}$, between 10$^{10}$ and 10$^{12}$ cm$^{-3}$ \citep[]{Kama16}. Using the assumed fractional abundance of dust grains of 1.3$\times$10$^{-12}$, we therefore obtain a range of values for $k_\mathrm{f}$.
In order to obtain the desorption temperature, we assume that $k_\mathrm{d}=k_\mathrm{f}$, as approximately equal quantities of the species will be in the gas and ice phases \citep[]{vantHoff2017}.% Therefore, one can write:

\begin{figure*}[!htp]
    \centering
    \includegraphics[width=\textwidth]{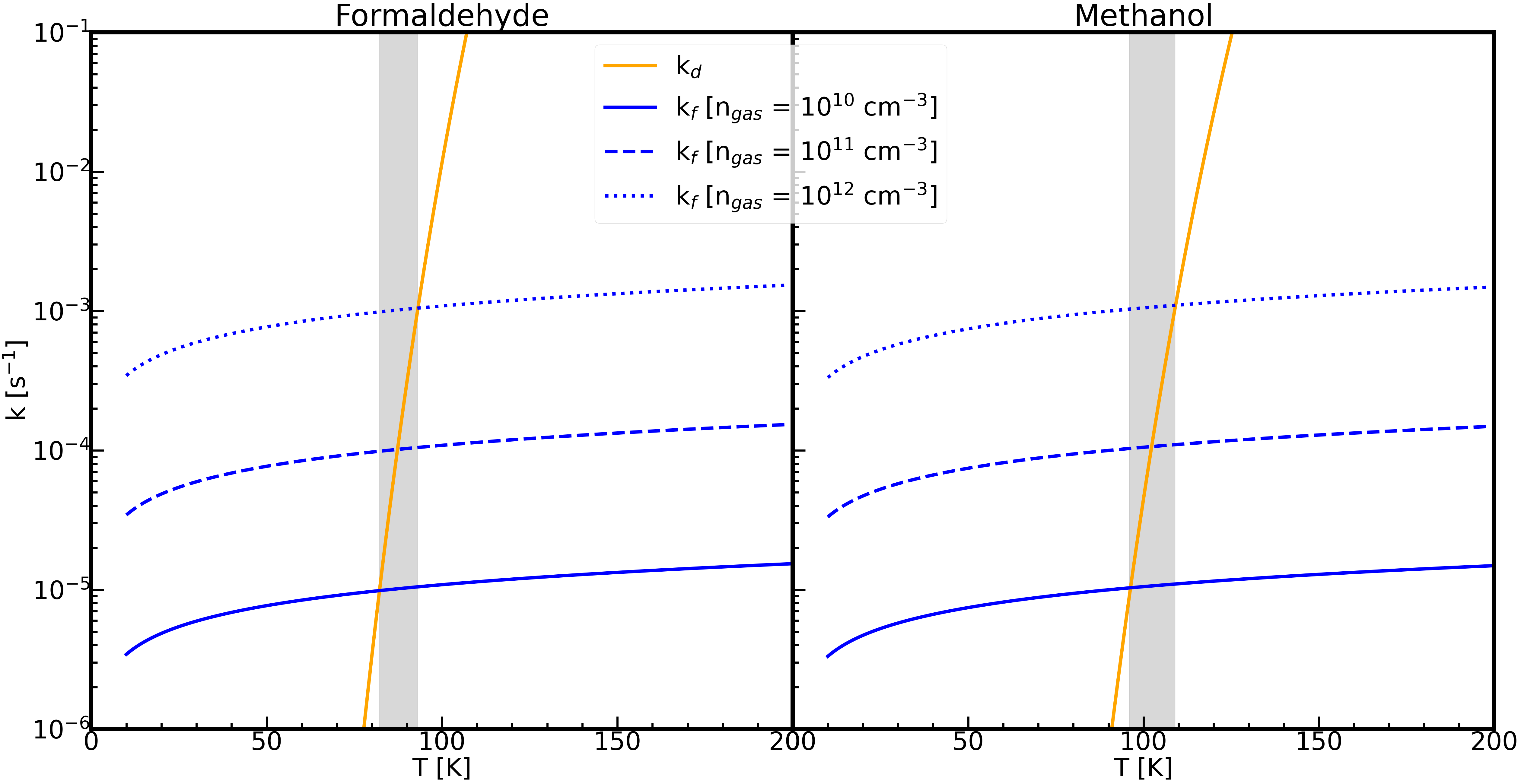}
    \caption{Plot showing the rate of desorption (orange) and gas density-dependent rates of freeze-out (blue) for \ce{H2CO} (left panel) and \ce{CH3OH} (right panel); the x-values of the intersections are equal to the desorption temperatures.}
    \label{fig.desorption}
\end{figure*}

%\begin{equation}
%\sqrt{\frac{2n_sE_d}{\pi^2m}}.exp\Big(-\frac{E_d}{T_{des}}\Big)=\sqrt{\frac{8k_BT_{des}}{\pi m}}.\sigma_gn_dS.
%\label{eq.tdes}
%\end{equation}
In order to obtain $T_\mathrm{des}$, we therefore plot $k_\mathrm{d}$(T) and $k_\mathrm{f}$(T) on a single axis (see Fig. \ref{fig.desorption}) and obtain the range of values for $T_\mathrm{des}$ as the temperature values where $k_\mathrm{d}$ intercepts the multiple $k_\mathrm{f}$ curves (according to our assumed values of $n_\mathrm{gas}$).
According to Fig. \ref{fig.desorption}, using the recommended binding energy for \ce{CH3OH} (3820~K; \citealt{Collings04}) leads to a higher range of desorption temperatures from 96$-$109 K % 124-140 K (UMIST)
, depending on the gas density. Meanwhile \ce{H2CO} would desorb at temperatures ranging between 82$-$93 K \citep[]{Collings04,Noble12,Penteado2017}%51-58 K (UMIST)
.
\clearpage
\section{Rotational Temperature Discussion}
\begin{figure*}[hbt!]
    \centering
    \includegraphics[width=0.8\textwidth]{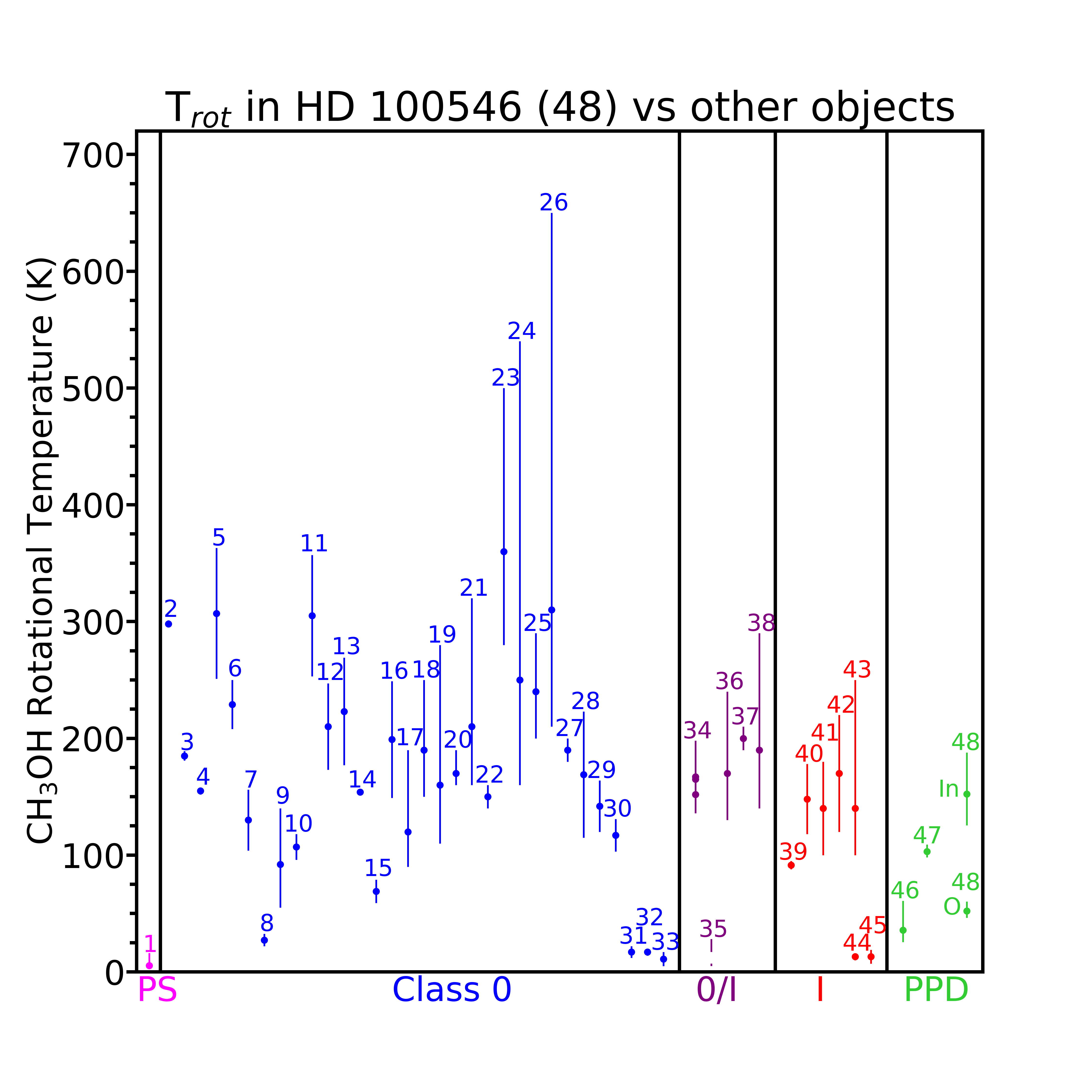}
    \caption{$T_\mathrm{rot}$ values for \ce{CH3OH} measured towards the inner and outer regions of HD~100546 compared to values measured towards objects of varying evolutionary stages. Numbers have been used to denote each object; please refer to Appendix G (Table \ref{tab.refs}) for the object and reference(s) associated with each number. Note that "PS" represents prestellar core values and "PPD" represents those of protoplanetary disks (both T Tauri and Herbig). % are noted in brackets where each appear in this caption. Prestellar (L1544 (1): \cite{Punanova2018}); Class 0 (G208.68-19.20N1 (2), G211.47-19.27S (3), G210.49-19.79W (4): \cite{Hsu2020}, IRAS 4A2 (A) and 2A (B): \cite{Taquet2015}, IRAS 16293-A (5, C) and B (D): \cite{Jorgensen20}, \cite{Persson2018}, \cite{Manigand2020}, Cep E-mm (E): \cite{Ospina-Zamudio2018}, HH212 (6, F): \cite{Lee2022}, Orion B9-SMM3 (G): \cite{Miettinen2016}, L1448-C (7), IRAS 2A1 (8), IRAS 4A2 (9), IRAS 4B (10), SerpM-S68N (11), SerpM-SMM4B (12), SerpS-MM18a (13), SerpS-MM18b (14), L1157 (15): \cite{Belloche2020}, Per-emb 26 (16), Per-emb 22A (17), Per-emb 22B (18), Per-emb 12B (19), Per-emb 13 (20), Per-emb 21 (21), Per-emb 18 (22), B1-bS (23), Per-emb 11A (24), Per-emb 11C (25), Per-emb 29 (26): \cite{Yang2021}, FIR6c-a (27), MMS9-a (28), MMS5 (29): \cite{Bouvier2022}, B1-c (30), IRAS 03245+3002 (31), L1014 IRS (32): \cite{Graninger2016}); Class 0/I (BHB07-11 (33) across three emission components: \cite{Vastel2022}, a sample of 14 Class 0/I proto-brown dwarfs (34): \cite{Riaz2023}, Per-emb 20 (35), Per-emb 44 (36), Per-emb 27 (37): \cite{Yang2021}); Class I (L1551-IRS5 (38): \cite{Mercimek2022}, SVS13-A VLA4A (39) and VLA4B (40): \cite{Bianchi2022}, Per-emb 35A (41): \cite{Yang2021}, B1-a (42), B5 IRS1 (43): \cite{Graninger2016}, DG Tau B (H): \cite{Garufi2020}, HL Tau (J): \cite{Garufi2022}, IRAS 040302+2247 (K): \cite{Podio2020}); Class II T Tauri objects (TW Hya (L, 44): Ilee et al. in prep., DG Tauri (M): \cite{Podio2019}); Class II Herbig objects (HD~169142 (N): \cite{Booth2023}, HD~163296 (O): \cite{Carney2019}); as well as a sample of 38 Solar System comets: \cite{Lippi2024}. The rotational temperature and ratio values for HD~100546 are denoted as 45 and P, respectively. 
    Separate $T_\mathrm{rot}$ measurements towards the inner and outer emission components of HD~100546 (48) are denoted "In" and "O", respectively.}
    \label{fig.trotcomp}
\end{figure*} 
\begin{table*}[hbt!]
    \caption{Associated letter and/or number along with reference(s) to each object in Figs. \ref{fig.coldenscomp} and \ref{fig.trotcomp}.}
    \resizebox{\textwidth}{!}{\begin{tabular}{c c c c c}
    \hline
    \hline
    Stage & Name & Number & Letter & Reference \\
    \hline
    \hline
    Prestellar & L1544 & 1 & --- & \cite{Punanova2018} \\
    \hline
    Class 0 & G208.68-19.20N1 & 2 & --- & \cite{Hsu2020} \\
    & G211.47-19.27S & 3 & --- & \cite{Hsu2020} \\
    & G210.49-19.79W & 4 & --- & \cite{Hsu2020} \\
    & IRAS 4A2 & 5 & A & \cite{Belloche2020}, \cite{Taquet2015} \\
    & IRAS 2A1 & 6 & B & \cite{Belloche2020}, \cite{Taquet2015} \\
    & IRAS 16293-A & 7 & C & \cite{Jorgensen20}, \cite{Persson2018}, \cite{Manigand2020} \\
    & IRAS 16293-B & --- & D & \cite{Jorgensen20}, \cite{Persson2018}, \cite{Manigand2020} \\
    & Cep E-mm & 8 & E & \cite{Ospina-Zamudio2018} \\
    & HH212 & 9 & F & \cite{Lee2022} \\
    & Orion B9-SMM3 & --- & G & \cite{Miettinen2016} \\
    & L1448-C & 10 & --- & \cite{Belloche2020} \\
    & IRAS 4B & 11 & --- & \cite{Belloche2020} \\
    & SerpM-S68N & 12 & --- & \cite{Belloche2020} \\
    & SerpM-SMM4B & 13 & --- & \cite{Belloche2020} \\
    & SerpS-MM18a & 14 & --- & \cite{Belloche2020} \\
    & SerpS-MM18b & 15 & --- & \cite{Belloche2020} \\
    & L1157 & 16 & --- & \cite{Belloche2020} \\
    & Per-emb 26 & 17 & --- & \cite{Yang2021} \\
    & Per-emb 22A & 18 & --- & \cite{Yang2021} \\
    & Per-emb 22B & 19 & --- & \cite{Yang2021} \\
    & Per-emb 12B & 20 & --- & \cite{Yang2021} \\
    & Per-emb 13 & 21 & --- & \cite{Yang2021} \\
    & Per-emb 21 & 22 & --- & \cite{Yang2021} \\
    & Per-emb 18 & 23 & --- & \cite{Yang2021} \\
    & B1-bS & 24 & --- & \cite{Yang2021} \\
    & Per-emb 11A & 25 & --- & \cite{Yang2021} \\
    & Per-emb 11C & 26 & --- & \cite{Yang2021} \\
    & Per-emb 29 & 27 & --- & \cite{Yang2021} \\
    & FIR6c-a & 28 & --- & \cite{Bouvier2022} \\
    & MMS9-a & 29 & --- & \cite{Bouvier2022} \\
    & MMS5 & 30 & --- & \cite{Bouvier2022} \\
    & B1-c & 31 & --- & \cite{Graninger2016} \\
    & IRAS 03245+3002 & 32 & --- & \cite{Graninger2016} \\
    & L1014 IRS & 33 & --- & \cite{Graninger2016} \\
    \hline
    Class 0/I & BHB07-11 & 34$^{(a)}$ & --- & \cite{Vastel2022} \\
    & 14 Proto-brown dwarfs & 35$^{(b)}$ & --- & \cite{Riaz2023} \\
    & Per-emb 20 & 36 & --- & \cite{Yang2021} \\
    & Per-emb 44 & 37 & --- & \cite{Yang2021} \\
    & Per-emb 27 & 38 & --- & \cite{Yang2021} \\
    \hline
    Class I & V883 Ori & 39 & --- & \cite{vantHoff2018}, \cite{Lee2019} \\
    & L1551-IRS5 & 40 & --- & \cite{Mercimek2022} \\
    & SVS13-A VLA4A & 41 & --- & \cite{Bianchi2022} \\
    & SVS13-A VLA4B & 42 & --- & \cite{Bianchi2022} \\
    & Per-emb 35A & 43 & --- & \cite{Yang2021} \\
    & B1-a & 44 & --- & \cite{Graninger2016} \\
    & B5 IRS1 & 45 & --- & \cite{Graninger2016} \\
    & DG Tau B & --- & H & \cite{Garufi2020} \\
    & HL Tau & --- & J & \cite{Garufi2022} \\
    & IRAS 040302+2247 & --- & K & \cite{Podio2020} \\
    \hline
    T Tauri & TW Hya & 46 & L & Ilee et al. (in prep). \\
    & DG Tauri & --- & M & \cite{Podio2019} \\
    \hline
    Herbig & HD~169142 & --- & N$^{(c)}$ & \cite{Booth2023} \\
    & HD~163296 & --- & O & \cite{Carney2019} \\
    & IRS 48 & 47 & P & \cite{vanderMarel2021}, \cite{Booth2024IRS}, \cite{Temmink2024} \\
    & HD~100546 & 48$^{(c)}$ & Q$^{(c)}$ & This work \\
    %\hline
    %Comets & 38 Solar System comets & --- & $^{(e)}$ & \cite{Lippi2024} \\ 
    \hline
    \hline
    \end{tabular}}
    \label{tab.refs}
    \footnotesize{$^{(a)}$$T_\mathrm{rot}$ values are measured for three emission components, each represented with a separate point on Fig. \ref{fig.trotcomp}\\
    $^{(b)}$Two $T_\mathrm{rot}$ values were measured; these are both shown in Fig. \ref{fig.trotcomp} as average values over the PBDs in which each component appears\\
    $^{(c)}$Separate values are measured towards inner and outer components\\
    %$^{(d)}$H$_2$CO is possibly optically thick which would overestimate CH$_3$OH/H$_2$CO\\}
    %$^{(e)}$All comet measurements with error bars are represented in Fig. \ref{fig.tempcomp} in orange, the median value is represented by a star}
    }
\end{table*}
Cold methanol has been detected as early as the prestellar stage, with a rotational temperature of $5.3^{+2.2}_{-11}$~K measured towards L1544 \citep{Punanova2018}.
Values towards Class 0 objects vary greatly according to literature, with values greater than 300 K measured towards embedded protostars in Perseus \citep{Yang2021} and values as low as 11 K measured towards deeply embedded protostars in other clouds \citep{Graninger2016}.
Generally speaking, however, the Class 0 measurements are higher than those measured towards L1544 and fall in line with the majority of measured values towards Class 0/I and I objects, which tend to lie in the approximate range of 100-300 K%, regardless of the presence of a hot corino
.
There are also outliers in these slightly more evolved objects, however.
For example, \cite{Riaz2023} measured a sample of 14 Class 0/I proto-brown dwarfs (PBDs) and found that, while a cold emission component with an average $T_\mathrm{rot}$ of 5-7 K exists in all of them, 78\% of the sample also exhibited a warmer component with an average measured $T_\mathrm{rot}$ of 17-28 K.
Both of these measurements, however, are lower than that generally found among other Class 0/I objects according to Fig. \ref{fig.trotcomp}.
Similarly, \cite{Graninger2016} measured a relatively low $T_\mathrm{rot}$ value of 13 K towards both of the Class I sources B1-a and B5 IRS1.
There is, therefore, perhaps some dependence of $T_\mathrm{rot}$ on the nascent cloud and/or the size of the object. It should further be noted that some of the observations presented in Table~\ref{tab.refs}, whether due to being single-dish or otherwise, fail to resolve the innermost regions in which thermal desorption is occurring.
%Class 0 objects have T$_\mathrm{rot}$ values typically ranging between approximately 130-200 K \citep[see recent work by e.g.,][]{Hsu2020,Manigand2020,Vastel2022}), regardless of the presence of a hot corino (as indicated by rich emission from COMs), while values up to approximately 170 K have been measured for protobinary objects \citep{Manigand2020,Vastel2022}. 
%The value measured towards the Class I protostar L1551-IRS5 is $148 \pm 30$~K \citep{Mercimek2022} and so is consistent with the lower range of the Class 0 measurements. 
Meanwhile, a lower $T_\mathrm{rot}$ value of $35.8^{+25.1}_{-10.3}$~K has recently been measured towards the T Tauri object TW Hya (Ilee et al. in prep.), while a value of $103^{+6}_{-5}$~K was measured towards the Herbig object IRS~48 \citep[]{vanderMarel2021}; intriguingly, this is lower than the value for $T_\mathrm{rot}$ measured in \ce{H2CO} towards this disk, which is the opposite trend to that seen in HD~100546. This trend in IRS~48 has been corroborated by \citet{Temmink2024}, although there is large scatter in these results due to the non-negligible effects of sub-thermal excitation.

As illustrated in Fig.~\ref{fig.trotcomp}, our results for the inner region of HD 100546 are consistent with the hot component of emission from the majority of the Class 0, 0/I and I objects.
Hot core/corino emission is also thought to arise from thermal desorption of ices on grain mantles when the surrounding core is heated to temperatures $\gtrsim 100$~K \citep[see, e.g.,][]{Jorgensen20}.
The rotational temperature of \ce{CH3OH} in the outer region of HD 100546 is more consistent with that measured for the disk around the T~Tauri star, TW Hya.  
The disk around TW~Hya is colder than than around HD~100546, with disk models fitted to the observations suggesting the absence of a CO snowline in the disk around HD~100546 \citep{Kama16}.
Models have predicted a small amount of CO freeze-out in the dust ring of HD~100546 (a factor of $\sim10-100$ depletion), however, this is not predicted to contribute significantly to the production of \ce{CH3OH} due to the low dust mass of this source \citep[]{Leemker24}. 
However, the similar $T_\mathrm{rot}$ for \ce{CH3OH} for the disk-integrated emission around TW~Hya and the outer component of emission in HD~100546 indicates a common chemical origin for gas-phase methanol that is likely driven by non-thermal desorption from icy grains.
\clearpage
\section{Chemical Modelling}
\begin{figure*}[!htp]
    \centering
    \includegraphics[width=0.4\textwidth]{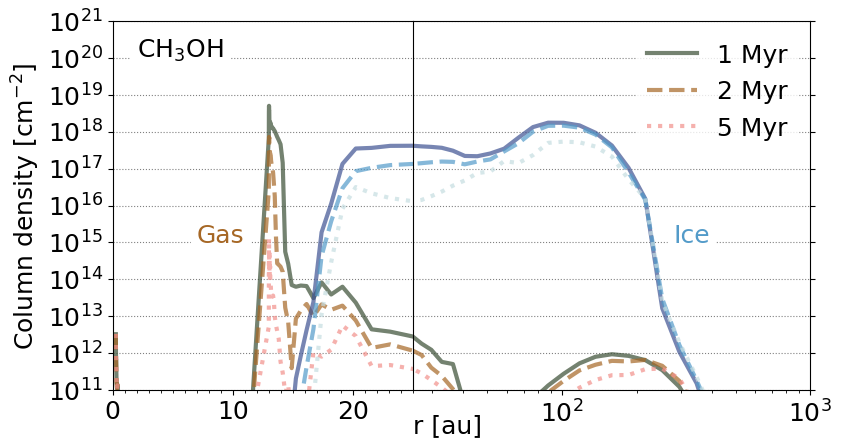}
    \includegraphics[width=0.4\textwidth]{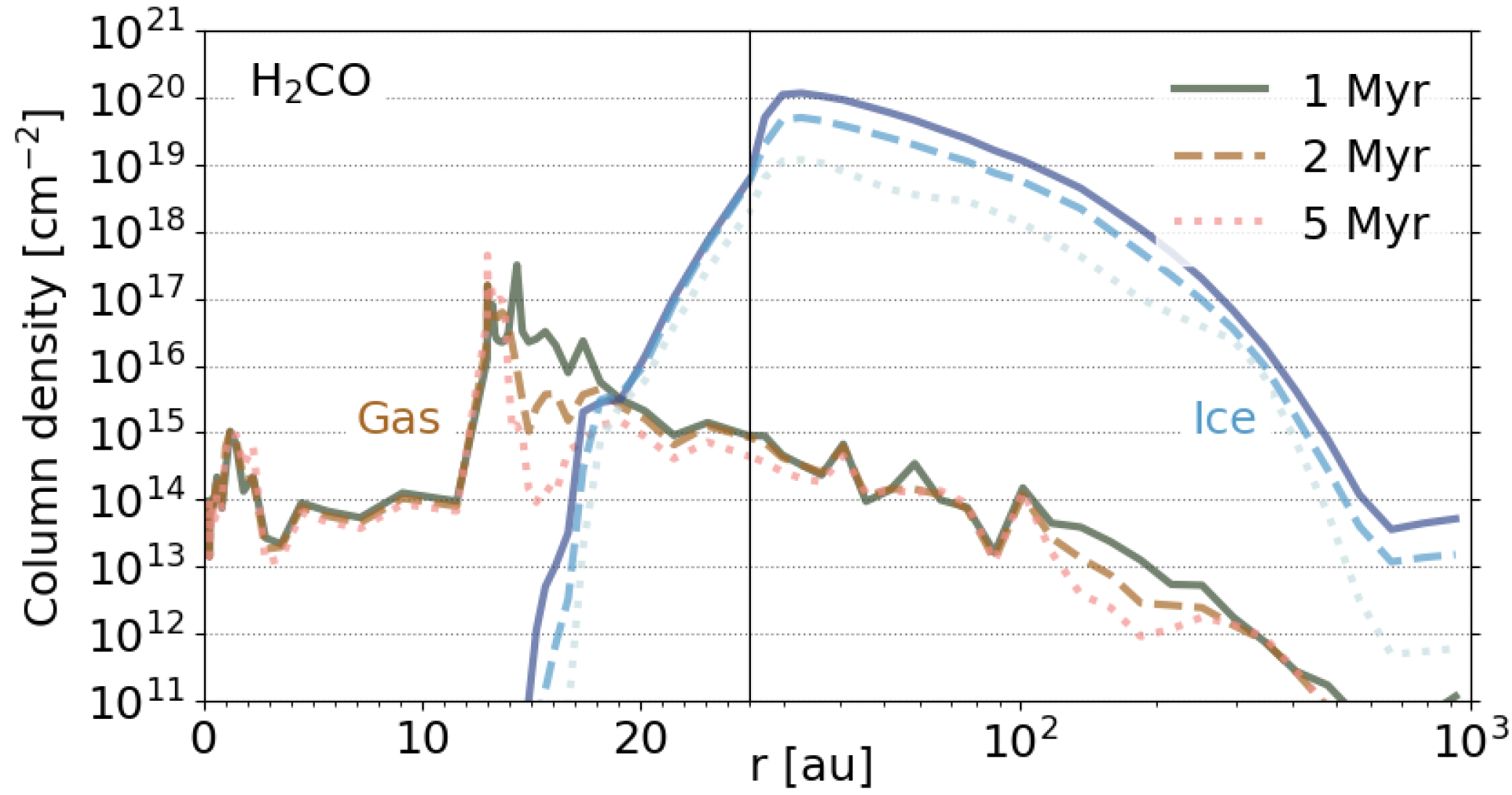}
    \includegraphics[width=0.4\textwidth]{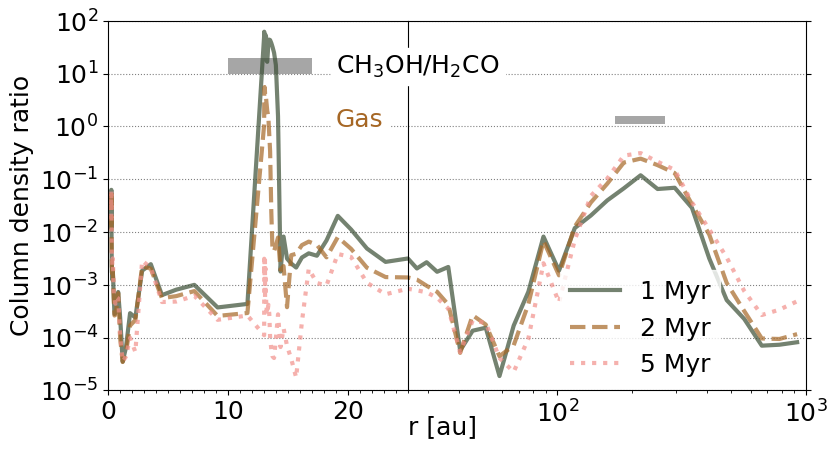}
    \includegraphics[width=0.4\textwidth]{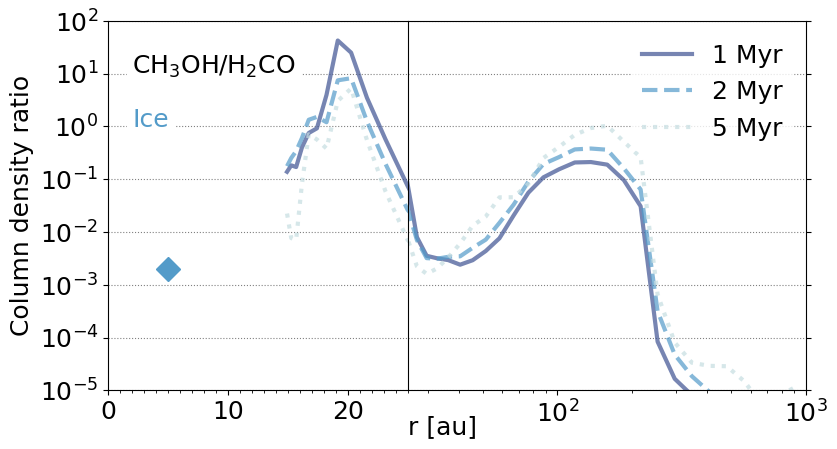}
    \caption{Top row: Column density calculated as a function of radius for the disk around HD~100546 for \ce{CH3OH} (left panel) and \ce{H2CO} (right panel) at 1 (solid), 2 (dashed) and 5 (dotted) Myr. Bottom row: \ce{CH3OH}/\ce{H2CO} column density ratio calculated for the gas (left panel) and ice (right panel) phase, at 1 (solid), 2 (dashed) and 5 (dotted) Myr. The brown colours denote gas phase molecules while the blue colours represent the ice phase. Note that the \ce{CH3OH}/\ce{H2CO} models plotted for the ice phase (bottom right panel) are restricted to radii $\ge 15$ au as the column density in the ice phase is negligible within this radius. The models within 25~au have been plotted on a linear scale to highlight the structure in the inner disk.
    The grey shaded regions on the bottom-left panel indicate the observed ratios derived in this work for the inner ($14.6^{+5.2}_{-4.6}$) and outer ($1.3^{+0.3}_{-0.2}$) components (note that the radial extent of the shaded regions is arbitrary and is set to guide the eye only). The diamond marker on the bottom-right panel indicates the initial abundance ratio for \ce{CH3OH}/\ce{H2CO} ice adopted in the chemical model ($2\times10^{-3}$).}
    \label{fig.modellingratiofull}
\end{figure*}

The gas-grain chemical model utilised to supplement our results couples a 2D physical abundance model for HD~100546, first published in \citet{Kama16}, with a reaction network based on the UMIST Database for Astrochemistry \citep[]{McElroy2013}, specifically described in \citet{Walsh2010,Walsh2012,Walsh2013,Walsh2014,Walsh15} and first implemented into the code in \citet{Drozdovskaya2014,Drozdovskaya2015}. The surface chemistry network for \ce{CH3OH} in particular has since been extended \citep[]{Chuang16} and the chemistry has been updated \citep[refer to][]{Walsh2018} to include fragmentation as a result of non-thermal photodesorption.The photodesorption yield of \ce{CH3OH} is 1.5$\times$10$^{-5}$ photon$^{-1}$, which takes into account the fragmentation upon the photodesorption of pure \ce{CH3OH} ice \citep[]{Bertin2016}, while the photodesorption yield assumed for \ce{H2CO} is 10$^{-3}$~photon$^{-1}$. In total, the reaction network consists of 709 species and 9441 reactions \citep[]{Booth2021}. The accretion and desorption rates \citep[]{Tielens1982} and the grain surface reaction rates \citep[]{Hasegawa1992,Garrod2008,Garrod2011} are taken from previous literature. The cosmic-ray ionisation rate (5$\times$10$^{-17}$~s$^{-1}$) along with gas and dust temperature are taken from \citet{Kama16}, while the gas-to-dust mass ratio is assumed to be 100 and the dust grains to have a radius of 0.1~$\mathrm{\mu}$m \citep[]{Booth2021}. Initial abundances are taken from from a single-point dark-cloud chemical model run for 1 Myr from atomic initial conditions, with n$_\mathrm{H}=$2$\times$10$^4$~cm$^{-3}$ and cosmic-ray ionisation rate$=$10$^{-17}$~s$^{-1}$ in order to simulate inheritance of ice from an earlier, colder evolutionary phase. The resultant initial fractional abundances of relevant species in the ice phase relative to H nuclei, as reported in \citet{Booth2021}, are listed in Table \ref{tab.abundances}. We extract the abundances at 1, 2 and 5 Myr and plot the resultant column densities and abundance ratio predictions in Figs.~\ref{fig.modellingratio} (radially binned) and \ref{fig.modellingratiofull} (original).
\begin{table}[!htp]
    \centering
    \caption{Initial fractional abundances of relevant species in the ice phase relative to H nuclei.}
    \begin{tabular}{c c}
       \hline
       Species & Initial Fractional Abundance \\
       \hline
       \ce{CH3OH} & 2.9$\times$10$^{-8}$ \\
       \ce{H2O} & 1.5$\times$10$^{-4}$ \\
       \ce{CO} & 3.6$\times$10$^{-5}$ \\
       \ce{CO2} & 6.7$\times$10$^{-7}$ \\
       \ce{CH4} & 4.2$\times$10$^{-6}$ \\
       \ce{NH3} & 1.4$\times$10$^{-6}$ \\
       \hline
    \end{tabular}
    \label{tab.abundances}
\end{table}
\begin{figure*}[hbt!]
    \centering
    \includegraphics[width=0.4\textwidth]{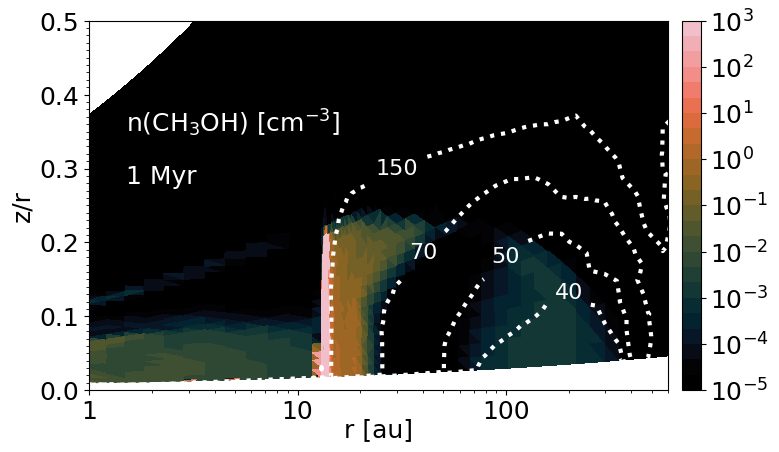}
    \includegraphics[width=0.4\textwidth]{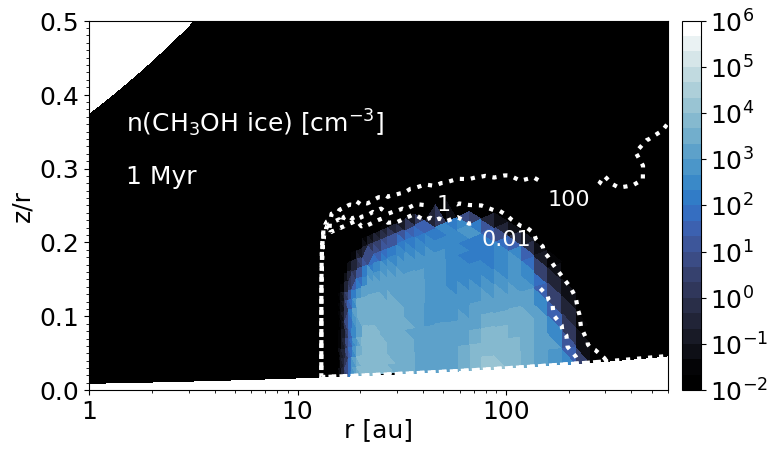}
    \includegraphics[width=0.4\textwidth]{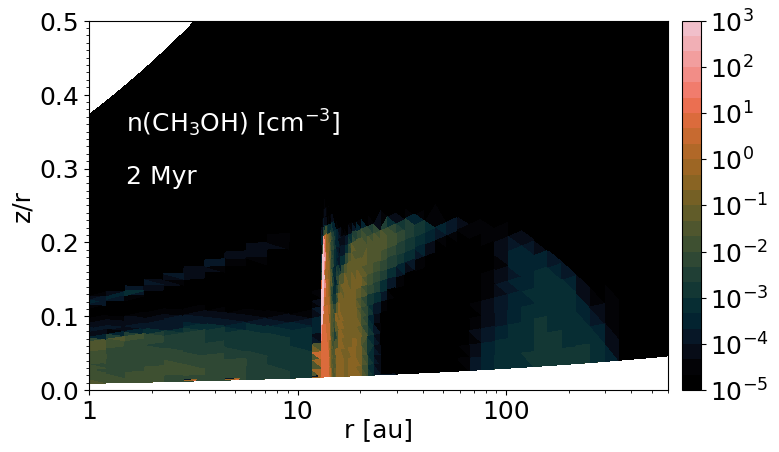}
    \includegraphics[width=0.4\textwidth]{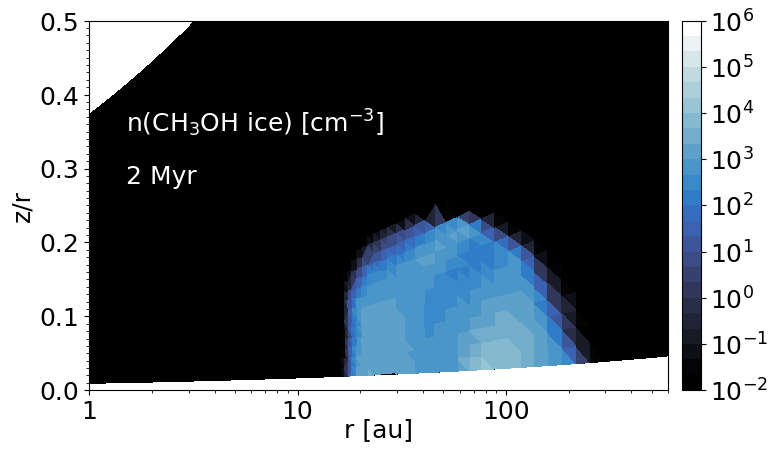}
    \includegraphics[width=0.4\textwidth]{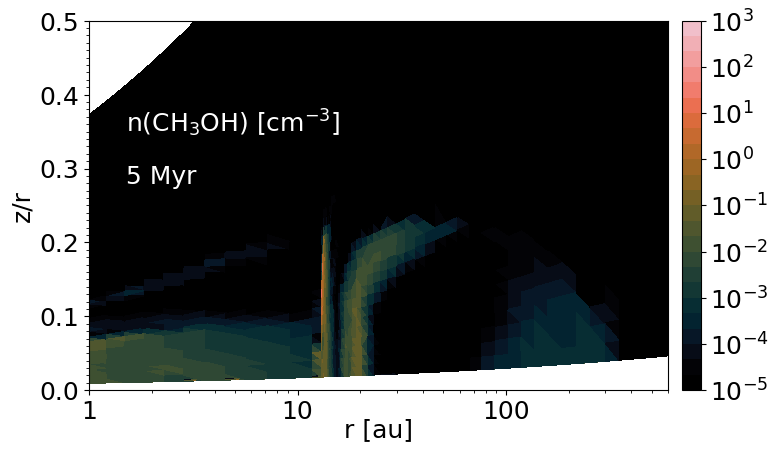}
    \includegraphics[width=0.4\textwidth]{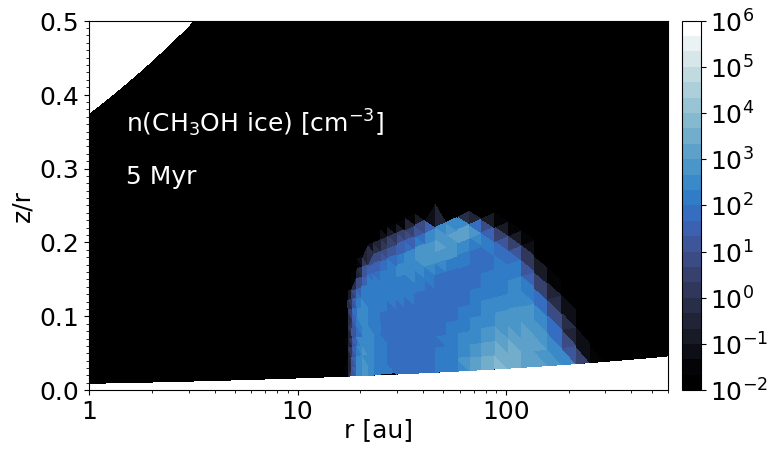}
    \caption{Number density (cm$^{-3}$) for gas-phase (left) and ice-phase (right) \ce{CH3OH} (left) at 1 (top), 2 (middle) and 5 (bottom) Myr as a function of disk radius, $r$ and height divided by the radius, $z/r$.
    The contours in the top-left panels show the gas temperature at 40, 50, 70 and 150~K.
    The contours in the top-right panel show the strength of the UV radiation field in the disk at 0.01, 1 and 100 times $G_0$, where $G_0 = 1.6\times 10^{-3}$~erg~cm$^{-2}$~s$^{-1}$ is the strength of the interstellar radiation field.}
    \label{fig.abundance_ch3oh}
\end{figure*}
\begin{figure*}
    \centering
    \includegraphics[width=0.4\textwidth]{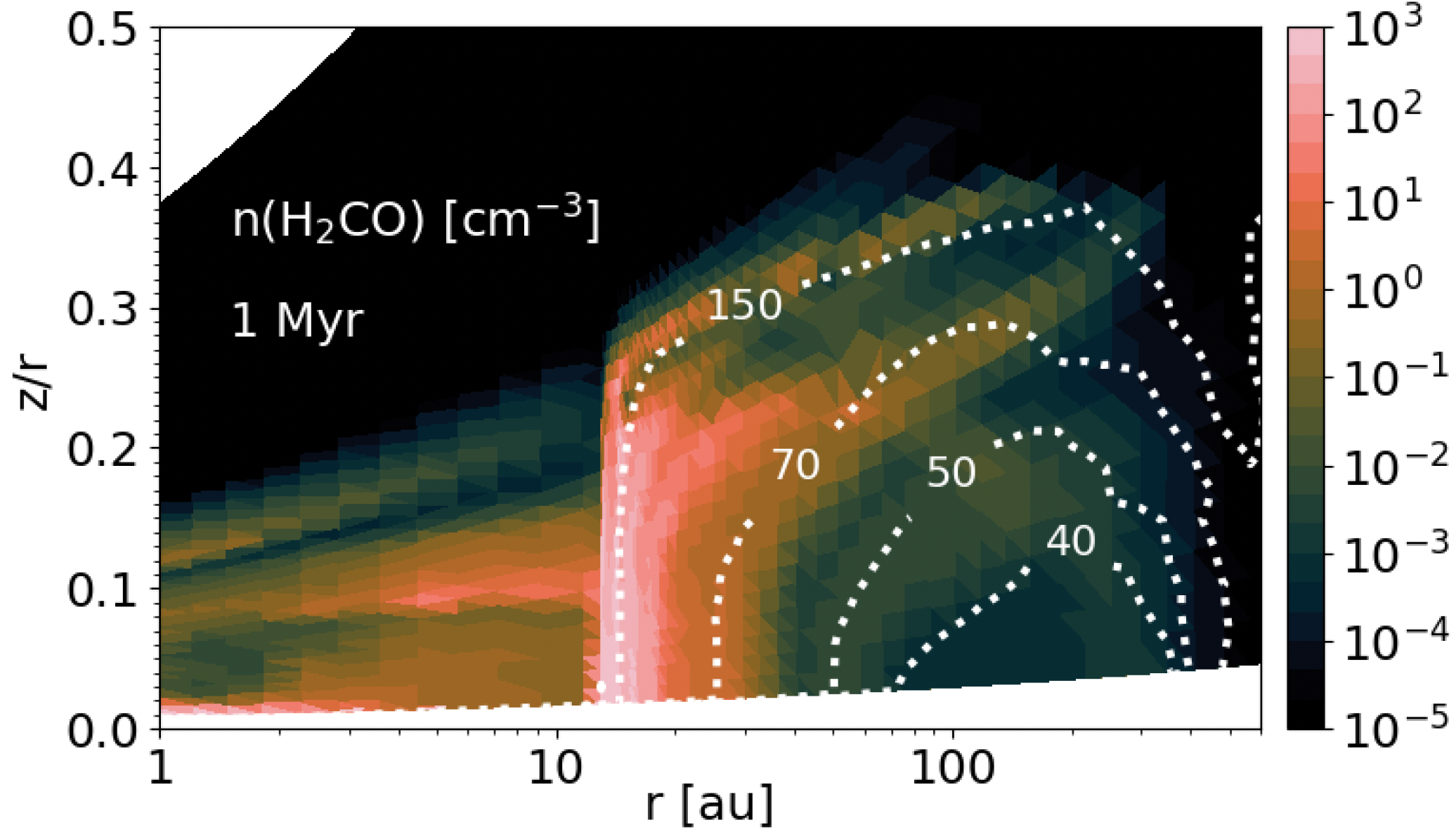}
    \includegraphics[width=0.4\textwidth]{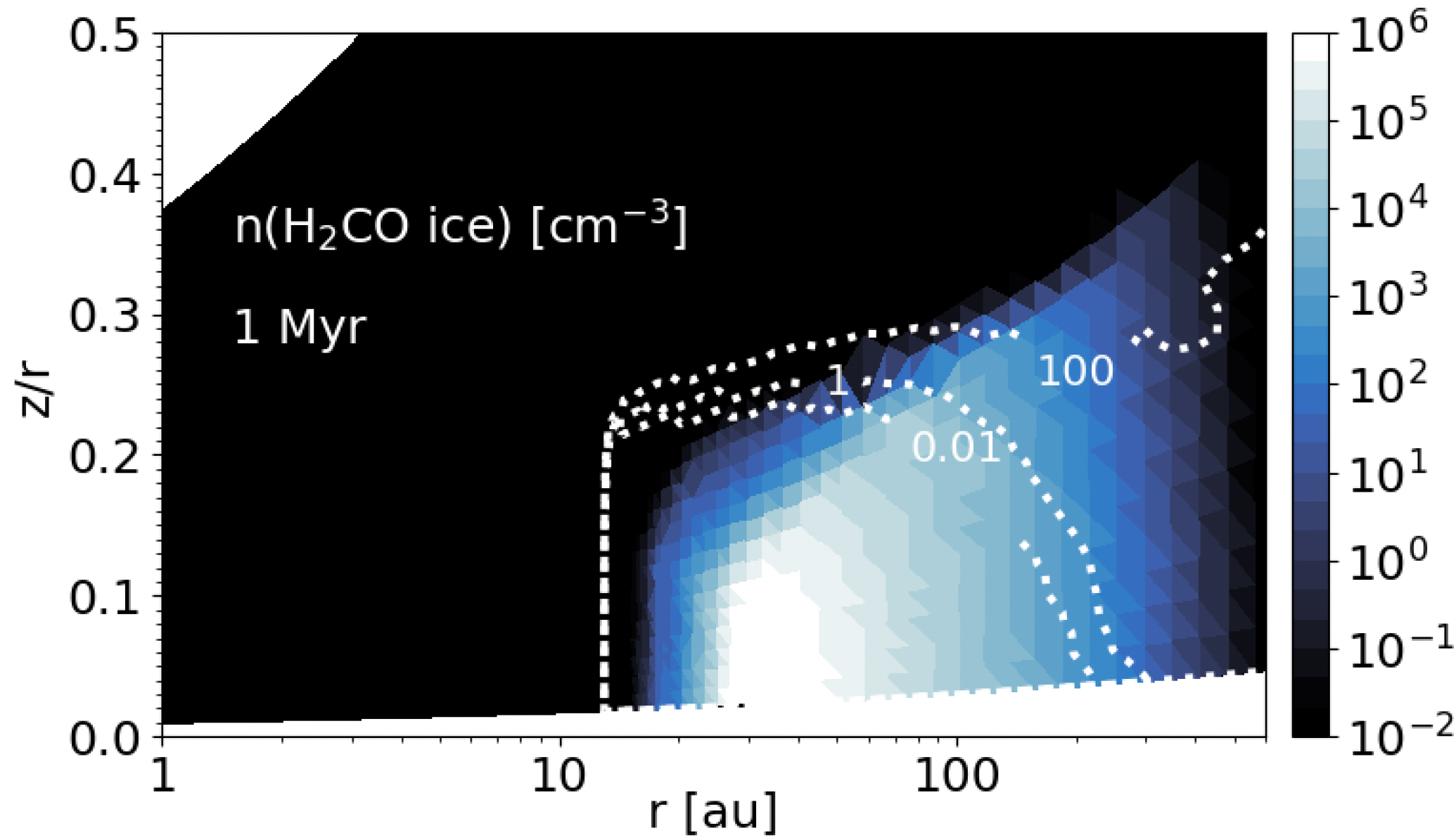}
    \includegraphics[width=0.4\textwidth]{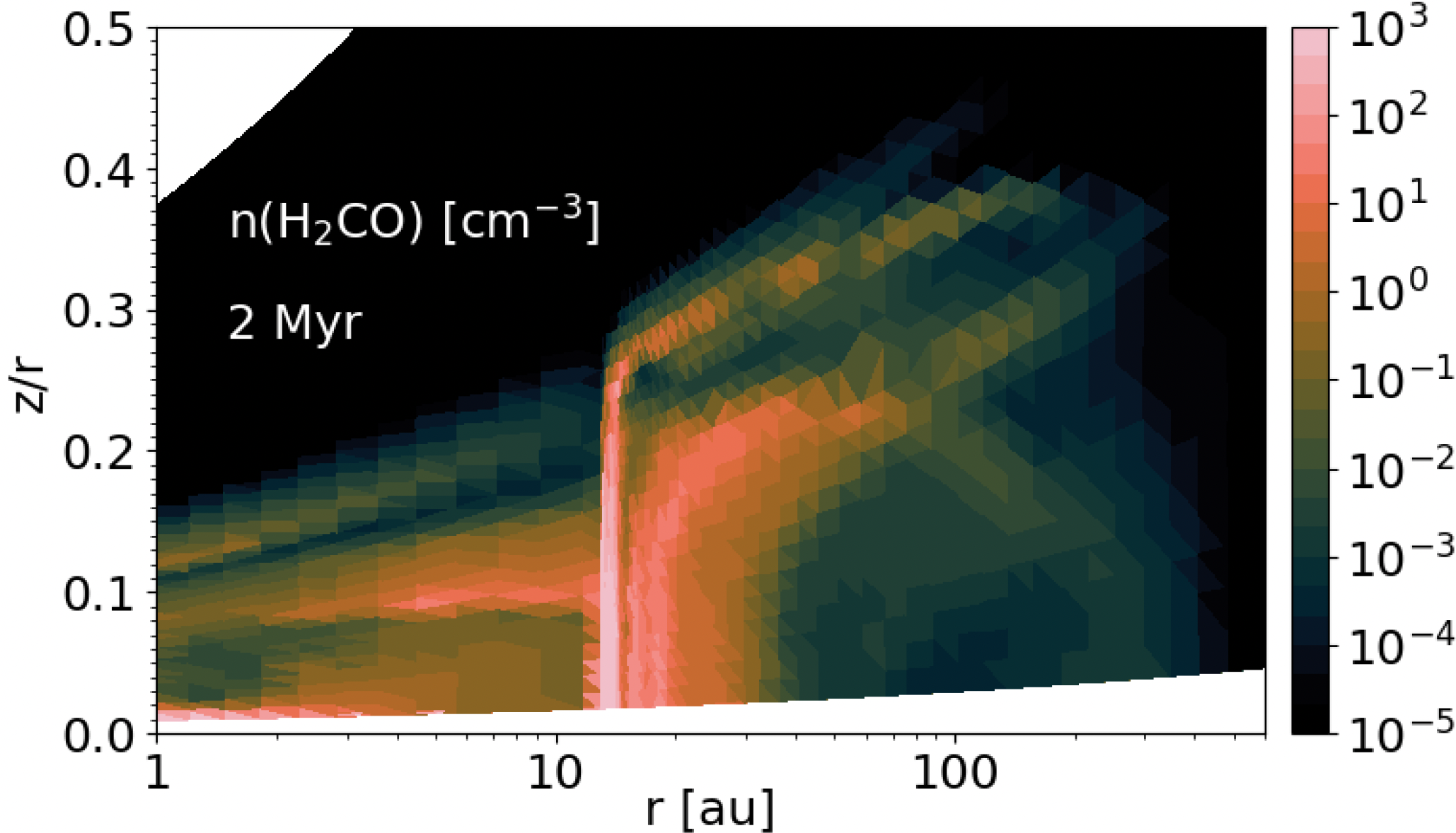}
    \includegraphics[width=0.4\textwidth]{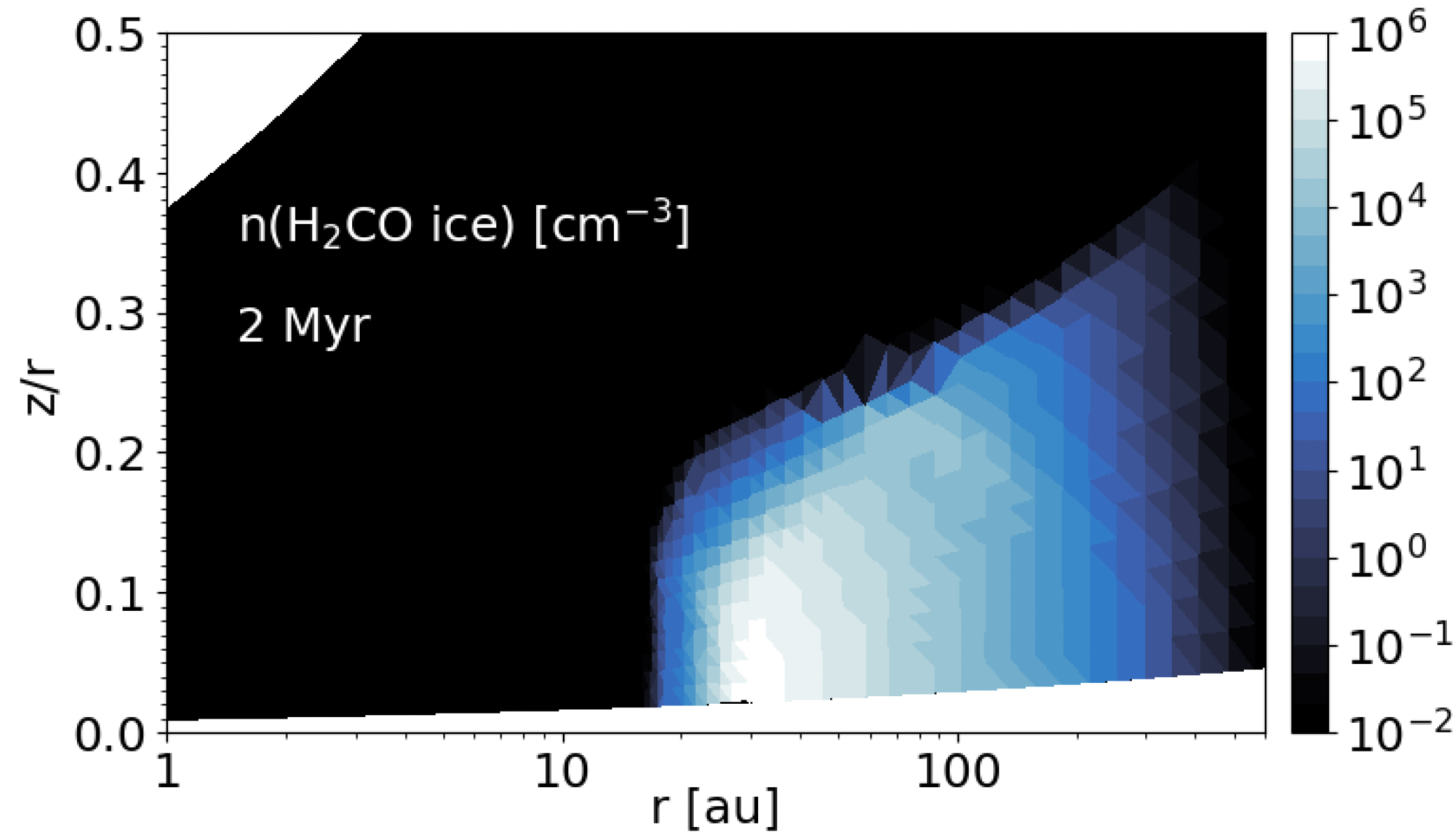}
    \includegraphics[width=0.4\textwidth]{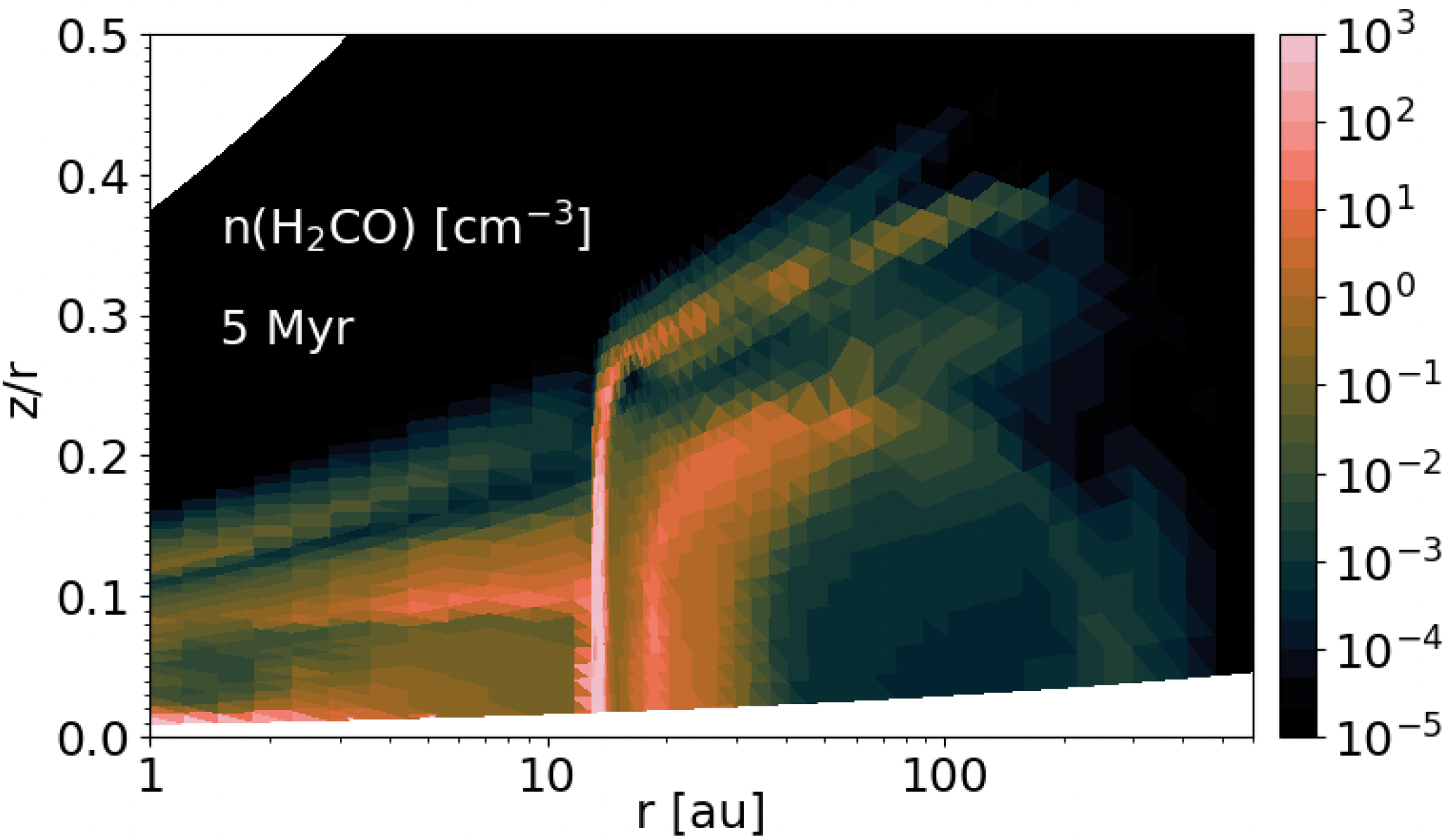}
    \includegraphics[width=0.4\textwidth]{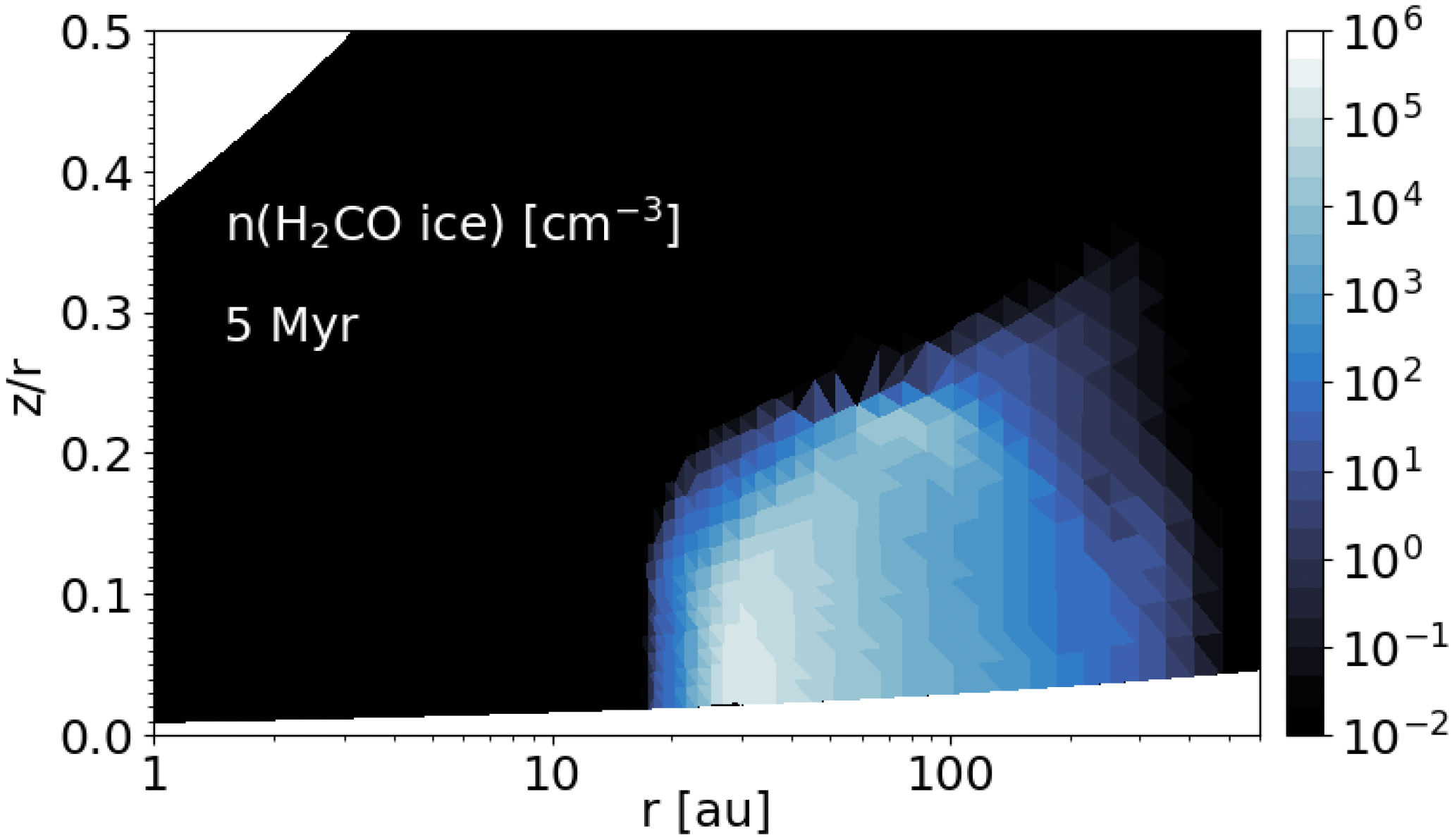}
    \caption{Number density (cm$^{-3}$) for gas-phase (left) and ice-phase (right) \ce{H2CO} (left) at 1 (top), 2 (middle) and 5 (bottom) Myr as a function of disk radius, $r$ and height divided by the radius, $z/r$. 
    The contours in the top-left panels show the gas temperature at 40, 50, 70 and 150~K.
    The contours in the top-right panel show the strength of the UV radiation field in the disk at 0.01, 1 and 100 times $G_0$, where $G_0 = 1.6\times 10^{-3}$~erg~cm$^{-2}$~s$^{-1}$ is the strength of the interstellar radiation field.}
    \label{fig.abundance_h2co}
\end{figure*}

\end{document}